\documentclass[letterpaper]{article}

\usepackage[draft]{aaai25}
\usepackage{times}
\usepackage{helvet}
\usepackage{courier}
\usepackage[hyphens]{url}
\usepackage{graphicx}
\usepackage{natbib}
\usepackage{caption}
\usepackage{booktabs}
\usepackage{tabularx}
\usepackage{longtable}
\usepackage{array}
\usepackage{enumitem}
\usepackage{amsmath}
\title{An Evaluation Framework for National AI Regulation}
\author{
    Kaushik Sanjay Prabhakar,\textsuperscript{\rm 1}
    Tarun Adarsh R S,\textsuperscript{\rm 2}
    Amal Dhivyan Gregory,\textsuperscript{\rm 2}
    Sreeparvathy Sajeev,\textsuperscript{\rm 2}
    Utkarsh Tomar,\textsuperscript{\rm 2}
    Avyay M Casheekar\textsuperscript{\rm 3}
}
\affiliations{
    \textsuperscript{\rm 1}University of Illinois Chicago\\
    \textsuperscript{\rm 2}Vellore Institute of Technology\\
    \textsuperscript{\rm 3}University of Michigan Law School\\
}

\begin{document}

\maketitle
\footnotetext[1]{$^*$Corresponding author: kprab@uic.edu}

\begin{abstract}
Governments use laws, institutions, funding programs and nonbinding guidance to shape how AI is developed and used. Comparing these national approaches is difficult. A binding rule and a detailed voluntary framework can address the same problem but create different duties. The resources needed to carry them out also differ by jurisdiction. This paper develops an evaluation framework for the documented design and implementation readiness of national AI policy. The comparison covers China, India, Japan, Singapore, South Korea, the United Kingdom and the United States. The European Union is included as a supranational comparator. The framework evaluates a versioned portfolio of official instruments rather than one prominent law or strategy. Its criteria ask whether the portfolio governs serious AI risks and whether responsible institutions can implement its commitments. They examine coverage across the AI lifecycle and the protections available to people affected by AI systems. Public benefit and responsible innovation remain a separate part of the assessment. Each sub-criterion is scored through ordered anchors and tied to the provision that supports the judgment. The protocol also records the source search, missing evidence, included instruments and cutoff date. The result is a traceable comparison of policy content that keeps category differences visible. It evaluates what a portfolio provides on paper. It does not estimate enforcement success or policy outcomes.
\end{abstract}

\begin{figure}[ht]
  \centering
  \includegraphics[width=\linewidth]{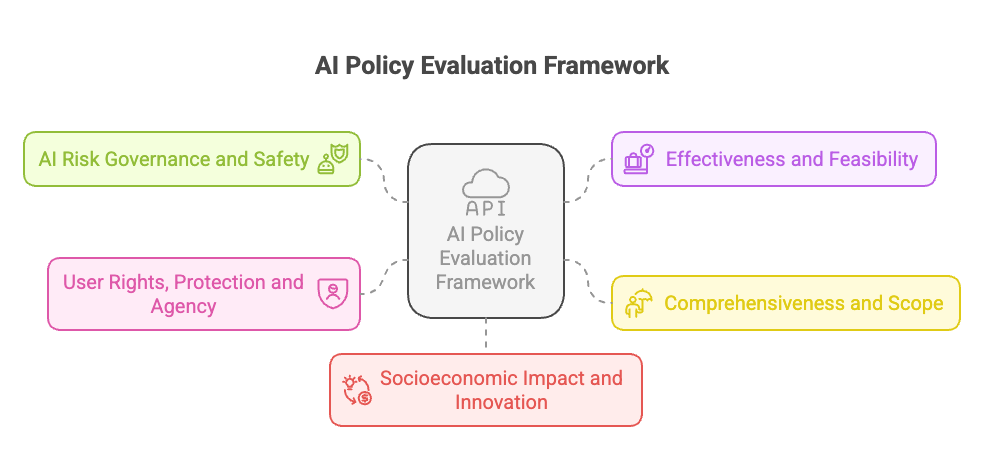}
  \caption{AI Policy Evaluation Framework}
  \label{fig:ai-policy-framework}
\end{figure}

\section{Introduction}
Artificial intelligence is changing how governments carry out ordinary work. Agencies already use systems for language translation and image recognition \cite{mohamed2024impact, li2022imagerecogresearch}. Other systems help detect fraud, process administrative data and allocate public resources \cite{obeng2024utilizing, mlfor-govt-data-analytics, rahman2024utilizing}. AI is also used in public services and in decisions that affect citizens directly \cite{ArtificialintelligenceforthepublicsectorresultsoflandscapingtheuseofAIingovernmentacrosstheEuropeanUnion, Algorithmic}. In healthcare it can support diagnosis and drug development \cite{healthcare}. In hiring it can screen applicants or assist workforce management \cite{TheImpactofArtificialIntelligenceReplacingHumansinMakingHumanResourceManagementDecisionsonFairness:ACaseofResumeScreening}. Transport systems use it for planning and increasingly for automated operation \cite{ArtificialIntelligenceforImprovingPublicTransport}. These uses place AI inside institutions whose mistakes can affect access to services, employment, mobility and individual rights.

Governments pursue AI because it can reduce routine administrative work and help agencies make use of large datasets. It can support policy analysis, improve the delivery of some services and help allocate limited staff or public resources \cite{ArtificialintelligenceforthepublicsectorresultsoflandscapingtheuseofAIingovernmentacrosstheEuropeanUnion, mlfor-govt-data-analytics, rahman2024utilizing}. Natural-language systems can make information easier to access, while forecasting and analytical tools can help officials examine problems that would otherwise be difficult to study at scale \cite{mohamed2024impact}. These are reasons to consider adoption. They are not results that follow automatically from deploying a model. The value of a system depends on the quality of its data, its performance in the intended setting and the process around its use. Officials must also be able to review its outputs and correct a decision when the system is wrong.

Economic conditions add pressure to adopt these technologies. Global corporate investment in AI reached \$252.3 billion in 2024. US private investment reached \$109.1 billion, compared with \$9.3 billion in China and \$4.5 billion in the United Kingdom. Over the same period, the cost of querying a system at a fixed level of performance fell more than 280-fold between late 2022 and late 2024 \cite{maslej2025aiindex}. Lower costs expand the number of organizations able to use capable models. They also shorten the time available for governments to develop procurement rules and oversight arrangements.

The technical frontier has changed as quickly as the economics. DeepSeek released V3 in December 2024 and R1 in January 2025. It reported 2.788 million H800 GPU hours for V3 pretraining, and R1 showed how post-training could produce strong reasoning performance \cite{deepseek2024v3, deepseek2025r1release}. By mid-2026, independent assessments placed leading Chinese systems about three to eight months behind the strongest closed US models. Frontier development in China had also broadened beyond one laboratory to systems from Moonshot AI, Zhipu AI and Alibaba \cite{nist2026caisideepseekv4, csis2026chinesemodels}. Agentic systems add a different problem. A model can act through tools, complete several steps and change an external system before a person reviews the result. Existing policy instruments often do not address the resulting questions about authority, monitoring and intervention directly \cite{staufer2026agentindex}.

The same systems create familiar public-law problems in a new form. A model trained on historical administrative records can reproduce discrimination or direct errors toward groups that already face barriers to public services. A system that uses health or identity records must protect the information it receives. Financial and employment records require the same care. When a model contributes to a consequential decision, the responsible institution must specify who reviews the output and who can override it. The affected person also needs a clear explanation and a way to challenge the result \cite{lim2024determinants, de2020artificialexplain}. Without these arrangements, an efficiency measure can weaken accountability instead of improving public administration.

Implementation poses practical problems even when the policy objective is clear. Agencies need staff who can procure and test an AI system before use. They need people who can monitor it after deployment and decide when it should be changed or retired. Reliable data practices and suitable technical infrastructure are also necessary. Funding must continue after the initial strategy is announced, and regulators with overlapping mandates need a workable way to coordinate. Workforce planning matters beyond technical teams because AI can change the work performed by public employees and by people in regulated industries \cite{OECD_2024}. Traditional rulemaking and procurement often move more slowly than the technology they are expected to govern.

Governments have responded with different policy instruments. Public investment and public-private partnerships can fund infrastructure or research that no single institution would undertake alone \cite{mikhaylov2018artificial}. Regulatory sandboxes allow a system to be tested in a controlled setting before wider deployment \cite{truby2022sandbox}. Technical standards and risk-management frameworks can state what evidence an organization should produce and how it should manage known risks \cite{cihon2019standards, nist_ai_rm_framework}. Procurement rules can make those expectations a condition of selling to the state. Tax policy, grants and training programs can encourage adoption or direct development toward public needs \cite{fischer2021ai-levers}.

These instruments cannot be treated as interchangeable. A statute can create a duty but leave its operation to later rules. A strategy can allocate resources without giving an affected person an enforceable right. Voluntary guidance can explain a practice in more detail than binding legislation, yet create no penalty when an organization ignores it. Institutional mandates and budgets determine whether any of these commitments can be carried out. National AI governance is therefore better understood as a policy portfolio than as one prominent law or strategy. Comparing only the best-known document can hide both the strengths and the gaps in the wider approach.

International coordination has not produced a single model. The Bletchley Declaration focused attention on frontier-system risk. The Seoul process added voluntary company commitments and cooperation among public institutions \cite{gov_uk2023bletchley, gov_uk2024seoul}. The United States and the United Kingdom did not sign the declaration issued at the 2025 Paris AI Action Summit \cite{techcrunch2025paris}. These events do not divide national approaches into two simple camps. They show that a comparative method must allow for different combinations of legal duties, voluntary practice and development policy.

The following sections develop such a method. Section~\ref{sec:background} first examines the technical and institutional setting of eight focal jurisdictions, then compares their policy portfolios with the units used by existing legal studies and readiness indices. Section~\ref{sec:eval-framework} defines a versioned national policy portfolio and explains how its provisions are scored. The resulting framework is designed to show what a policy provides on paper, how clearly it provides it and which evidence supports the judgment. It does not treat a documented commitment as proof of compliance or social benefit.

\section{Background and Related Work}
\label{sec:background}

\subsection{Cross-National Comparison of AI Capabilities}\label{sec:capabilities_context}

\subsubsection{Metrics for Comparison.}

\begin{figure}[ht]
  \centering
  \includegraphics[width=\linewidth]{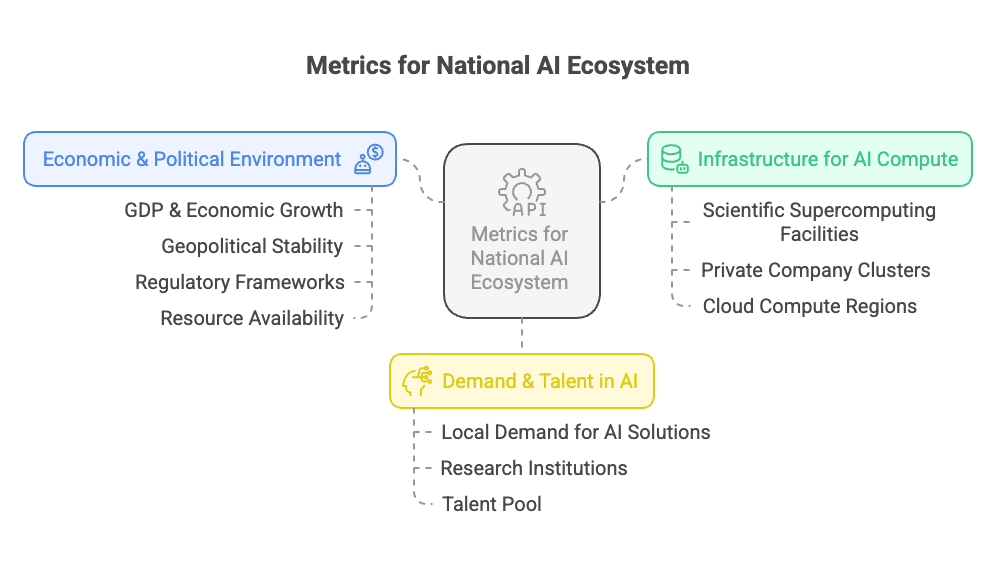}
  \caption{Components of a national AI ecosystem}
  \label{fig:metrics-ai-ecosystem}
\end{figure}

National capability affects the resources available for AI policy and the setting in which that policy operates. A government with substantial public compute can support independent testing or domestic research directly. A government that depends on imported hardware and foreign cloud providers faces different constraints. Domestic semiconductor firms and an experienced technical workforce can also change what the state is able to build, inspect or regulate. None of these conditions determines the quality of regulation. They show which commitments are feasible and where external dependence can hinder implementation.

The focal jurisdictions are China, India, Japan, Singapore, South Korea, the United Kingdom and the United States. The European Union is included as a supranational comparator. Together they provide marked differences in technical capacity and economic conditions. They also differ in talent and regulatory approach \cite{yolanda, sou}. Figure~\ref{fig:metrics-ai-ecosystem} organizes this context around compute infrastructure, the economic and political environment and domestic demand for AI expertise. These dimensions do not form a second index and do not contribute to the policy score. They help explain why the same policy commitment can be easier to carry out in one jurisdiction than another.

\textbf{Infrastructure for AI Compute.} Computing capacity is divided into three forms because each provides a different kind of access. A national supercomputer is not equivalent to a private training cluster. Neither is equivalent to a commercial cloud region.
\begin{itemize}
    \item \textit{Scientific Supercomputing Facilities.} Public agencies, universities and national laboratories operate high-performance computing centers. These facilities can make expensive scientific workloads available to researchers or public bodies that could not fund them alone. Peak performance and accelerator counts describe the machine. Access rules show who can use it. A place in a published ranking does not establish that the system is suitable for model training or open to outside researchers.
    \item \textit{Private Company Clusters.} Companies operate these facilities for model development and commercial services. A large cluster can support frontier research and deployment at scale. Its location can also affect domestic expertise and energy demand. Public sources rarely disclose a complete inventory or show how fully the hardware is used. The comparison therefore reports documented facilities and commitments rather than presenting them as national totals.
    \item \textit{Cloud Compute Regions.} A cloud region is a local group of data centers through which a provider offers computing services. Local service can reduce latency and help an organization meet a data-residency rule. It also lets smaller users obtain computing resources without building a facility. The presence of a region does not show the price or scale of available AI accelerators. A general cloud region should not be treated as a large accelerator cluster.
\end{itemize}

\textbf{Economic and Political Environment.} Economic capacity and external relations shape national AI development. The comparison records these conditions as context. It does not assume that a large economy has better policy or reduce political conditions to a single stability rank.

\begin{itemize}
    \item \textit{GDP and Economic Growth.} GDP shows the scale of an economy, while growth records its recent direction. A larger fiscal and industrial base can support public infrastructure or research funding. It can also fund regulatory staff and transition assistance. These measures do not show how much a government spends on AI or whether that spending produces sound policy.

    \item \textit{Political and External Context.} Trade policy and export controls can determine whether firms can obtain advanced semiconductors. International relations can also affect foreign investment and research partnerships. The field reports these concrete constraints together with relevant national policy. The reviewed sources do not support one comparable stability measure across all jurisdictions.

    \item \textit{Regulatory Frameworks.} AI-specific instruments often operate beside general law and sectoral regulation. Executive policy can add standards or voluntary guidance. Legal force is recorded separately because a detailed recommendation and an enforceable duty can describe the same conduct while creating different obligations.

    \item \textit{Resource Availability.} AI development depends on electricity and semiconductors. It also requires suitable data centers and network infrastructure. These physical inputs affect cost and scale. Their presence does not prove practical access because prices, trade restrictions or competing demand can still limit use.

\end{itemize}

\textbf{Demand and Talent in AI.} Domestic demand affects which systems are developed and where pressure for adoption arises. Research institutions and the available workforce determine whether those systems can be built and tested locally. They also affect governance because the state needs its own technical and domain expertise.

\begin{itemize}
    \item \textit{Local Demand for AI Solutions.} Adoption and procurement show where organizations are already using AI. Investment provides another signal of expected demand. The field considers major private sectors as well as public administration. These signals can reveal pressure for clear rules, but they do not form a comparable market-size estimate unless the sources use the same definitions and period.
    \item \textit{Research Institutions.} Universities and public laboratories can train specialists or provide shared infrastructure. Research networks can also support testing and independent expertise. Listing these institutions does not by itself measure the quality of their research or public access to it.

    \item \textit{Talent Pool.} Workforce estimates and education programs indicate the available supply of AI expertise. Migration and reported shortages show where that supply is under pressure. The relevant pool extends beyond researchers and engineers. Officials, lawyers, auditors and domain specialists are also needed to govern deployed systems. Sources define AI occupations differently, so their estimates are not combined into one ranking.

\end{itemize}

\subsubsection{Analysis of Selected Jurisdictions.}
The broader inventory in Appendix~\ref{app:computeinfra} covers thirteen jurisdictions. Eight were selected for the policy comparison. The selection was purposive because the framework needed to be examined against different levels of technical capacity and different forms of governance. The United States and China have large research and commercial ecosystems, but organize regulatory authority differently. The European Union provides a supranational legal model built around risk classification. The United Kingdom relies on existing sectoral regulators. Japan and Singapore make extensive use of coordination and industry-facing guidance. South Korea combines development policy with a comprehensive AI statute. India joins rapid capability development and digital public infrastructure with data protection law and an evolving AI governance structure. This set is not a representative sample of governments and cannot establish global prevalence.

Only these eight jurisdictions form the policy comparison. Tables~\ref{tab:ai_ecosystem_comparison} and \ref{tab:ai_ecosystem_comparison_continued} summarize the context for that comparison. Policy status claims are current through August 15, 2026. Historical datasets retain their stated reference years. The tables deliberately keep policy context separate from the later score. A jurisdiction is not rewarded in the policy framework merely because it has more compute, a larger economy, or a larger AI workforce.

\renewcommand{\arraystretch}{1.05}
\begin{table*}[tp]
\centering
\footnotesize
\setlength{\tabcolsep}{4pt}
\begin{tabularx}{\textwidth}{|p{1.7cm}|X|X|X|}
\hline
\textbf{Country} & \textbf{Infrastructure for AI Compute} & \textbf{Economic and Political Environment} & \textbf{Demand and Talent in AI} \\
\hline
Singapore & The government committed S\$270 million to expand the National Supercomputing Centre and related training. Commercial providers also operate accelerator capacity and local cloud regions \cite{singaporesuper, singaporecloud}. & The economy grew 4.8\% in 2025. Budget measures committed S\$1 billion over five years to AI and S\$150 million to the Enterprise Compute Initiative. These investments operate alongside nonbinding governance tools \cite{mti2026gdp2025, wong2024budgetai, disg2025eci, singaporeverify}. & The national strategy identifies finance, healthcare, logistics, and public services as areas for wider use. NUS, NTU, and public research bodies support research and training. The strategy targets 15,000 AI practitioners by 2028, and Singapore issued an agentic AI framework in January 2026 \cite{smartnationNationalStrategy, imda2026mgfagentic}. \\
\hline
Japan & ABCI 3.0 became publicly available in January 2025 with 6,128 NVIDIA H200 GPUs and a stated peak of 6.22 EFLOPS at FP16. Fugaku and private or cloud capacity extend the compute base \cite{nvidia2025abci3, arxivABCI30}. & A large economy and sustained technology investment support public and private AI programs. Human-centered principles and sectoral guidance now operate alongside the 2025 AI Promotion Act, which relies on guidance, information requests, and public disclosure rather than private-sector penalties \cite{deloitteJapanEconomic, japan2025aiact, govonline2025aipromotionact}. & Manufacturing, healthcare, and robotics create demand. The University of Tokyo, AIST, and other institutions use shared infrastructure for research, while government and university programs support AI education and training \cite{sphericalinsightsJapanArtificial, nvidiaJapanEnhances}. \\
\hline
India & The PARAM series and National Supercomputing Mission provide public high-performance computing. The IndiaAI compute portal reported more than 38,000 GPUs onboarded by March 2026, alongside commercial cloud regions \cite{indiasupercomp, meity2026indiaaicompute, indiacloudcom}. & Nominal GDP was estimated at about \$4.15 trillion in 2025, with 6.6\% real growth projected for fiscal year 2025--26. The IndiaAI Mission operates alongside the 2023 data-protection statute, 2025 rules, and voluntary AI Governance Guidelines \cite{imf2025indiaarticleiv, meity2025dpdprulesnotified, meity2025aigovernanceguidelines}. & Demand spans healthcare, agriculture, education, and urban services. IITs, IIITs, IISc, public agencies, and research hubs support development and training, while sources continue to report unmet demand for advanced skills \cite{bcgUnlockingAIs, res, kwe, indiaaiIndiasTalent}. \\
\hline
United Kingdom & ARCHER2, CSD3, and Isambard-AI provide scientific and AI computing capacity. London, Cambridge, Oxford, Manchester, and Edinburgh host additional commercial and research clusters \cite{ukriARCHER2, nvidiaNVIDIAAccelerates, bristol2025isambardai, railswareWhereLondon}. & Nominal GDP reached about £3.03 trillion in 2025. The government adopted all 50 recommendations of the AI Opportunities Action Plan and retained a regulator-led approach built around nonstatutory principles, the AI Security Institute, and regulatory sandboxes \cite{ons2026ukgdp2025, dsit2025aiopportunitiesplan, dsit2025aisecurityinstitute, parliament2026aiblueprint}. & Oxford, Cambridge, Imperial College London, University College London, and the Alan Turing Institute support research and government collaboration. UKRI doctoral funding, industry partnerships, and international recruitment form part of the training system \cite{nationalhealthexecutiveReportHighlights, analyticsinsightUniversitiesLeading}. \\
\hline
\end{tabularx}
\caption{Cross-National AI Ecosystem Comparison}
\label{tab:ai_ecosystem_comparison}
\end{table*}

\begin{table*}[tp]
\centering
\footnotesize
\setlength{\tabcolsep}{4pt}
\begin{tabularx}{\textwidth}{|p{1.7cm}|X|X|X|}
\hline
\textbf{Country} & \textbf{Infrastructure for AI Compute} & \textbf{Economic and Political Environment} & \textbf{Demand and Talent in AI} \\
\hline
United States & El Capitan, Frontier, and Aurora ranked second through fourth on the June 2026 TOP500 list. Commercial capacity includes large GPU clusters and custom accelerators from Google, AWS, and Cerebras. The National AI Research Resource pilot provides shared resources for academic and public-interest research \cite{top5002026june, georgetownNAIRRPilot, googlecloud2026ironwood, aws2026trainium3, cerebras2026wse3}. & Nominal GDP was about \$30.5--30.8 trillion in 2025. Executive Order 14179 and the July 2025 AI Action Plan emphasize innovation, infrastructure, and national leadership, while export controls on advanced chips changed several times during 2025 \cite{bea2026gdp2025, whitehouse2025eo14179barriers, whitehouse2025aiactionplan, bis2025aidiffusionrescission}. & MIT, Stanford, CMU, federally funded institutes, and industry laboratories support a large research base. The United States remains a major destination for researchers and engineers, while immigration policy affects recruitment and retention \cite{aiindex2025, ifpStrengtheningAmericas}. \\
\hline
European Union & EuroHPC coordinates shared public supercomputing systems across member states. The InvestAI initiative targets €200 billion in public and private AI investment, including support for large shared facilities \cite{eurohpc, eucommission2025investai, eucommission2025aicontinent}. & GDP was €16.22 trillion in 2024. The AI Act entered into force in 2024, and Regulation 2026/1744 revised parts of its implementation schedule. Collective digital policy and member-state implementation make the EU a supranational comparator rather than a single national environment \cite{eurostat2024gdp, eu2024aiactforce, eu2026omnibusregulation, eucommission2026digitaldecade}. & Universities, national laboratories, and the ELLIS network support research across member states. EU programs support adoption in manufacturing, finance, healthcare, and public services, while private AI investment remains below the US level \cite{europaArtificialIntelligence, europa, stanfordhai2026index}. \\
\hline
South Korea & Samsung and SK Hynix supply memory used in AI systems. Naver, Kakao, domestic cloud providers, and a government-industry plan for more than 260,000 GPUs support public and private compute \cite{skscientific, skprivate, nvidia2025koreainfrastructure}. & Nominal GDP was about \$1.86--1.87 trillion in 2025. The AI Basic Act took effect on January 22, 2026, and distinguishes high-impact AI by sector and consequence from high-performance AI defined partly through a $10^{26}$ operation threshold \cite{imf2026gdpkorea, korea2025aibasicact, jon2026korea}. & Healthcare, finance, manufacturing, urban planning, and public services create demand. KAIST, Seoul National University, and industrial laboratories support research, patents, education, and workforce development \cite{georgetownAssessingSouth}. \\
\hline
China & LineShine ranked first on the June 2026 TOP500 list at 2.198 exaFLOPS. Huawei's Atlas 900 A3 SuperPoD connects up to 384 Ascend chips, while domestic cloud providers and the East Data West Computing program support a distributed compute base \cite{top5002026lineshine, huawei2025superpod, alibabacloud2026regions, tencentcloud2026regions, chinaresource}. & Final revised GDP for 2024 was 134.81 trillion yuan, with reported real growth of 5.0\%. The August 2025 AI Plus initiative, synthetic-content rules, and AI-specific amendments to the Cybersecurity Law supplement existing platform and data regulation \cite{nbs2024gdpfinal, chinagdp, chinastatecouncil2025aiplus, cac2025aigclabeling, chinanpc2025cybersecurityamendment}. & Universities, public institutes, and laboratories at DeepSeek, Moonshot AI, Zhipu AI, Alibaba, and other firms support a large research base. China produced 23.2\% of AI publications in 2023, while independent evaluations placed its leading systems months behind the strongest closed US systems in 2026 \cite{stanfordhai2025chinashare, nist2026caisideepseekv4, csis2026chinesemodels}. \\
\hline
\end{tabularx}
\caption{Cross-National AI Ecosystem Comparison continued}
\label{tab:ai_ecosystem_comparison_continued}
\end{table*}
\setlength{\tabcolsep}{6pt}
\renewcommand{\arraystretch}{1.0}

\subsection{Cross-National Discussion of AI Policy Approaches}
\label{sec:policy_approaches}

The eight portfolios differ in legal force and institutional structure. They also place different weight on development policy and protective regulation. For this reason, the comparison evaluates a portfolio instead of searching for one national AI law. The discussion below identifies the differences that shape the framework. Appendix~\ref{app:crossnational} provides the fuller instrument-by-instrument comparison.

\textbf{European Union}.
The AI Act entered into force on August 1, 2024 \cite{eu_ai_act_source, eu2024aiactforce}. It prohibits specified practices and places obligations on high-risk systems. Other provisions apply to systems subject to transparency duties and to general-purpose AI models. The duty depends on the system and on the actor's role. A provider of a high-risk system must establish risk management and data-governance arrangements. It must also document the system, provide for human oversight and monitor performance after market entry. General-purpose AI provisions require documentation and information for downstream providers. They also address copyright compliance, with additional duties for models that present systemic risk.

Implementation is phased. Prohibited-practice and AI-literacy provisions applied from February 2025. Governance and general-purpose AI provisions applied from August 2025, alongside a voluntary Code of Practice \cite{euaioffice2024, eu2025codeofpractice}. Regulation 2026/1744 later revised the schedule for several high-risk obligations \cite{euconsilium2026omnibusagreement, eu2026omnibusregulation}. The portfolio now extends beyond the binding Act to standards and implementing work. Guidance and the Code of Practice are meant to help providers demonstrate compliance. Enforcement is divided between the AI Office, the Commission and institutions at member-state level. The model gives rights and risk controls a clear legal basis, but its operation depends on coordination and adequate institutional capacity \cite{EPRS2022}. It also depends on the completion of technical standards. Legal force must therefore be assessed separately from clarity, resources and implementation.

\textbf{United States}.
The United States has no comprehensive federal AI statute. Executive policy and existing agency authority carry much of the federal approach. Technical standards, federal procurement and research programs add further instruments, while state law supplies many binding rules for specific uses. Under the previous administration, Executive Order 14110 directed federal action on AI safety and security. It also addressed civil rights, procurement and government use. The Blueprint for an AI Bill of Rights stated nonbinding principles on safety and discrimination. Its other principles concerned privacy, notice and access to a human alternative \cite{us_executive_order, us_ai_bill_of_rights}.

The federal direction changed in January 2025. Executive Order 14110 was revoked. Executive Order 14179 then directed a new plan focused on US leadership and the removal of barriers to AI development \cite{whitehouse2025eo14148rescissions, whitehouse2025eo14179barriers}. The July 2025 AI Action Plan contains more than 90 actions. They address innovation and infrastructure as well as government adoption and international policy \cite{whitehouse2025aiactionplan}. NIST continues to maintain the voluntary AI Risk Management Framework, while sectoral agencies use powers derived from existing law. States remain an important source of binding rules in particular domains \cite{nist_ai_rm_framework, statescoop2025moratoriumeo14365}. Access to advanced chips has changed as well. The AI Diffusion Rule was rescinded before it took effect, and controls concerning China changed again later in 2025 \cite{bis2025aidiffusionrescission, uscommerce2025h20controls}. A score based on one executive order would miss this dispersion and the change over time. The unit must be a federal portfolio frozen at a stated cutoff. Subnational law should either be excluded or analyzed under a separately declared scope.

\textbf{United Kingdom}.
The United Kingdom retains the regulator-led approach set out in its National AI Strategy and 2023 AI White Paper \cite{uk_ai_strategy_2021, uk_ai_white_paper_2023}. The White Paper asks existing regulators to interpret five cross-sector principles within their current mandates. Safety, security and robustness form the first principle. The next two concern transparency and explainability, then fairness. The last two address accountability and governance, followed by contestability and redress. The approach uses the domain knowledge of bodies such as the Information Commissioner's Office and the Financial Conduct Authority. The Competition and Markets Authority has a separate role within the same model. No single general AI regulator was created.

The government published the AI Opportunities Action Plan on January 13, 2025, and accepted all 50 recommendations \cite{dsit2025aiopportunitiesplan}. The plan focuses on the capacity needed for wider adoption. It addresses compute and data access, together with talent and the ability of the state to use AI. It does not create a comprehensive AI statute. In February 2025, the AI Safety Institute became the AI Security Institute and focused its public mandate on serious security risks \cite{dsit2025aisecurityinstitute}. Later policy continued to emphasize coordination and controlled experimentation \cite{parliament2026aiblueprint}. A regulator-led model can adapt rules to sector-specific harms, but only if the regulators have suitable authority and resources. Consistency across their interpretations also matters. The portfolio therefore makes interagency coordination and the limits of existing mandates central to assessment.

\textbf{Singapore}.
Singapore uses nonbinding frameworks alongside sectoral law. The Model AI Governance Framework gives organizations practices for internal governance and human involvement. It also addresses operations management and communication. AI Verify provides a testing and reporting structure intended to make organizational claims easier to examine \cite{singaporeverify}. Separate frameworks address generative AI and, from January 2026, agentic AI \cite{imda2024mgfgenai, imda2026mgfagentic}.

This approach can update more quickly than legislation and can give firms operational guidance before a consensus exists on binding rules. It also depends heavily on voluntary adoption and does not itself create penalties or a general right of redress. A useful evaluation must therefore distinguish substantive detail from legal force. Singapore can score highly for specifying a practice while receiving a different score on whether that practice is mandatory. Monitoring and enforcement remain separate questions.

\textbf{Japan}.
Japan combines the Society 5.0 strategy with human-centered principles and sectoral guidance. The 2025 AI Promotion Act adds a statutory basis to that approach \cite{japan_society50, japan_ai_strategy, japan2025aiact}. The broader strategy links AI development to demographic pressure and productivity. It also presents AI as a way to improve public services and meet social needs. Human-centered principles supply a normative basis, while ministries and sectoral bodies translate them into guidance for particular settings.

The AI Promotion Act was promulgated on June 4, 2025, and fully entered into force on September 1, 2025. It authorizes national planning and research support. The government can request information, issue guidance and disclose noncompliance. The Act does not establish private-sector fines or criminal penalties \cite{govonline2025aipromotionact}. Japan therefore shows why statutory status and coercive enforcement are different features. Its portfolio also makes socioeconomic objectives relevant to the framework because development and social use are part of the policy's stated purpose.

\textbf{India}.
India combines development programs with digital public infrastructure. Data protection law and voluntary AI guidance add protective elements. The 2018 National Strategy for Artificial Intelligence emphasizes inclusive growth. Its priority areas include healthcare and agriculture, as well as education and mobility. It also addresses smart cities and infrastructure \cite{india_national_ai_strategy}. The IndiaAI Mission funds compute and datasets. Other programs support skills, startups and domestic model development.

The implementing rules for the Digital Personal Data Protection Act were notified in November 2025, with major duties scheduled to apply in phases \cite{india_dpdp_act, meity2025dpdprulesnotified}. The November 2025 AI Governance Guidelines propose a light-touch architecture and new forms of institutional coordination. They also propose an incident database and an AI safety institute \cite{meity2025aigovernanceguidelines}. The portfolio therefore separates binding personal-data duties from broader AI recommendations and capacity-building programs. It also shows why implementation feasibility matters. A provision can be well stated while still depending on institutional authority and technical expertise. Infrastructure and continuing resources must be assessed separately.

\textbf{Republic of Korea}.
The Republic of Korea enacted the AI Basic Act in January 2025, and the Act took effect on January 22, 2026 \cite{korea2025aibasicact, jon2026korea}. The statute combines industrial support and governance obligations. It distinguishes high-impact AI used in specified consequential domains from high-performance AI defined partly through a $10^{26}$ operation threshold. These classifications are not interchangeable and trigger different provisions. The Act also imposes transparency duties for specified generative AI outputs and requires domestic representation for certain foreign providers.

The Personal Information Protection Act supplies rights and duties involving personal data and automated decisions. Nonbinding national ethics standards add principles involving human dignity and public benefit. They also address appropriate use \cite{pipa_korea, nia_ai_ethics}. Korea's portfolio therefore contains a horizontal AI statute and general data-protection law. Development policy and soft-law principles add further provisions. Evaluating the statute alone would miss protections and implementation conditions found elsewhere. Collapsing high-impact and high-performance AI would also misstate the Act's scope.

\textbf{China}.
China governs public-facing AI services through several linked instruments. Industrial strategy operates alongside platform regulation. Content rules also draw on cybersecurity and data law. The 2022 deep-synthesis provisions and 2023 generative AI measures impose duties on providers. They address training data and security assessment, together with user protection and content governance. Labeling duties also apply \cite{cac2022aiguidelines, cac2023generativeai}. The scope of the two instruments differs. Each score should therefore identify the service and the provision that supports it rather than refer generally to Chinese AI regulation.

Joint labeling measures took effect in September 2025 and require explicit and implicit markers for specified synthetic content \cite{cac2025aigclabeling}. The State Council issued the AI Plus initiative in August 2025, and amendments to the Cybersecurity Law added AI-specific provisions from January 2026 \cite{chinastatecouncil2025aiplus, chinanpc2025cybersecurityamendment}. No standalone national AI law had been enacted by the cutoff. The resulting model can add targeted rules quickly and connect AI governance to established cybersecurity and data institutions. It can also make the boundaries of the overall regime harder to see because obligations are distributed across instruments and enforcement information is not equally public across sectors. The portfolio illustrates why source coverage and institutional responsibilities must accompany any score.

\textbf{Emerging Governance Challenges}.
Several developments cut across the national differences. International work on frontier AI risk advanced through the Bletchley Declaration and the Seoul Frontier AI Safety Commitments \cite{gov_uk2023bletchley, gov_uk2024seoul}. The United States and United Kingdom then declined to sign the declaration issued at the 2025 Paris summit \cite{techcrunch2025paris}. National institutions also changed their mandates and terminology. These shifts show that international coordination can produce shared practices without producing one stable national model. The framework therefore considers participation and interoperability as evidence of harmonization. It also considers practical cooperation, rather than treating agreement with one declaration as the only relevant measure.

Governance has also moved toward general-purpose and advanced models. The EU created dedicated general-purpose AI obligations. The United States responded through evaluation and procurement rather than one general model law. Infrastructure policy and export controls form another part of its response. The United Kingdom assigned advanced-system research and evaluation to a specialized institute. These instruments differ in legal form and purpose, but each responds to capabilities that can affect many downstream sectors. Category 1 therefore asks whether the portfolio covers the relevant systems and risks. Credit does not depend on the use of a particular label such as frontier or general-purpose AI.

Agentic systems create a newer coverage problem. Deployed agents can take multi-step actions through tools, yet public safety disclosure remains limited \cite{staufer2026agentindex}. Singapore issued a dedicated government framework for agentic AI in January 2026 \cite{imda2026mgfagentic}. Other portfolios can still govern an agent through general risk duties or sectoral law. Human-oversight and cybersecurity provisions may also apply. The absence of the label therefore does not prove the absence of coverage. The relevant question is whether the portfolio addresses the capability and the operations it permits. This distinction motivates separate criteria for advanced-system safety and lifecycle coverage. It also explains the need to ask whether a policy is reviewed as technical conditions change.

\subsection{Related Work}
\label{sec:related_work}

The preceding comparison shows why an evaluation method must identify what it is evaluating. Some studies describe a national AI ecosystem. Others compare legal models or ask whether a government is ready to adopt AI. Work on responsible AI often includes social conditions and observed implementation. Company rubrics assess organizational commitments. Each unit supports a different conclusion. A country can have strong technical capacity and a thin regulatory portfolio. A statute can state detailed duties while the institution responsible for them lacks staff or authority. A readiness index can identify enabling conditions without showing which legal provision creates a particular obligation.

Comparative legal studies explain how regulatory philosophies and instruments differ across the United States, the United Kingdom, the European Union and China \cite{Related1, chun2024comparative}. One line of work uses an ``art, craft and science'' account to explain how policy develops. More recent comparisons examine the distinct regulatory philosophy of each jurisdiction. These studies clarify the institutional and political choices behind a legal model. Their usual purpose is to organize legal difference, not to provide an operational scoring protocol that another evaluator can apply to the same defined portfolio. They also show why legal design cannot be inferred from a country's technological capacity.

Meta-frameworks and research on actionable AI principles address a related problem \cite{Related4, Related3}. They connect ethical commitments to regulatory process and institutional responsibility. They also explain why a principle needs implementation support and some means of adaptation. This work is mainly conceptual or procedural. It does not always state how an evaluator should score a provision that appears in several instruments. Nor does it necessarily separate a detailed nonbinding recommendation from a less detailed legal duty. Those distinctions become unavoidable in a cross-national score.

Several national assessments use a broader unit than the one adopted here. UNESCO's Readiness Assessment Methodology examines law and social conditions through a national multi-stakeholder process \cite{unesco2023ram}. Economic and educational conditions are included, together with scientific and technical capacity. Its purpose is to support national diagnosis and policy dialogue. That breadth matters because governance depends on more than formal policy. It also means that a UNESCO readiness assessment and a score based on policy text answer different questions.

Large cross-national indices make this difference clearer. The 2026 Global Index on Responsible AI covers 135 countries and jurisdictions. It uses 68,138 data points across 38 indicators and includes evidence about implementation and civil-society participation \cite{adams2026girai}. The AGILE Index covers 40 countries through four pillars, 17 dimensions and 43 indicators. It reports constraints as well as scores \cite{zeng2025agile}. The 2025 Government AI Readiness Index covers 195 governments. It asks whether a government can use AI for public benefit by examining government capacity, the technology sector and data infrastructure \cite{oxfordinsights2025gari}. These projects show that structured cross-national evaluation is feasible. They also reveal differences in state capacity that a document review cannot capture.

The present framework draws a narrower boundary. A readiness measure can give weight to the strength of the technology sector or the supply of skills. It can consider public-sector adoption and the availability of useful data. A responsible AI index can include observed implementation and civil-society evidence. The policy score developed here asks what a declared portfolio provides and how specifically it provides it. Economic capacity is reported as context, not converted into a higher policy score. A readiness ranking is not treated as evidence that an individual legal duty exists. The score also does not infer an outcome from the words of a policy.

Corporate grading rubrics provide a closer methodological analogy. They use tiered criteria to assess frontier safety frameworks and make the basis of a grade visible \cite{alaga2024gradingrubricaisafety, saferai2025evaluating}. This allows a reader to distinguish a specific commitment from a vague statement. Their unit, however, is a company's voluntary safety framework. A national portfolio can include legislation and executive policy. It can also include a strategy, an institutional mandate or a procurement rule. Guidance and a budget commitment can provide further evidence. The same practice can therefore appear as a legal duty in one jurisdiction and as a voluntary recommendation in another.

Domain-specific frameworks provide much of the substantive material needed for a national rubric. Healthcare governance supplies detailed questions about high-stakes use \cite{tehai}. Privacy and security frameworks clarify data and system safeguards \cite{priv}. Research on legal effectiveness identifies features needed for a rule to operate \cite{impli}. Work on system evaluation and organizational readiness contributes further evidence questions \cite{mapping, digital}. These frameworks often provide more depth within their domain than a national comparison can. They do not attempt to connect that depth to the entire national portfolio, where infrastructure and enforcement matter alongside individual rights. Advanced-system risk and economic transition also need to remain visible.

The unresolved problem is therefore methodological, not a lack of governance concepts. A national policy evaluation needs a declared unit so that the evaluator cannot choose only the most favorable document. Each score needs a citation to the provision that supports it. The method must separate the detail of a commitment from its legal status because those attributes can move independently. Category results are also necessary. Otherwise a high total can conceal a weak result for safety, rights or institutional capacity.

Policy change creates another problem. A new law can replace an executive instrument. Later guidance can supply details that were absent when the law was adopted. A budget can fund an institution only after its mandate has been announced. Court and regulatory decisions can change the practical scope of the same text. A score without a cutoff can therefore combine instruments that were never in force at the same time.

Missing evidence must also be handled directly. Governments publish official material in different languages and with different levels of detail. An unsuccessful search is not always evidence that a provision is absent. Treating it as absence would penalize a portfolio because it is harder to access, even when the underlying policy is unknown. The method therefore records the source search and creates a separate not-scorable state. It also reports how much of the rubric could be scored.

National AI policy crosses several fields because its problems do. A principles-based review can miss enforcement and resources. A technical-risk review can miss individual rights or distributional effects. An innovation index can overlook the controls needed for high-consequence systems. The framework in Section~\ref{sec:eval-framework} brings these questions into one instrument but reports their results separately. Appendix~\ref{app:frameworkreview} records the main sources used in the design synthesis and explains how their units differ from the one used here.

\section{Evaluation Framework}\label{sec:eval-framework}

\subsection{Framework Development and Design}
The framework was developed through a purposive design synthesis rather than a systematic review. That choice follows from the research aim. The purpose is to construct and explain an evaluation instrument, not to estimate how often a concept appears in a fixed body of literature. Technical AI safety research was used to identify problems of misuse and control. It also informed the treatment of testing, intervention and longer-term resilience \cite{hendrycks2023overviewcatastrophicairisks, slattery2025airiskrepositorycomprehensive, ji2025aialignmentcomprehensivesurvey}. Existing governance and readiness frameworks supplied concepts concerning rights and institutions. They also informed the treatment of enforcement, participation and policy change. Public instruments from the eight focal jurisdictions showed how these concepts appear in actual laws and strategies. Standards, budgets and official guidance provided further examples.

The synthesis then translated each concern into a question that could be answered from public policy evidence. A general concern about model misuse, for example, is too broad to score. It becomes a question about what testing is required and who is responsible for it. Separate questions ask whether the policy controls access and requires incident reporting. Enforcement is examined on its own. Accountability is treated in the same way. Responsibility allocation is distinct from independent oversight, while liability and redress require their own evidence. This translation is necessary because naming a risk does not show that a government has created an operating response.

Candidate concepts were grouped by the policy problem they addressed, then checked against the eight portfolios. An item was divided when it joined features that can differ in practice. The detail of a rule, for example, can be strong even when its legal force is weak. Items were merged when they required the same evidence or would count one provision twice. A sub-criterion was retained only when it named a distinct feature that an evaluator could identify and justify from the declared sources. Appendix~\ref{app:frameworkreview} records the principal frameworks used in this process.

The design gives sustained attention to serious AI risk without treating safety as the whole of national policy. Category 1 examines advanced systems and misuse. It also addresses testing, human intervention and the resilience of supporting infrastructure. The remaining categories ask whether a policy can be implemented and whether its scope is adequate. They preserve individual rights and the distributional effects of AI as separate concerns. Development policy and responsible innovation therefore remain visible without being allowed to compensate silently for weak safety or rights provisions.

The scoring system supports comparison and diagnosis. It is not intended to create a false sense of precision. Ordered anchors distinguish an absent provision from a general statement. They then distinguish a partial mechanism from a clear operating provision and from a fully specified response. Category and criterion results carry more information than the overall mean. The number remains useful because it forces the evaluator to state which evidence changes the judgment. It also permits the same portfolio to be scored again under the same rule.

Every score must be traceable to a source, and every assessment must state when the portfolio was frozen. A later implementing rule can change the meaning of an earlier result. The creation of an institution or a change in its mandate can do the same. Withdrawal of an instrument also matters. Each such change produces a new portfolio version. The older result remains interpretable because its sources and cutoff are preserved.

The unit of analysis is therefore a versioned national policy portfolio. It includes official instruments that were in force or had been adopted by the cutoff. Depending on the declared scope, the instruments can be laws or regulations. Strategies and executive instruments can also form part of the portfolio, as can official implementation guidance. Institutional mandates and budget commitments are included when they supply evidence for a criterion. The score sheet lists every instrument and gives its date or version. Subnational instruments are excluded unless the assessment declares a different scope. The European Union is treated as a supranational comparator. Capability measures in Section~\ref{sec:capabilities_context} provide context but do not affect the score.

The protocol requires evidence for each judgment and keeps legal force separate from substantive detail. It reports category results so that a single total does not hide where a portfolio is strong or weak. It also supports rescoring when the portfolio changes. These choices make an assessment auditable. They do not establish that the rubric is reliable or valid in every legal and administrative setting. The instrument must still be tested by multiple evaluators. Its sensitivity to weighting should also be examined, and later work must compare the score with evidence about implementation.

\subsection{Framework Structure}
\label{framework-structure}
Table~\ref{tab:framework_criteria_list} shows the framework's five categories and 25 criteria. Appendix~\ref{complete-eval-framework} supplies the sub-criteria and the evidence prompts beneath them. This hierarchy separates a broad policy objective from the provision an evaluator can locate in a portfolio. A category names a main dimension of governance. Its criteria divide that dimension into distinct questions, while the sub-criteria provide the unit of scoring. The listed indicators direct the evidence search. They do not receive separate scores. The default scheme gives equal weight to each category. Section~\ref{scoring-methodology} explains the remaining weights.

\subsubsection{AI Risk Governance and Safety.}
This category evaluates provisions for high-consequence risks from current and advanced AI systems. Safety standards ask whether the portfolio identifies the systems and risks subject to stronger control. The standards should also explain what testing is required and connect that requirement to an accepted method or responsible institution. Red teaming and auditing examine evaluation before and after deployment. The rubric considers whether reviewers have enough independence and access to do the work, then asks whether an adverse finding leads to remediation. Human oversight concerns both authority and technical means. A responsible person must be able to pause or restrict a system, correct its operation or retire it when continued use creates unacceptable risk.

The remaining criteria address the boundaries and longer-term conditions of safe operation. System interfacing covers access control and containment. It also considers connected services, tool use and the security of the surrounding supply chain. Long-term governance examines misuse and concentration. Infrastructure resilience asks whether critical services can continue and whether an institution can track risks that persist beyond one deployment cycle. Sub-criterion 1.2 covers advanced or general-purpose systems regardless of the terminology used. A policy can receive credit without using terms such as AGI or frontier AI when its operative provisions cover the relevant systems and risks. Criterion 5 concerns resilience over longer horizons. Criterion 15 asks a different question about monitoring technical change and updating the policy. Keeping them separate reduces double counting.

\begin{table*}[htbp]

    \centering

    \small 
    \begin{tabularx}{\textwidth}{@{}>{\raggedright\arraybackslash}X >{\raggedright\arraybackslash}X >{\raggedright\arraybackslash}X@{}}

        \toprule
        \multicolumn{3}{@{}l}{\textbf{Category 1. AI Risk Governance and Safety}} \\ \addlinespace
        1. Safety Standards & 2. Red Teaming and Auditing & 3. Human Oversight and Interventions \\
        4. System Interfacing and Delimitation & 5. Long-Term AI Governance and Infrastructure Resilience & \\
        \midrule
        \multicolumn{3}{@{}l}{\textbf{Category 2. Effectiveness and Feasibility}} \\ \addlinespace
        6. Institutional Capacity and Authority & 7. Clarity and Specificity & 8. Enforceability \\
        9. Measurability & 10. Resource Requirements & 11. Incentive Alignment \\
        \midrule
        \multicolumn{3}{@{}l}{\textbf{Category 3. Comprehensiveness and Scope}} \\ \addlinespace
        12. Coverage of AI Lifecycle & 13. Range of Risks Addressed & 14. Stakeholder Inclusion \\
        15. Adaptability to Technological Advancements & 16. International Harmonization & \\
        \midrule
        \multicolumn{3}{@{}l}{\textbf{Category 4. User Rights, Protection and Agency}} \\ \addlinespace
        17. Fairness \& Non-discrimination & 18. Transparency \& Explainability & 19. Privacy \& Data Protection \\
        20. Accountability \& Oversight & & \\
        \midrule
        \multicolumn{3}{@{}l}{\textbf{Category 5. Socioeconomic Impact and Innovation}} \\ \addlinespace
        21. Societal Benefit & 22. Innovation & 23. Economic Growth \\
        24. Social Equity & 25. Public Trust & \\
        \bottomrule
    \end{tabularx}
    \caption{Overview of AI Policy Evaluation Framework Criteria}
    \label{tab:framework_criteria_list}

\end{table*}

\subsubsection{Effectiveness and Feasibility.}
This category assesses the provisions that make implementation possible. A policy can state a strong objective while leaving no institution responsible for carrying it out. Institutional capacity therefore begins with a clear mandate and suitable authority. The responsible body also needs staff, expertise and stable funding. Its independence and ability to coordinate with other bodies are assessed separately. Clarity asks whether regulated actors and affected people can determine the scope of a rule. They should be able to identify the duty and any exception. Enforceability examines the rule's legal status and the means by which a violation can be detected. It also considers corrective action, sanctions and routes for review.

Measurability asks whether implementation can be tracked. A measurable policy states its objective and identifies a suitable baseline. It sets an indicator and reporting period, then assigns a body to collect or publish the result. Feasibility also depends on resources. The policy should account for administrative cost and technical infrastructure, as well as workforce needs and burdens placed on regulated actors. Incentive alignment asks whether tools such as procurement or grants reward compliance and responsible development. Liability and market access can serve a similar function. The word effectiveness refers here to design and implementation readiness before an outcome is observed. A score in this category is not evidence that the policy caused an outcome.

\subsubsection{Comprehensiveness and Scope.}
This category evaluates whether the portfolio leaves a major part of AI development or use outside its coverage without explanation. Lifecycle coverage begins with research and design. It continues through data collection and training, then examines evaluation and deployment. Monitoring after deployment matters as much as incident response and later modification. The final stage is decommissioning. No single instrument must cover the entire lifecycle, but the portfolio should allocate responsibility across it.

The range-of-risks criterion begins with technical failure and deliberate misuse. It then considers discrimination and privacy loss. Security, labor effects and environmental cost require separate attention because they arise through different mechanisms. Stakeholder inclusion asks who can contribute while a policy is made and put into effect. It also considers later review and access to redress, with particular attention to affected groups and expertise outside government or industry. International harmonization concerns standards and cross-border enforcement where national action is insufficient. Research cooperation and technical interoperability can support the same aim. Adaptability asks whether the portfolio monitors technical change and reviews its assumptions. Credit depends on a defined update process, not the use of any particular technical label.

\subsubsection{User Rights, Protection and Agency.}
This category assesses protections for people affected by AI systems and by decisions made with them. Fairness and non-discrimination require a way to assess unequal treatment. The portfolio should also address mitigation and continuing monitoring, then provide a remedy when harm occurs. Transparency is not the same as explainability. A public description of a system serves a different purpose from the information needed to understand a particular decision. The affected person and the regulator often need different information. Privacy and data protection begin with lawful collection and limits on purpose. Security and controls on later sharing are also relevant. Access, correction and deletion rights determine what the individual can do after data have been collected.

Accountability asks who is responsible when work is divided between a developer and a deployer. Vendors and public bodies can have separate duties, while an individual official can retain authority over the decision. The portfolio should also provide independent oversight and a route for complaint or appeal. Liability and meaningful redress are examined rather than assumed. These protections overlap in practice but are scored separately. A portfolio can provide notice without an explanation. It can grant data rights without allowing a person to challenge an automated decision. It can also name a responsible institution without providing an effective remedy.

\subsubsection{Socioeconomic Impact and Innovation.}
This category examines what the portfolio seeks to enable and who is expected to benefit. Societal benefit asks whether a priority use responds to an identified public need. It then considers access and evaluation, together with the mechanism by which public value is expected to arise. Innovation policy can support research or shared infrastructure. Economic policy can support startups and technology transfer, while competition policy affects who can enter the market. These criteria do not reward growth language alone. A higher anchor requires a program and a responsible institution, supported by resources or comparable implementation detail.

Workforce measures include training and support for people whose work changes. Labor protection is a separate concern, as is the ability of a public institution to hire and retain expertise. Social equity asks whether access is distributed across regions and groups. It also considers whether vulnerable groups bear a greater share of the risk. Public trust is assessed through concrete measures such as communication and consultation. Transparency and credible assurance can contribute, but a general promise that the public will accept AI does not. The category scores commitments and implementation provisions documented in the portfolio. It does not score later economic growth or changes in public confidence. Distributional outcomes also require separate evidence.

\subsection{Scoring Methodology and Aggregation.}
\label{scoring-methodology}
An evaluator assigns one score to each sub-criterion. The listed indicators are evidence prompts for that judgment. They are not separate items that receive separate scores. Each score sheet records the cited provision, a short rationale, and any source-search notes.

Before scoring begins, the portfolio and the search protocol are frozen. Official consolidated text is preferred. Implementing guidance is reviewed in the original language or through a documented translation. A secondary source can help locate an instrument, but an available primary source should support the score. The record identifies every document searched and the date of the search. It also states the terms used to locate relevant provisions.

The process begins with an inventory rather than a score. The evaluator first states the jurisdictional level and cutoff. The declared scope explains which types of instrument are included and how subnational law is treated. It also gives the language procedure. Each instrument is then recorded under its official title with its date and legal status. A source link or official publication reference completes the entry. This inventory prevents an assessment from combining superseded and current provisions. It also prevents an evaluator from adding a favorable instrument only after a weak result becomes visible.

The portfolio is then reviewed against every sub-criterion. A keyword search can locate a candidate provision, but the surrounding text determines whether it applies and which actor it binds. Later instruments must be checked for amendment or replacement. The score cites the operative text and explains why it satisfies the selected anchor. When several instruments jointly support the score, the rationale explains what each one contributes. This evidence trail does not remove judgment. It makes the judgment reviewable.

\subsubsection{Sub-criterion Assessment on a 1--5 Scale.}
The five anchors are ordered descriptions. The numbers support aggregation, but they do not establish equal intervals between adjacent anchors.

\begin{enumerate}
    \item \textbf{Absent.} A defined search of the included sources finds no relevant provision.
    \item \textbf{Minimal/Vague.} The portfolio acknowledges the issue but gives no specific commitment or responsible actor. It provides little operating detail.
    \item \textbf{Basic/Partial.} The portfolio includes a concrete provision, but its scope or authority is incomplete. Important operating detail is missing.
    \item \textbf{Substantial/Clear.} The portfolio includes clear provisions that cover the main evidence prompts and identify how the commitment operates.
    \item \textbf{Comprehensive/Specified.} The portfolio covers the evidence prompts in detail. It specifies the responsible actor and procedure, together with scope and review where relevant. Enforcement is stated when the sub-criterion requires it.
\end{enumerate}

The evaluator selects the highest anchor whose description is supported as a whole. One indicator cannot by itself justify a 5, and the indicators are not averaged as independent questions. The term comprehensive describes coverage and specification under the particular sub-criterion. It does not mean that the policy is universally optimal or that implementation will succeed.

If the declared search does not allow the evaluator to determine whether a provision exists, the item is marked \textit{not scorable}. It is not assigned a 1. The item is excluded from its parent mean and the remaining child weights are renormalized. The result states the share of sub-criteria that received a score. A score of 1 is used only when the declared source search supports a finding of absence.

If every child of a criterion or category is not scorable, that parent is also not scorable. The analysis does not report an overall score when an entire category is not scorable.

This distinction matters when official information is incomplete. It also matters when a document is unavailable in a language that can be assessed reliably or when delegated guidance cannot be located. Absence is a finding about the declared source set. Not scorable is a finding about the limits of the assessment. The coverage rate and search notes keep missing evidence from disappearing inside an apparently precise total.

\subsubsection{Weighted Aggregation}
Let $s_{cjk}$ denote the score for sub-criterion $k$ under criterion $j$ in category $c$. The overall score is
\begin{equation}
S = \sum_{c=1}^{5} w_c
    \sum_{j \in J_c} w_{j \mid c}
    \sum_{k \in K_{cj}} w_{k \mid cj}s_{cjk}.
\end{equation}
Weights sum to one at each parent. The default gives every category a weight of 20\%. Criteria receive equal weight within their category, and sub-criteria receive equal weight within their criterion. Category 2 has six criteria. Its table displays each weight as 16.67\%, but the calculation uses the exact value $1/6$. These defaults replace the unspecified weights in Appendix~\ref{complete-eval-framework}.

An analysis can use custom weights only when they are declared before scoring and justified with the research question. The default result must also be reported. Results produced under different weighting or omission rules are not directly comparable.

Equal weighting is a transparent baseline, not a claim that every governance question has identical social value. It prevents an evaluator from tuning the weights after seeing which jurisdiction benefits. A targeted study can assign different weights, but it should also show the result under the default. Sensitivity analysis is especially important when totals are close. It is also needed when missing evidence causes substantial renormalization.

\subsubsection{Final Score Interpretation (1-5 Scale)}
The output reports the overall score and the result for each category. Criterion scores provide a more detailed view. The record also states coverage and includes the instrument list with its cutoff. The total summarizes alignment with the rubric. It is not a causal estimate. A high result for innovation does not erase a low result for safety or rights, and the category scores keep that difference visible.

Comparisons should begin with the category results and evidence notes rather than the total alone. Two portfolios can reach the same overall score through different strengths. One can rely on legal force, while another provides more detailed safety provisions or development policy. Rights and institutional capacity can differ even when the totals match. A small numerical difference should not be treated as meaningful until evaluator agreement is known and the result has been tested against alternative weights and missing-data rules.

When two evaluators score the same portfolio, agreement should be computed before they discuss their differences. Weighted Cohen's kappa is suitable for the ordered scale \cite{cohen1968weightedkappa}. The study must declare whether the calculation uses linear or quadratic weights. More than two evaluators can be assessed through an ordinal form of Krippendorff's alpha \cite{krippendorff2018contentanalysis}. Disagreements can then be resolved, while the record retains the pre-consensus scores and the agreed rationale.

Every update creates a new portfolio version rather than silently replacing the old score. Annual reassessment is a useful default, with an additional update after a major legal or institutional change. This procedure permits comparison over time while preserving the evidence behind each score.

The same protocol can support an expert panel or a commissioned audit. It can also be used in a public research project. In every case, evaluators should score independently before discussion. Consensus can resolve interpretive differences for the final record. The pre-consensus scores remain necessary because they show reliability and identify anchors that require revision.

\section{Discussion and Future Work}
\textbf{Contribution.} Existing work provides comparative legal analysis and measures national readiness. Governance indices examine broad country conditions, while company rubrics grade voluntary commitments. Detailed frameworks also exist for particular sectors. The remaining difficulty is to turn their useful concepts into a traceable comparison of national policy portfolios. The instruments in such a portfolio do not share the same legal status or scope. They also differ in the amount of operating detail they provide. The framework narrows the task to documented policy design and implementation readiness. It asks what the portfolio provides and which source supports that finding. It then asks who is responsible and how the provision is meant to operate.

The category structure preserves differences that a single list of principles can hide. A technical safety obligation is not equivalent to an institution with enforcement authority. An individual right serves another purpose, while an innovation program addresses the development side of the portfolio. Category 1 examines technical risk governance and the resilience needed to sustain it. Category 2 asks whether the institutions named in the policy can act. It also examines authority and resources, together with the incentives and measurement needed for implementation. Category 3 concerns coverage. It tests whether the portfolio follows AI across its lifecycle and responds when the technology changes.

The remaining categories keep the social purpose of policy visible. Category 4 examines the protection and agency of people affected by AI. It requires more than a statement of ethical values because rights need a responsible institution and a route to redress. Category 5 considers public benefit and economic policy. It also asks how workforce change and unequal access are addressed. Reporting these categories separately prevents a strong development program from compensating silently for weak rights or safety provisions.

Narrower frameworks remain useful. A healthcare framework can examine the evidence needed for clinical use in greater detail than a national rubric \cite{tehai}. Work on privacy or market regulation can do the same within its domain \cite{impli}. Those approaches cannot by themselves show whether the national portfolio creates a cross-sector oversight body or gives it adequate resources. They can also omit a general incident-reporting rule or a process for reviewing emerging risk. A broad governance index has the opposite limitation. It can reveal national conditions without identifying the provision that creates a right or duty. The present rubric connects these levels while keeping its claim limited to policy content.

\textbf{Legal force in context.} Sub-criterion 8.4 illustrates why scope and legal status must be recorded. If the declared portfolios are limited to the named AI-specific instruments, the EU AI Act and the Republic of Korea's AI Basic Act are binding primary legislation. Each would receive a 5 for legal status and force. Singapore's Model AI Governance Framework is detailed official guidance, but it is nonbinding. It would receive a 2 on this sub-criterion \cite{eu_ai_act_source, korea2025aibasicact, singaporeverify}. These scores answer one question about legal force. They do not compare substantive quality or practical influence. They say nothing about the enforcement record or the overall portfolio score.

The example also shows why legal force cannot stand in for policy quality. Singaporean guidance can describe an operating practice more clearly than a broadly worded statute. A binding law can create authority while postponing important obligations or leaving details to later standards. A strategy can allocate substantial resources without creating an individual right. Treating all official documents as equivalent would ignore these differences. Treating statutory force as a complete measure would make the opposite error.

\textbf{Using and interpreting the framework.} A complete assessment begins by defining the portfolio. The record states the cutoff and jurisdictional level before any score is assigned. It identifies the source types and translation procedure, then lists the included instruments. Each sub-criterion receives one score supported by cited evidence and a short explanation. Independent evaluators apply the same search and scoring rules before they discuss a disagreement. The final record retains their initial scores and the agreement statistic. It also retains the resolved score and explanation.

Equal weighting provides a common starting point for a general comparison. A narrower research question can justify different weights. A study of AI in public administration, for example, can place greater weight on procurement and institutional authority. Rights and access to redress can also receive more weight in that setting. A study of advanced-model governance can instead emphasize testing and intervention. Misuse and system resilience would receive closer attention. The analysis should still report the default result and disclose the custom weights. Otherwise a reader cannot tell whether a difference comes from the policy or from the priorities built into the calculation.

Interpretation should begin with the category and criterion results. The overall score is convenient, but the supporting evidence lies below it. A low institutional-capacity result can explain why an enforcement provision has no responsible body. Limited resources can also constrain monitoring or prevent a policy from being updated. These relationships matter, but one weak provision should not be counted several times without distinct evidence. The score sheet should support each sub-criterion separately. The written analysis can then explain how the weaknesses interact.

The framework distinguishes several kinds of incompleteness because they imply different responses. A score of 1 means that the declared search supports the absence of a provision. A score of 2 means that the portfolio acknowledges the issue without explaining how the response will operate. A not-scorable entry means that the evidence did not support either conclusion. An absent provision can require a new commitment. A vague provision can require implementing detail. A not-scorable item can require better publication or a reliable translation before any substantive judgment is justified.

Detailed application will take time. A complete portfolio can spread one policy question across legislation and guidance issued by different bodies. Amendments and superseded versions must be checked before the operative text is identified. The evaluator then has to apply a large rubric without treating a keyword match as sufficient evidence. This cost limits rapid assessment, but much of it follows from the breadth of national AI governance rather than from the scoring scale itself. A public instrument inventory and reusable evidence record could reduce later work without replacing expert judgment.

\textbf{Limits of document review.} A detailed rubric reduces unstructured judgment but cannot eliminate interpretation. Evaluators can disagree about whether a provision is sufficiently specific. They may also disagree about whether two instruments jointly satisfy one anchor. A legal term can have a different practical meaning across jurisdictions even when the translation looks similar. Independent scoring and agreement statistics expose these disagreements. Legal and linguistic expertise is especially important when the rubric addresses fairness or accountability. Institutional independence also needs to be understood in its national setting.

The framework favors features that can be documented. An explicit duty is easier to score than an informal practice. A named institution and published procedure leave a clearer evidence trail than coordination that occurs inside government. A stated budget or review period can be assessed in the same way. This preference improves auditability, but it can underrate soft law that strongly shapes conduct. It can also underrate a public institution that works effectively without detailed published rules. The opposite error is possible as well. A carefully drafted law can receive a strong design score even when it is weakly enforced. Effectiveness and feasibility are therefore limited to design and readiness. The category does not measure an outcome.

Information availability creates a related limit. Governments publish official instruments in different forms and languages. Some implementation evidence remains inside agencies or regulated firms. A secondary source can reveal that a missing instrument exists, but using it in place of official text can introduce error. The search protocol and not-scorable state make this problem visible. Coverage reporting shows how much of the rubric rests on evidence. None of these measures can create information that is not public.

The rubric also contains normative choices. It gives weight to advanced-system safety and enforceable rights. Institutional capacity, public participation and equitable access also affect the score. Responsible innovation remains part of the assessment. Another study could reasonably choose different weights or refine an anchor for a particular legal system. A revised rubric should be versioned and justified rather than presented as a neutral correction. It should also be tested across different administrative traditions before small score differences are treated as substantive rankings.

\textbf{Future work.} The immediate empirical task is to apply a frozen version of the rubric to complete portfolios using multiple evaluators. That study should publish the inventory of instruments and the source-search record. Evidence extracts and completed score sheets would allow another evaluator to inspect each judgment. Initial scores should be preserved alongside the agreement statistic and the final consensus result. Sensitivity analysis would show whether the conclusions depend on the default weights or missing-evidence rule. Together these materials would reveal which anchors work consistently and which require revision.

A maintained public implementation could make later assessments easier to audit. It should preserve each version of the rubric and each dated national portfolio. Official sources could be linked to examples that show why a provision satisfies a particular anchor. A change log would explain later revisions. Expert review would remain necessary because a collaborative repository cannot determine legal scope or factual accuracy by vote. Its main value would be traceability and reuse, not automated scoring.

Specialized modules could add depth for foundation models or agentic systems. Public-sector procurement could use a module of its own. Healthcare and finance also raise domain questions that the national core cannot address fully. A module should not change the core score unless that change is declared in advance. Stability in the core is needed for comparison over time, while the modules can respond to applications that require more specific evidence.

Outcome validation is a separate long-term task. Later studies can compare policy-design scores with enforcement actions and compliance records. Incident data and institutional budgets can show whether the stated mechanisms operate. Public complaints and adoption data provide other forms of evidence, while social or economic outcomes answer still broader questions. Such work could test whether the documented features measured here predict implementation or public benefit. It requires longitudinal evidence and a causal design that the present content-based framework does not provide.

\section{Conclusion}
National AI governance cannot be understood through the most prominent law or strategy alone. A binding provision can depend on later guidance and on an institution with enough authority to apply it. A development strategy can supply resources without creating an enforceable right. A detailed voluntary framework can shape practice while leaving compliance to the organization. The comparison of eight jurisdictions shows these elements being combined in different ways.

The European Union and the Republic of Korea place comprehensive AI statutes at the center of their portfolios. The United States distributes federal policy across executive action and existing agencies, while the United Kingdom asks sectoral regulators to apply common principles. Singapore and Japan rely more heavily on guidance and coordination, together with measures that promote adoption. India joins development policy and data protection with an emerging AI governance structure. China's approach is spread across rules for platforms and content, supported by cybersecurity and data law. These differences make the portfolio, rather than one document, the appropriate unit of analysis.

The framework defines that unit as a national portfolio frozen at a stated cutoff. Its 25 criteria ask whether serious risks are governed and whether the stated response can be implemented. They also examine whether the policy's scope is adequate. Separate categories preserve protections for affected people and the development side of AI policy. Every sub-criterion requires a cited provision and an ordered judgment. Category and criterion results show how the final score was reached. The coverage report identifies missing evidence, while the instrument list and cutoff make later rescoring possible.

The result is a method for evaluating documented design and implementation readiness. A strong score does not establish that an institution has enforced the policy or that a regulated organization complies with it. It also does not prove an improvement in safety or individual rights. Innovation and public welfare require outcome evidence of their own. The next empirical step is to apply one frozen version of the rubric to complete portfolios using independent evaluators. Publishing the evidence and initial judgments will allow reliability and sensitivity to be tested. The instrument should then be revised only where application reveals a genuine ambiguity or omission.

\clearpage
\onecolumn
\setlength{\tabcolsep}{3pt}
\small
\sloppy

\appendix
\section{Compute Infrastructure Review}
\label{app:computeinfra}
This thirteen-jurisdiction inventory informed the purposive selection in Section~\ref{sec:capabilities_context}. It provides context only and does not enter the policy score.
\begin{longtable}{|p{0.65cm}|p{1.7cm}|p{4.18cm}|p{4.18cm}|p{4.18cm}|}
\hline
\textbf{No.} & \textbf{Country} & \textbf{Scientific Supercomputing Facilities} & \textbf{Private Company Clusters} & \textbf{Cloud Compute Regions} \\
\hline
\endfirsthead

\multicolumn{5}{c}%
{{\tablename\ \thetable{} -- continued from previous page}} \\
\hline
\textbf{No.} & \textbf{Country} & \textbf{Scientific Supercomputing Facilities} & \textbf{Private Company Clusters} & \textbf{Cloud Compute Regions} \\
\hline
\endhead

\hline \multicolumn{5}{|r|}{{Continued on next page}}
\\ \hline
\endfoot
\caption{Infrastructure for AI Compute by Country}
\label{tab:compute-infrastructure-review}
\endlastfoot

1 & Thailand & The TARA system reports a theoretical peak near 500 teraFLOPS from 4,136 Intel Xeon CPU cores and 28 NVIDIA Tesla V100 GPUs \cite{cmuTARAYear}.
& Siam.AI and other firms have announced private AI infrastructure investments \cite{bangkokpostSiamAIInvest}.
& Alibaba Cloud and other large providers offer cloud services in Thailand \cite{cloudcomputingnewsAlibabaCloud}. \\
\hline

2 & Malaysia & YTL's AI Cloud in Johor uses NVIDIA systems for commercial and research workloads \cite{malaysuper}.
& YTL and cloud providers have announced private data-center and accelerator capacity in Malaysia \cite{malaysuper}.
& AWS and Microsoft have announced or opened Malaysian cloud regions \cite{malaycloud}. \\
\hline

3 & Vietnam & Viettel and research institutions have developed computing facilities for scientific and AI workloads \cite{vietsuper}.
& Vingroup supports AI research and data infrastructure through VinAI and VinBigdata \cite{vietprivate1,vietprivate2}.
& Foreign and domestic providers offer cloud services, while the cited AWS presence was an edge location rather than a full region \cite{vietcloud}. \\
\hline

4 & Philippines & The National Center for AI Research was established to support applied research and public-sector projects \cite{philipsuper}.
& Converge ICT and Supermicro announced data-center infrastructure for AI workloads \cite{philipprivate}.
& The reviewed cloud source describes an expanding domestic data-center market \cite{philipcloud}. \\
\hline

5 & Indonesia & Indosat and NVIDIA announced AI cloud infrastructure using high-density accelerator systems \cite{indonesiasuper1,indonesiasuper2}.
& GoTo and other technology firms operate private infrastructure for consumer and business services \cite{indonesiaprivate}.
& AWS, Google Cloud, and Microsoft operate or have announced cloud regions in Indonesia \cite{indonesiacloud}.\\
\hline

6 & Singapore & The government committed S\$270 million to expand the National Supercomputing Centre and related training \cite{singaporesuper}.
& Commercial providers operate private cloud and accelerator capacity in Singapore \cite{singaporecloud}.
& AWS, Microsoft Azure, and Google Cloud operate Singapore cloud regions \cite{singaporecloud}. \\
\hline

7 & Japan & ABCI 3.0 became publicly available in January 2025 with 6,128 NVIDIA H200 GPUs and a stated peak of 6.22 EFLOPS at FP16 \cite{nvidia2025abci3, arxivABCI30}.
& SoftBank, Fujitsu, NEC, and other firms operate or have announced private compute clusters \cite{SoftBank}.
& Domestic and global providers operate Japanese cloud regions. The comparison does not equate commercial regions with scientific supercomputers. \\
\hline

8 & South Korea & Samsung and SK Hynix supply memory used in AI systems, while the national infrastructure plan includes public and private compute capacity \cite{skscientific, nvidia2025koreainfrastructure}.
& Naver, Kakao, Samsung, and SK Hynix operate research or compute facilities \cite{skprivate}.
& AWS, Google Cloud, Microsoft Azure, KT Cloud, and NHN Cloud offer services in South Korea \cite{skcloud}. \\
\hline

9 & India & The PARAM series and National Supercomputing Mission provide public high-performance computing capacity \cite{indiasupercomp}.
& Large technology and telecommunications firms operate private infrastructure for AI and data services \cite{tcs}.
& AWS, Google Cloud, Microsoft Azure, and Oracle operate Indian cloud regions \cite{indiacloudcom}. \\
\hline

10 & UK & ARCHER2, CSD3, and Isambard-AI provide scientific and AI computing capacity \cite{ukriARCHER2,nvidiaNVIDIAAccelerates, bristol2025isambardai}.
& London, Cambridge, Oxford, Manchester, and Edinburgh host commercial AI firms and research clusters \cite{railswareWhereLondon}.
& AWS, Microsoft Azure, Google Cloud, and specialist providers operate UK data centers and cloud services \cite{datacentremagazineCoreWeaveWorking}.
\\
\hline
11 & USA &
On the June 2026 TOP500 list, El Capitan, Frontier, and Aurora ranked second through fourth behind China's LineShine system \cite{top5002026june}. The National AI Research Resource pilot provides shared resources for academic and public-interest research \cite{georgetownNAIRRPilot}.
&
Commercial capacity includes NVIDIA GPU clusters and custom accelerators such as Google's Ironwood TPU, AWS Trainium, and Cerebras wafer-scale systems \cite{nvidianews2026rubin, googlecloud2026ironwood, aws2026trainium3, cerebras2026wse3}.
&
AWS, Google Cloud, and Microsoft Azure operate multiple US regions with GPU and proprietary accelerator services \cite{aaaiComputeNorth, aws2026trainium3, googlecloud2026ironwood}. \\

\hline
12 & European Union &
EuroHPC coordinates shared public supercomputing systems, including LUMI in Finland and facilities in Germany and France \cite{eurohpc, lumi, julich, jeanzay}.
&
European telecommunications and technology firms operate private data centers, and global firms maintain facilities in several member states \cite{atos, telekomDeutscheTelekom, metaeurope}.
&
AWS, Google Cloud, and Microsoft Azure hold a large share of the European cloud market, with regions across several member states \cite{eurocloudcomp1}.
\\
\hline

13 & China & LineShine at the Shenzhen National Supercomputing Center ranked first on the June 2026 TOP500 list at 2.198 exaFLOPS \cite{top5002026lineshine}.
& Huawei's Atlas 900 A3 SuperPoD connects up to 384 Ascend chips \cite{huawei2025superpod}.
& Alibaba Cloud and Tencent Cloud document regions across China. The East Data West Computing program coordinates national computing hubs and data-center clusters \cite{alibabacloud2026regions, tencentcloud2026regions, chinaresource}. \\
\bottomrule
\end{longtable}

The table below records selected economic, political, and external conditions from sources that use different measures. The entries provide context and are not a comparable stability ranking.

\begin{longtable}{|p{0.6cm}|p{1.7cm}|p{3.2cm}|p{3.2cm}|p{2.9cm}|p{3.2cm}|}
\hline
\textbf{No.} & \textbf{Country} & \textbf{GDP and Economic Growth} & \textbf{Political and External Context} & \textbf{Regulatory Frameworks} & \textbf{Resource Availability} \\
\hline
\endfirsthead

\multicolumn{6}{c}%
{{\tablename\ \thetable{} -- continued from previous page}} \\
\hline
\textbf{No.} & \textbf{Country} & \textbf{GDP and Economic Growth} & \textbf{Political and External Context} & \textbf{Regulatory Frameworks} & \textbf{Resource Availability} \\
\hline
\endhead

\hline \multicolumn{6}{|r|}{{Continued on next page}} \\ \hline

\endfoot

\hline
\caption{Economic and Political Environment by Country}
\label{tab:economic-and-political-env-country}
\endlastfoot

1 & Thailand & The Eastern Economic Corridor and national plans identify advanced technology as an economic priority \cite{thailandgdp}.
& The national AI strategy and Eastern Economic Corridor provide the selected policy context \cite{thailandstability}.
& Industry 4.0 programs operate alongside data-protection and cybersecurity rules \cite{thailandregulatory}.
& NSTDA supports technical training and research \cite{thailandresource}. \\
\hline

2 & Malaysia & The economy and national development plans support investment in digital infrastructure \cite{malaysiaimf}.
& The cited investment analysis and the NVIDIA-YTL project show continued foreign and domestic infrastructure commitments \cite{morganstanleymalaysia, midaNvidiaYTLTieup}.
& Data-protection, consumer, transaction, and cybersecurity rules form the relevant regulatory base \cite{malaysiaregulation}.
& Solar projects and the YTL Green Data Center add power and data-center capacity \cite{midamalaysiasolar, ytlCorporationBerhad}. \\
\hline

3 & Vietnam & Rapid economic growth and public digital-infrastructure programs support technology investment \cite{vietnameconomy, vietnamdis}.
& National digital-transformation policy provides the selected policy context \cite{vietnamstability}.
& The National Digital Transformation Roadmap operates alongside developing data and AI rules \cite{vietnamregulatory, vietnamdigital}.
& Reviewed sources document foreign partnerships and domestic research institutions, but do not provide a comparable estimate of high-end compute access \cite{vietnamresource1,vietnamresource2}. \\
\hline

4 & Philippines & The cited forecast reported post-pandemic growth near 5--6\% \cite{phil_gdp}.
& No comparable quantitative stability measure was retained for this review.
& The National AI Roadmap 2.0 supports research, adoption, and responsible-use guidance \cite{phil_regulatory}.
& Renewable-energy development can support data-center expansion, though the cited sources do not quantify AI-specific supply \cite{philip_resource_1, philip_resource_2}. \\
\hline

5 & Indonesia & The cited forecast placed the digital economy above \$130 billion by 2025 \cite{indonesiagdp}.
& No comparable quantitative measure was retained for this field.
& The Personal Data Protection Law and national ethics guidance provide the main horizontal rules relevant to AI \cite{indonesiaregulatory,indonesiaregulatory2}.
& The cited source describes domestic energy and infrastructure resources but does not isolate AI use \cite{indonesiaresource}. \\
\hline

6 & Singapore & The economy grew 4.8\% in 2025. Budget measures committed S\$1 billion over five years to AI and S\$150 million to the Enterprise Compute Initiative \cite{mti2026gdp2025, wong2024budgetai, disg2025eci}.
& No comparable quantitative measure was retained for this field.
& Data-protection and cybersecurity rules operate alongside nonbinding AI frameworks. A dedicated agentic AI framework followed in January 2026 \cite{singaporeregulation, imda2026mgfagentic}.
& Singtel and the National Supercomputing Centre provide commercial and public compute capacity \cite{singaporeresource1, singaporeresource2}. \\
\hline

7 & Japan & A large economy and sustained technology investment support public and private AI programs \cite{deloitteJapanEconomic}.
& G7 participation and international digital-policy partnerships shape the external setting \cite{japnministryofecon}.
& Human-centered principles and sectoral guidance now operate alongside the 2025 AI Promotion Act \cite{japan2025aiact, govonline2025aipromotionact}.
& ABCI and the private clusters listed above are the documented resources used in this comparison.
\\
\hline

8 & South Korea & Nominal GDP was about \$1.86--1.87 trillion in 2025. The government also announced large AI financing and compute programs \cite{imf2026gdpkorea, chosunSouthKorea, nvidia2025koreainfrastructure}.
& Trade relationships and semiconductor supply-chain partnerships shape the external setting \cite{tradeSouthKorea}.
& The AI Basic Act took effect on January 22, 2026. It distinguishes high-impact AI by sector and consequence from high-performance AI defined partly by a $10^{26}$ operation threshold \cite{korea2025aibasicact, jon2026korea}.
& Samsung and SK Hynix provide semiconductor manufacturing capacity relevant to AI systems \cite{tradeSouthKorea}. \\
\hline

9 & India & Nominal GDP was estimated at about \$4.15 trillion in 2025, with 6.6\% real growth projected for fiscal year 2025--26 \cite{imf2025indiaarticleiv}.
& No comparable quantitative measure was retained for this field.
& The national strategy and IndiaAI Mission operate alongside the 2023 data-protection statute, 2025 rules, and 2025 voluntary AI Governance Guidelines \cite{meity2025dpdprulesnotified, meity2025aigovernanceguidelines}.
& National skilling programs, research institutions, and commercial cloud services support AI development \cite{indiaresourceavail}. \\
\hline

10 & UK & Nominal GDP reached about £3.03 trillion in 2025. The government adopted the AI Opportunities Action Plan, and firms announced investment under the 2025 technology agreement \cite{ons2026ukgdp2025, dsit2025aiopportunitiesplan, cnbc2025techprosperitydeal}.
& G7 and other international partnerships support research and policy coordination \cite{ukreg2}.
& The Data Protection Act and sectoral law operate alongside nonstatutory AI principles. The AI Security Institute and AI Growth Lab add evaluation and sandbox functions \cite{ukreg1, dsit2025aisecurityinstitute, parliament2026aiblueprint}.
& Public research funding, Innovate UK, and private investment support compute and talent \cite{ukresource1,ukresource2}. \\
\hline

11 & USA & Nominal GDP was about \$30.5--30.8 trillion in 2025. Large technology firms directed substantial capital spending toward data centers and specialized hardware \cite{bea2026gdp2025}.
& Export controls on advanced chips changed several times in 2025, including rescission of the AI Diffusion Rule before it took effect \cite{bis2025aidiffusionrescission, uscommerce2025h20controls}.
& Executive Order 14179 and the July 2025 AI Action Plan replaced the prior executive-order framework. The CHIPS Act continues to support domestic semiconductor production \cite{whitehouse2025eo14179barriers, whitehouse2025aiactionplan, cfrWhatCHIPS}.
& The public and private facilities listed above are the documented resource base used in this comparison.\\

\hline
12 & European Union & GDP was €16.22 trillion in 2024. InvestAI targets €200 billion in public and private AI investment, while 2025 private investment remained well below the US level \cite{eurostat2024gdp, eucommission2025investai, stanfordhai2026index}.
& As a supranational comparator, the EU entry describes collective digital policy rather than a single national political environment \cite{gaiax2025trustframework, eucommission2026digitaldecade}.
& The AI Act applies in phases. Regulation 2026/1744 revised parts of the implementation schedule \cite{eu2024aiactforce, eu2026omnibusregulation}.
& EuroHPC, the Connecting Europe Facility, and member-state energy systems support infrastructure across the Union \cite{euresourceavail2}. \\
\hline

13 & China & Final revised GDP for 2024 was 134.81 trillion yuan, with reported real growth of 5.0\% \cite{nbs2024gdpfinal, chinagdp}.
& The 2024 Global Peace Index ranked China 88th of 163 jurisdictions \cite{chinastability}.
& The 2017 development plan, 2025 AI Plus initiative, synthetic-content labeling rules, and amended Cybersecurity Law form part of the national portfolio \cite{chinaregulatory1, chinastatecouncil2025aiplus, cac2025aigclabeling, chinanpc2025cybersecurityamendment}.
& The East Data West Computing program directs national computing hubs and data-center clusters toward eastern and western regions \cite{chinaresource}.
 \\

\end{longtable}

The entries below summarize signals reported by the cited sources. They do not establish causal effects or provide a common measure of demand or talent across jurisdictions.

\begin{longtable}{|p{0.65cm}|p{1.7cm}|p{4.18cm}|p{4.18cm}|p{4.18cm}|}
\hline
\textbf{No.} & \textbf{Country} & \textbf{Local Demand for AI solutions} & \textbf{Research Institutions} & \textbf{Talent Pool} \\
\hline
\endfirsthead

\multicolumn{5}{c}%
{{\tablename\ \thetable{} -- continued from previous page}} \\
\hline
\textbf{No.} & \textbf{Country} & \textbf{Local Demand for AI solutions} & \textbf{Research Institutions} & \textbf{Talent Pool} \\
\hline
\endhead

\hline \multicolumn{5}{|r|}{{Continued on next page}} \\ \hline
\endfoot

\hline

\caption{Demand and Talent in AI by Country}
\label{tab:demand-and-talent-country}
\endlastfoot

1 & Thailand & The cited reporting describes investment in domestic AI services and interest among Thai businesses \cite{bangkokpostSiamAIInvest}.
& NSTDA and universities support AI research and collaboration with industry \cite{opengovasiaThailandNational}.
& Government and university programs provide training in AI and data science \cite{opengovasiaThailandNational}.
\\
\hline

2 & Malaysia & Reviewed sources describe AI adoption in finance, healthcare, and manufacturing as part of wider digitalization efforts \cite{malaysiaregulation}.
& The Malaysia Genome Institute and MRANTI support research and commercialization in advanced technologies \cite{malaysiakiniLaunchingMyGenom}.
& Education programs seek to expand the workforce, while reported shortages remain in AI, data analytics, and cybersecurity skills \cite{peda}.
\\
\hline

3 & Vietnam & PhoBERT documents work on Vietnamese-language models, while the reviewed sources also describe applications in healthcare, transportation, and finance \cite{arxivPhoBERTPretrained}.
& VinAI and university research groups conduct AI research and collaborate with overseas institutions \cite{nhandanVietnamNational}.
& Universities and companies fund AI training, while the reviewed source reports shortages in advanced expertise \cite{opengovasiaBridgingTalent}. \\
\hline

4 & Philippines & The national AI roadmap and the National Center for AI Research identify healthcare, agriculture, and manufacturing as areas for AI development \cite{dtiLaunchesNational}.
& The roadmap links the center with universities and industry partners for research and knowledge transfer \cite{dtiLaunchesNational}.
& The roadmap includes reskilling and workforce development intended to address reported talent shortages \cite{dtiLaunchesNational}. \\
\hline

5 & Indonesia & The reviewed report identifies fintech, electronic commerce, and healthcare as areas of AI adoption and places AI within the ``Making Indonesia 4.0'' agenda \cite{seavet}.
& Bandung Institute of Technology, the University of Indonesia, and other institutions conduct AI and data-science research \cite{seavet}.
& The report identifies gaps in workforce readiness and describes public programs in digital skills and STEM education \cite{seavet}. \\
\hline

6 & Singapore & The National AI Strategy identifies finance, healthcare, logistics, and public services as areas for wider AI use \cite{smartnationNationalStrategy}.
& The National University of Singapore, Nanyang Technological University, and public research organizations support research and industry collaboration \cite{smartnationNationalStrategy}.
& The strategy sets a target of tripling the number of AI practitioners to 15,000 by 2028 and describes scholarships and training programs \cite{smartnationNationalStrategy}. \\
\hline

7 & Japan & The reviewed market report identifies manufacturing, healthcare, and robotics as major areas of AI adoption \cite{sphericalinsightsJapanArtificial}.
& The University of Tokyo, AIST, and other institutions use the AI Bridging Cloud Infrastructure for research \cite{nvidiaJapanEnhances}.
& Government and university programs support AI education and research training \cite{nvidiaJapanEnhances}. \\
\hline

8 & South Korea & The reviewed report describes AI use in healthcare, finance, manufacturing, urban planning, and public services \cite{georgetownAssessingSouth}.
& KAIST, Seoul National University, and other institutions contribute to AI research and patenting \cite{georgetownAssessingSouth}.
& National programs fund AI education and workforce training in response to reported shortages \cite{georgetownAssessingSouth}.
\\
\hline

9 & India & Reviewed sources describe AI use and demand in healthcare, agriculture, education, and urban services, alongside public digitalization programs and startup investment \cite{bcgUnlockingAIs, indiaaiUndefined}.
& IITs, IIITs, NITs, public agencies, and research hubs conduct AI research and support collaboration with industry \cite{res, kwe, aseaniitArtificialIntelligence}.
& National and state programs provide AI training, including ``Responsible AI for Youth,'' while industry sources continue to report unmet demand for advanced skills \cite{indiaaiIndiasTalent,drishtiiasResponsibleYouth}.
\\
\hline

10 & UK & The cited sources document AI activity in healthcare, financial services, and urban applications \cite{nationalhealthexecutiveReportHighlights}.
& Oxford, Cambridge, Imperial College London, University College London, and the Alan Turing Institute conduct AI research and collaborate with government and industry \cite{nationalhealthexecutiveReportHighlights, analyticsinsightUniversitiesLeading}.
& UKRI doctoral funding, company partnerships, and the Global Talent Visa form part of the country's AI training and recruitment system \cite{nationalhealthexecutiveReportHighlights}. \\
\hline

11 & USA & The 2025 AI Index reports extensive private investment and commercial model development, while public procurement includes defense and health applications \cite{aiindex2025, indiatimesPentagonsInitiatives}.
& Universities and federally funded AI institutes support research and collaboration with industry \cite{aiindex2025}.& The United States attracts researchers and engineers from abroad, while immigration policy affects recruitment and retention \cite{ifpStrengtheningAmericas}. \\

\hline
12 & European Union & EU institutions and member-state programs support AI adoption across manufacturing, finance, healthcare, and public services \cite{europaArtificialIntelligence}.
& Universities, national laboratories, and the ELLIS network support AI research across member states \cite{europa}.& Universities and digital-skills programs train AI workers, but available sources use different occupational definitions and should not be combined into one workforce count \cite{frenusRevolutionGerman}. \\
\hline
13 & China & Baidu, Alibaba, Tencent, government agencies, and state-owned enterprises procure or provide AI services in domestic markets \cite{chinalocaldemand1}.
& Supercomputing centers, universities, research institutes, and company laboratories contribute to AI research \cite{chinaresearchuni}.
& The 2025 AI Index reports that China produced 23.2\% of AI publications in 2023 \cite{stanfordhai2025chinashare}. MacroPolo reports that 11\% of China-educated elite AI researchers in its 2024 and 2025 conference sample worked in China, compared with 16\% in its 2019 sample \cite{macropolo2025talenttracker3}. \\

\end{longtable}

\clearpage
\section{Cross-National Policy Overview}\label{app:crossnational}
The tables report the status of the reviewed instruments through August 15, 2026. They describe what the cited sources state and do not infer causal effects.

\begin{longtable}{|p{2.1cm}|p{2.1cm}|p{3.5cm}|p{3.5cm}|p{3.5cm}|}
\hline
\textbf{Policy Instrument} & \textbf{Covered Actors} & \textbf{Documented Provisions} & \textbf{Implications for AI Systems} & \textbf{Implementation Status and Limitations} \\
\hline
\endfirsthead

\multicolumn{5}{c}%
{{\tablename\ \thetable{} -- continued from previous page}} \\
\hline
\textbf{Policy Instrument} & \textbf{Covered Actors} & \textbf{Documented Provisions} & \textbf{Implications for AI Systems} & \textbf{Implementation Status and Limitations} \\
\hline
\endhead

\hline \multicolumn{5}{|r|}{{Continued on next page}} \\ \hline
\endfoot

\hline
\caption{European Union AI Policy Overview}
\label{tab:eu_ai_policy}
\endlastfoot

European Union Artificial Intelligence Act (2024) \cite{eu_ai_act_source, EPRS2022} & AI Developers, Businesses, Public Sector, End Users &
\begin{enumerate}[leftmargin=*]
    \item Prohibits specified AI practices
    \item Imposes duties on providers and deployers of high-risk systems
    \item Adds transparency duties for specified systems
    \item Regulates general-purpose AI models
    \item Provides conformity assessment, documentation, oversight, and penalties
\end{enumerate} &
\begin{enumerate}[leftmargin=*]
    \item High-risk systems must meet applicable risk-management, data, documentation, oversight, accuracy, robustness, and cybersecurity duties
    \item Specified practices are prohibited
    \item General-purpose AI providers have separate documentation and risk duties
\end{enumerate} &
\begin{enumerate}[leftmargin=*]
    \item Entered into force in August 2024
    \item Prohibited-practice and general-purpose AI provisions began applying in February and August 2025 \cite{eu2024aiactforce}
    \item Regulation 2026/1744 revised several application dates \cite{eu2026omnibusregulation}
\end{enumerate} \\
\hline

General Data Protection Regulation (GDPR, 2018)   \cite{gdpr_eu} & Organizations, AI Developers, Individuals &
\begin{enumerate}[leftmargin=*]
    \item Requires a lawful basis for processing personal data
    \item Data minimization
    \item Rights concerning solely automated decisions
    \item Access to meaningful information about the logic involved where applicable
    \item Conditions for transfers outside the European Economic Area
\end{enumerate} &
\begin{enumerate}[leftmargin=*]
    \item Applies when model development or use processes personal data
    \item Truly anonymized data falls outside the Regulation
    \item Article 22 limits specified decisions based solely on automated processing
\end{enumerate} &
\begin{enumerate}[leftmargin=*]
    \item Binding and directly applicable since 2018
    \item Enforced by national supervisory authorities
    \item Does not provide a general AI approval regime
\end{enumerate} \\
\hline

Digital Services Act (DSA, 2022)
 \cite{dsa_eu} & Online Platforms, Consumers, Regulators &
\begin{enumerate}[leftmargin=*]
    \item Recommender-system transparency
    \item Advertising transparency
    \item Systemic-risk assessment and mitigation for very large services
    \item Notice, complaint, and redress procedures for content moderation
\end{enumerate} &
\begin{enumerate}[leftmargin=*]
    \item Applies to platform systems used for recommendation, advertising, and content moderation
    \item Requires risk mitigation by very large online platforms and search engines
    \item Does not impose a general ban on models that amplify harmful content
\end{enumerate} &
\begin{enumerate}[leftmargin=*]
    \item Binding platform regulation
    \item Obligations depend on service type and size
    \item Enforced by the Commission and national Digital Services Coordinators
\end{enumerate} \\
\hline
\end{longtable}

\begin{longtable}{|p{2.1cm}|p{2.1cm}|p{3.5cm}|p{3.5cm}|p{3.5cm}|}
\hline
\textbf{Policy Instrument} & \textbf{Implementing Body} & \textbf{Purpose and Provisions} & \textbf{Implications for AI Systems} & \textbf{Implementation Status and Limitations} \\
\hline
\endfirsthead

\multicolumn{5}{c}%
{{\tablename\ \thetable{} -- continued from previous page}} \\
\hline
\textbf{Policy Instrument} & \textbf{Implementing Body} & \textbf{Purpose and Provisions} & \textbf{Implications for AI Systems} & \textbf{Implementation Status and Limitations} \\
\hline
\endhead

\hline \multicolumn{5}{|r|}{{Continued on next page}} \\ \hline
\endfoot

\hline
\caption{United States AI Policies Overview} \label{tab:us_ai_policy_concise}
\endlastfoot

Executive Order 14110 (2023), \textit{revoked January 20, 2025}
  \cite{us_executive_order, whitehouse2025eo14148rescissions} & Executive Office of the President &
\begin{enumerate}[leftmargin=*]
    \item Directed federal action on safety, security, privacy, civil rights, research, and workforce issues
    \item Used Defense Production Act authority for specified reporting
    \item Directed agencies to develop standards and guidance
\end{enumerate} &
\begin{enumerate}[leftmargin=*]
    \item Required specified developers to report training and safety information
    \item Directed evaluations and standards for several high-impact uses
\end{enumerate} &
\begin{enumerate}[leftmargin=*]
    \item Revoked on January 20, 2025
    \item No longer forms the operative federal executive-order framework
\end{enumerate} \\
\hline

Executive Order 14179 (Jan.\ 2025) \& America's AI Action Plan (July 2025)
  \cite{whitehouse2025eo14179barriers, whitehouse2025aiactionplan} & Executive Office of the President and federal agencies &
\begin{enumerate}[leftmargin=*]
    \item Reorients federal policy toward US AI leadership
    \item Directs revision of the NIST AI RMF
    \item Renames the AI Safety Institute as the Center for AI Standards and Innovation (CAISI)
    \item Organized around innovation, infrastructure, and international diplomacy
\end{enumerate} &
\begin{enumerate}[leftmargin=*]
    \item Removes or revises several prior federal requirements
    \item Directs work on infrastructure, standards, procurement, exports, and security evaluations
\end{enumerate} &
\begin{enumerate}[leftmargin=*]
    \item Operative federal policy at the cutoff
    \item The AI Diffusion Rule was rescinded in May 2025 before taking effect
    \item State AI laws remain outside this federal portfolio
\end{enumerate} \\
\hline

AI Bill of Rights Blueprint (2022)
  \cite{us_ai_bill_of_rights} & White House OSTP &
\begin{enumerate}[leftmargin=*]
    \item Nonbinding principles on safe systems, discrimination, privacy, notice, explanation, and human alternatives
    \item Includes technical companion guidance
\end{enumerate} &
\begin{enumerate}[leftmargin=*]
    \item Provides design and governance guidance for automated systems
    \item Creates no approval requirement or penalty
\end{enumerate} &
\begin{enumerate}[leftmargin=*]
    \item Published as a White House blueprint
    \item Nonbinding and not a source of legal authority
\end{enumerate} \\
\hline

National AI Initiative Act (2020)  \cite{us_national_ai_initiative_act} & US Congress &
\begin{enumerate}[leftmargin=*]
    \item Establishes a coordinated federal AI research and development program
    \item Supports research institutes, education, standards, and interagency coordination
\end{enumerate} &
\begin{enumerate}[leftmargin=*]
    \item Authorizes programs and coordination rather than model-specific approval rules
\end{enumerate} &
\begin{enumerate}[leftmargin=*]
    \item Binding statutory basis for federal programs
    \item Does not create a comprehensive private-sector AI regime
\end{enumerate} \\
\hline

AI Risk Management Framework (NIST AI RMF) (2023)  \cite{nist_ai_rm_framework} & NIST &
\begin{enumerate}[leftmargin=*]
    \item Voluntary framework for managing AI risks across the lifecycle
    \item Organized around Govern, Map, Measure, and Manage functions
\end{enumerate} &
\begin{enumerate}[leftmargin=*]
    \item Provides a common risk-management vocabulary and process
    \item Creates no independent approval requirement
\end{enumerate} &
\begin{enumerate}[leftmargin=*]
    \item Voluntary unless another instrument incorporates it
    \item Does not itself authorize penalties or suspend systems
\end{enumerate} \\
\hline

California Consumer Privacy Act and California Privacy Rights Act \cite{ccpa_update_2023} & California State Government &
\begin{enumerate}[leftmargin=*]
    \item Grants access, deletion, correction, and opt-out rights for covered personal information
    \item Adds limits and rights for sensitive personal information
\end{enumerate} &
\begin{enumerate}[leftmargin=*]
    \item Applies to covered businesses that process personal information in AI development or use
    \item Does not create a general AI training-data consent rule
\end{enumerate} &
\begin{enumerate}[leftmargin=*]
    \item Binding state law with threshold and exemption rules
    \item Enforced by the California Privacy Protection Agency and Attorney General
    \item Not part of the federal portfolio score unless the declared scope includes subnational law
\end{enumerate} \\
\hline
\end{longtable}

\begin{longtable}{|p{2.1cm}|p{2.1cm}|p{3.5cm}|p{3.5cm}|p{3.5cm}|}
\hline
\textbf{Policy Instrument} & \textbf{Implementing Body} & \textbf{Purpose} & \textbf{Documented Provisions} & \textbf{Implementation Status and Limitations} \\
\hline
\endfirsthead

\multicolumn{5}{c}%
{{\tablename\ \thetable{} -- continued from previous page}} \\
\hline
\textbf{Policy Instrument} & \textbf{Implementing Body} & \textbf{Purpose} & \textbf{Documented Provisions} & \textbf{Implementation Status and Limitations} \\
\hline
\endhead

\hline \multicolumn{5}{|r|}{{Continued on next page}} \\ \hline
\endfoot

\hline
\caption{Overview of Singapore AI Policies}
\label{tab:singapore_ai_policy}
\endlastfoot

Artificial Intelligence in Healthcare Guidelines (AIHGle) & Ministry of Health (MOH), Health Sciences Authority (HSA), Integrated Health Information Systems (IHiS) &
\begin{enumerate}[leftmargin=*]
    \item Guide safe development and use of AI in healthcare
\end{enumerate} &
\begin{enumerate}[leftmargin=*]
    \item Recommends validation, monitoring, fairness, responsibility, explainability, and patient-centered design \cite{moh_health_sciences_authority}
    \item Addresses data governance and cybersecurity
\end{enumerate} &
\begin{enumerate}[leftmargin=*]
    \item Guidance rather than a standalone approval statute
    \item Separate medical-device and professional rules can create binding duties
\end{enumerate} \\
\hline

Personal Data Protection Act (PDPA) and Sector-Specific Guidelines & Personal Data Protection Commission (PDPC) &
\begin{enumerate}[leftmargin=*]
    \item Protect personal data processed by organizations
    \item Provide nonbinding AI governance guidance
\end{enumerate} &
\begin{enumerate}[leftmargin=*]
    \item Requires compliance with consent, purpose, notification, access, correction, protection, retention, and transfer duties where applicable
    \item The Model AI Governance Framework provides separate organizational guidance \cite{singaporeverify}
\end{enumerate} &
\begin{enumerate}[leftmargin=*]
    \item The PDPA is binding and technology neutral
    \item The Model AI Governance Framework is nonbinding
\end{enumerate} \\
\hline

MAS FEAT Principles & Monetary Authority of Singapore (MAS) &
\begin{enumerate}[leftmargin=*]
    \item Promote fairness, ethics, accountability, and transparency in AI use within financial sector
\end{enumerate} &
\begin{enumerate}[leftmargin=*]
    \item Sets fairness, ethics, accountability, and transparency principles \cite{mas_feat}
    \item Supports assessment methods for financial institutions
\end{enumerate} &
\begin{enumerate}[leftmargin=*]
    \item Sectoral supervisory guidance rather than a general AI statute
\end{enumerate} \\
\hline

Healthcare Cybersecurity Essentials Guidelines & Ministry of Health (MOH) &
\begin{enumerate}[leftmargin=*]
    \item Protect healthcare AI systems and IT infrastructure against cyber threats
\end{enumerate} &
\begin{enumerate}[leftmargin=*]
    \item Recommends access control, protection, detection, response, recovery, and staff practices \cite{healthcare_cybersecurity}
\end{enumerate} &
\begin{enumerate}[leftmargin=*]
    \item General healthcare cybersecurity guidance rather than an AI-specific rule
\end{enumerate} \\
\hline

Human Biomedical Research Act (HBRA) & Ministry of Health (MOH) &
\begin{enumerate}[leftmargin=*]
    \item Ensure ethical compliance and data privacy in biomedical AI research
\end{enumerate} &
\begin{enumerate}[leftmargin=*]
    \item Requires review, consent, and protection for covered human biomedical research \cite{human_biomedical_research_act}
\end{enumerate} &
\begin{enumerate}[leftmargin=*]
    \item Binding for research within its statutory scope
    \item Not a general approval regime for AI models or clinical products
\end{enumerate} \\
\hline
\end{longtable}

\begin{longtable}{|p{2cm}|p{3cm}|p{3cm}|p{3cm}|p{3.5cm}|}
\hline
\textbf{Policy Instrument} & \textbf{Covered Actors} & \textbf{Documented Provisions} & \textbf{Implications for AI Systems} & \textbf{Implementation Status and Limitations} \\
\hline
\endfirsthead

\multicolumn{5}{c}%
{{\tablename\ \thetable{} -- continued from previous page}} \\
\hline
\textbf{Policy Instrument} & \textbf{Covered Actors} & \textbf{Documented Provisions} & \textbf{Implications for AI Systems} & \textbf{Implementation Status and Limitations} \\
\hline
\endhead

\hline \multicolumn{5}{|r|}{{Continued on next page}} \\ \hline
\endfoot

\hline
\caption{United Kingdom AI Policy Portfolio} \label{tab:uk_ai_regulations_concise}
\endlastfoot

UK IPO AI and IP Consultation and Government Response, 2021 and 2022 \cite{uk_ipo_ai_ip_2021} &
\begin{enumerate}[leftmargin=*]
  \item Creators, rights holders, researchers, AI developers, and inventors
\end{enumerate}
&
\begin{enumerate}[leftmargin=*]
  \item Considers copyright protection for computer-generated works, text and data mining, and patent protection for AI-devised inventions
  \item The 2022 response retained existing protection for computer-generated works and did not change patent inventorship rules
  \item The response proposed a broader text and data mining exception, which was not implemented
\end{enumerate}
&
\begin{enumerate}[leftmargin=*]
  \item The consultation and response address how existing intellectual property law applies to AI development and output
  \item They do not amend the underlying law
\end{enumerate}
&
\begin{enumerate}[leftmargin=*]
  \item Consultation opened in October 2021 and the government response was published in June 2022
  \item This is intellectual property policy rather than an AI safety or approval regime
\end{enumerate}
\\ \hline

UK AI White Paper  \cite{uk_ai_white_paper_2023} &
\begin{enumerate}[leftmargin=*]
  \item Existing regulators and the organizations within their sectoral remits
  \item AI developers and deployers whose activities fall within those remits
\end{enumerate}
&
\begin{enumerate}[leftmargin=*]
  \item Uses existing law and existing regulators rather than a single horizontal AI regulator
  \item Sets five cross-sector principles covering safety, transparency, fairness, accountability, and contestability
  \item Proposes central coordination and support for regulators
\end{enumerate}
&
\begin{enumerate}[leftmargin=*]
  \item Regulators apply the principles to AI uses within their existing powers and sectoral context
  \item The framework does not create a general licensing or conformity-assessment regime
\end{enumerate}
&
\begin{enumerate}[leftmargin=*]
  \item Nonstatutory policy framework whose application initially depended on each regulator's remit and discretion
  \item The October 2025 regulatory blueprint retained the regulator-led approach \cite{parliament2026aiblueprint}
\end{enumerate}
\\ \hline

National AI Strategy \cite{uk_ai_strategy_2021} &
\begin{enumerate}[leftmargin=*]
  \item Government, research institutions, businesses, and the workforce
\end{enumerate}
&
\begin{enumerate}[leftmargin=*]
  \item Sets ten-year priorities for investment, research, skills, adoption, and governance
  \item Organizes the strategy around ecosystem needs, diffusion across the economy, and national and international governance
\end{enumerate}
&
\begin{enumerate}[leftmargin=*]
  \item Provides strategic direction for public programs and later policy instruments
  \item Does not create conduct duties for individual AI systems
\end{enumerate}
&
\begin{enumerate}[leftmargin=*]
  \item Published in September 2021 as a strategy rather than legislation
  \item The 2025 AI Opportunities Action Plan supplements its infrastructure, adoption, and skills agenda
\end{enumerate}
\\ \hline

AI Opportunities Action Plan \cite{dsit2025aiopportunitiesplan} &
\begin{enumerate}[leftmargin=*]
  \item UK government and delivery bodies
  \item Regulators and public sector organizations
  \item AI developers, researchers, and investors
\end{enumerate}
&
\begin{enumerate}[leftmargin=*]
  \item Recommends AI Growth Zones and expanded sovereign compute
  \item Proposes better access to public data and stronger adoption capacity
  \item Addresses skills, research, procurement, and AI security institutions
\end{enumerate}
&
\begin{enumerate}[leftmargin=*]
  \item Expands the infrastructure and institutions available for AI development and use
  \item Does not itself set binding conduct rules for particular model classes
\end{enumerate}
&
\begin{enumerate}[leftmargin=*]
  \item Published and accepted by the government in January 2025
  \item Implementation depends on later budgets, programs, and institutional decisions
  \item The plan is a strategy document rather than legislation
\end{enumerate}
\\ \hline

UK Algorithmic Transparency Standard \cite{uk_algo_transparency_2021} &
\begin{enumerate}[leftmargin=*]
  \item All government departments
  \item Arm's-length bodies that deliver public or frontline services or interact directly with the public
  \item Other public-sector bodies can use the standard voluntarily
\end{enumerate}
&
\begin{enumerate}[leftmargin=*]
  \item Provides a standard record for an algorithmic tool's purpose, ownership, data, risks, human oversight, and review
  \item Requires publication of records for covered organizations and tools
\end{enumerate}
&
\begin{enumerate}[leftmargin=*]
  \item Mandatory scope covers tools that significantly influence a decision with public effect or interact directly with the public
\end{enumerate}
&
\begin{enumerate}[leftmargin=*]
  \item Mandatory across central government since 2024
  \item The scope and exemptions policy was published in December 2024 and the guidance was updated in May 2025
\end{enumerate}
\\ \hline
\end{longtable}

\begin{longtable}{|p{2cm}|p{3cm}|p{3cm}|p{3cm}|p{3.5cm}|}
\hline
\textbf{Policy Instrument} & \textbf{Covered Actors} & \textbf{Documented Provisions} & \textbf{Implications for AI Systems} & \textbf{Implementation Status and Limitations} \\
\hline
\endfirsthead

\multicolumn{5}{c}%
{{\tablename\ \thetable{} -- continued from previous page}} \\
\hline
\textbf{Policy Instrument} & \textbf{Covered Actors} & \textbf{Documented Provisions} & \textbf{Implications for AI Systems} & \textbf{Implementation Status and Limitations} \\
\hline
\endhead

\hline \multicolumn{5}{|r|}{{Continued on next page}} \\ \hline
\endfoot

\hline
\caption{India AI Policy Portfolio} \label{tab:india_ai_regulations_concise}
\endlastfoot

National Strategy for Artificial Intelligence, 2018 \cite{niti_aayog_ai_strategy} &
\begin{enumerate}[leftmargin=*]
  \item Government departments
  \item Healthcare, agriculture, education, urban services, and mobility sectors
  \item Startups and small and medium enterprises
  \item Research institutions
\end{enumerate}
&
\begin{enumerate}[leftmargin=*]
  \item Identifies five priority sectors and an inclusive-growth objective
  \item Proposes research centers, shared data platforms, skilling, and public-private collaboration
  \item Discusses privacy, security, transparency, accountability, and bias as responsible-AI concerns
\end{enumerate}
&
\begin{enumerate}[leftmargin=*]
  \item Sets priorities for public programs and sector adoption
  \item Does not create approval, audit, or penalty rules for AI systems
\end{enumerate}
&
\begin{enumerate}[leftmargin=*]
  \item Nonbinding national strategy rather than legislation
  \item Later programs and the 2025 AI Governance Guidelines supplement its development and governance proposals
\end{enumerate}
\\ \hline

Digital Personal Data Protection Act 2023 and Rules 2025 \cite{dpdp_act_2023, meity2025dpdprulesnotified} &
\begin{enumerate}[leftmargin=*]
  \item Data fiduciaries and processors
  \item People whose digital personal data are processed
  \item The Data Protection Board and central government
\end{enumerate}
&
\begin{enumerate}[leftmargin=*]
  \item Requires a lawful purpose based on consent or a listed legitimate use
  \item Provides rights to information, correction, erasure, and grievance redress
  \item Sets additional duties for significant data fiduciaries
  \item Permits the government to restrict transfers to notified countries or territories
\end{enumerate}
&
\begin{enumerate}[leftmargin=*]
  \item Applies when AI training or deployment processes digital personal data within its scope
  \item Does not create a separate approval or audit regime for AI systems
\end{enumerate}
&
\begin{enumerate}[leftmargin=*]
  \item The 2025 Rules began a staged implementation of the Act
  \item Several substantive duties commence on later dates under that schedule
  \item The regime is binding data protection law rather than AI-specific regulation
\end{enumerate}
\\ \hline

Revised MeitY Advisory on AI Deployment, March 15, 2024 \cite{azb2024meityadvisory} &
\begin{enumerate}[leftmargin=*]
  \item Intermediaries and online platforms
  \item Providers that make AI tools available through those services
\end{enumerate}
&
\begin{enumerate}[leftmargin=*]
  \item Links AI use to existing intermediary due diligence obligations
  \item Directs platforms to address unlawful content, discrimination, and electoral integrity
  \item Calls for warnings on outputs from under-tested or unreliable systems
  \item Addresses identification of synthetic content used as misinformation or deepfakes
\end{enumerate}
&
\begin{enumerate}[leftmargin=*]
  \item Does not require prior government permission to release an under-tested model
  \item Places the stated duties on intermediaries and platforms rather than all AI developers
\end{enumerate}
&
\begin{enumerate}[leftmargin=*]
  \item Superseded the March 1 advisory that had required prior permission
  \item Its scope and legal force remain less clear than those of legislation or notified rules
\end{enumerate}
\\ \hline

India AI Governance Guidelines, 2025 \cite{meity2025aigovernanceguidelines} &
\begin{enumerate}[leftmargin=*]
  \item Central and state government bodies
  \item Sector regulators
  \item AI developers, deployers, and users
\end{enumerate}
&
\begin{enumerate}[leftmargin=*]
  \item Sets principles for trustworthy, people-centered, and innovation-supporting governance
  \item Recommends a coordinating governance group and a technical policy committee
  \item Recommends risk assessment, incident reporting, and shared technical tools
\end{enumerate}
&
\begin{enumerate}[leftmargin=*]
  \item Uses existing law and sector oversight as the base for AI governance
  \item Supports proportionate obligations and regulatory sandboxes
\end{enumerate}
&
\begin{enumerate}[leftmargin=*]
  \item Issued as national guidance in November 2025
  \item Recommendations require later institutional and sector-specific implementation
  \item The document does not itself create a comprehensive AI statute
\end{enumerate}
\\ \hline
\end{longtable}

\begin{longtable}{|p{2cm}|p{3cm}|p{3cm}|p{3cm}|p{3.5cm}|}
\hline
\textbf{Policy Instrument} & \textbf{Covered Actors} & \textbf{Documented Provisions} & \textbf{Implications for AI Systems} & \textbf{Implementation Status and Limitations} \\
\hline
\endfirsthead

\multicolumn{5}{c}%
{{\tablename\ \thetable{} -- continued from previous page}} \\
\hline
\textbf{Policy Instrument} & \textbf{Covered Actors} & \textbf{Documented Provisions} & \textbf{Implications for AI Systems} & \textbf{Implementation Status and Limitations} \\
\hline
\endhead

\hline \multicolumn{5}{|r|}{{Continued on next page}} \\ \hline
\endfoot

\hline
\caption{Japan AI Regulations Overview} \label{tab:japan_ai_regulations_concise}
\endlastfoot

Act on Promotion of Research and Development and Utilization of AI-Related Technologies, 2025 \cite{japan2025aiact} &
\begin{enumerate}[leftmargin=*]
  \item National and local government
  \item Research institutions and businesses
  \item AI developers, providers, and users
\end{enumerate}
&
\begin{enumerate}[leftmargin=*]
  \item Establishes an AI Strategy Headquarters and a national basic plan
  \item Assigns responsibilities for research, adoption, safety, and international cooperation
  \item Authorizes information requests, guidance, and public disclosure when misuse harms rights or interests
\end{enumerate}
&
\begin{enumerate}[leftmargin=*]
  \item Relies mainly on planning, guidance, voluntary cooperation, and disclosure
  \item Does not create a general licensing regime for AI models
\end{enumerate}
&
\begin{enumerate}[leftmargin=*]
  \item Promulgated on June 4, 2025
  \item Fully in force from September 1, 2025
  \item Does not prescribe private-sector fines or criminal penalties for noncompliance
\end{enumerate}
\\ \hline

Society 5.0 Vision \cite{japan_society50} &
\begin{enumerate}[leftmargin=*]
  \item Government, researchers, and industry
  \item Public service and infrastructure sectors
\end{enumerate}
&
\begin{enumerate}[leftmargin=*]
  \item Defines a human-centered society that integrates physical and digital systems
  \item Connects technological development with economic and social goals
\end{enumerate}
&
\begin{enumerate}[leftmargin=*]
  \item Provides strategic direction for AI and related digital technologies
\end{enumerate}
&
\begin{enumerate}[leftmargin=*]
  \item A high-level national vision rather than a binding conduct standard
\end{enumerate}
\\ \hline

Social Principles of Human-Centric AI, 2019 \cite{japan_ai_strategy} &
\begin{enumerate}[leftmargin=*]
  \item Government, business, researchers, and the public
\end{enumerate}
&
\begin{enumerate}[leftmargin=*]
  \item Addresses human dignity, education, privacy, security, fair competition, fairness, accountability, transparency, and innovation
\end{enumerate}
&
\begin{enumerate}[leftmargin=*]
  \item Supplies principles for later strategies and business guidance
\end{enumerate}
&
\begin{enumerate}[leftmargin=*]
  \item Nonbinding principles with no direct penalty mechanism
\end{enumerate}
\\ \hline

AI Guidelines for Business, Version 1.0 \cite{meti2024aibusinessguidelines} &
\begin{enumerate}[leftmargin=*]
  \item AI developers, providers, and business users
\end{enumerate}
&
\begin{enumerate}[leftmargin=*]
  \item Consolidates earlier guidance for AI governance and risk management
  \item Addresses safety, fairness, privacy, security, transparency, and accountability
\end{enumerate}
&
\begin{enumerate}[leftmargin=*]
  \item Provides lifecycle guidance that organizations can adapt to their roles and risks
\end{enumerate}
&
\begin{enumerate}[leftmargin=*]
  \item Published jointly by METI and MIC in April 2024
  \item Guidance rather than binding legislation
\end{enumerate}
\\ \hline
\end{longtable}

\begin{longtable}{|p{2.1cm}|p{2.1cm}|p{3.5cm}|p{3.5cm}|p{3.5cm}|}
\hline
\textbf{Policy Instrument} & \textbf{Covered Actors} & \textbf{Documented Provisions} & \textbf{Implications for AI Systems} & \textbf{Implementation Status and Limitations} \\
\hline
\endfirsthead

\multicolumn{5}{c}%
{{\tablename\ \thetable{} -- continued from previous page}} \\
\hline
\textbf{Policy Instrument} & \textbf{Covered Actors} & \textbf{Documented Provisions} & \textbf{Implications for AI Systems} & \textbf{Implementation Status and Limitations} \\
\hline
\endhead

\hline \multicolumn{5}{|r|}{{Continued on next page}} \\ \hline
\endfoot

\hline
\caption{South Korea AI Policy Overview} \label{tab:sk_ai_policy} \\
\endlastfoot

Framework Act on the Development of Artificial Intelligence and the Creation of a Foundation for Trust \cite{korea2025aibasicact, jon2026korea} & Government bodies, AI business operators, and people affected by AI services &
\begin{enumerate}[leftmargin=*]
    \item Establishes national planning and coordination institutions
    \item Sets transparency duties for generative and high-impact AI
    \item Requires safety measures for high-performance AI
    \item Sets safeguards for high-impact AI and encourages fundamental-rights impact assessment
\end{enumerate} &
\begin{enumerate}[leftmargin=*]
    \item High-impact AI is identified by sector, use, and risk to life, safety, or fundamental rights
    \item High-performance AI is a separate class tied in part to training compute of at least $10^{26}$ operations
    \item Covered systems remain deployable when the applicable duties are met
\end{enumerate} &
\begin{enumerate}[leftmargin=*]
    \item Enacted on January 21, 2025 and effective from January 22, 2026
    \item Subordinate rules specify operational details and the high-performance compute threshold
    \item The Act combines binding duties with promotion and support measures
\end{enumerate} \\
\hline

Personal Information Protection Act, revised 2023 \cite{pipa_korea} & Personal information controllers and data subjects &
\begin{enumerate}[leftmargin=*]
    \item Sets lawful-basis, purpose, security, and data-subject-rights requirements
    \item Article 37-2 provides rights concerning qualifying decisions made through a completely automated system
    \item Provides a right to request an explanation of an automated decision
\end{enumerate} &
\begin{enumerate}[leftmargin=*]
    \item Applies when AI development or use processes personal information within the Act's scope
    \item Requires specified responses when a person objects to or requests an explanation of a covered automated decision
\end{enumerate} &
\begin{enumerate}[leftmargin=*]
    \item Binding general data-protection law
    \item The automated-decision provisions took effect through the 2023 amendments and implementing decree
    \item Does not create a general AI safety approval regime
\end{enumerate} \\
\hline

National Guidelines for AI Ethics, 2020 \cite{nia_ai_ethics} & Government, developers, providers, users, and other AI stakeholders &
\begin{enumerate}[leftmargin=*]
    \item Organizes the guidance around human dignity, public good, and fitness for purpose
    \item Sets ten requirements covering human rights, privacy, diversity, data management, accountability, safety, transparency, solidarity, and related duties
\end{enumerate} &
\begin{enumerate}[leftmargin=*]
    \item Provides a nonbinding reference for the development and use of AI
\end{enumerate} &
\begin{enumerate}[leftmargin=*]
    \item Released by the Ministry of Science and ICT in December 2020
    \item Creates no direct penalty or approval mechanism
\end{enumerate} \\
\hline
\end{longtable}

\begin{longtable}{|p{2.1cm}|p{2.1cm}|p{3.5cm}|p{3.5cm}|p{3.5cm}|}
\hline
\textbf{Policy Instrument} & \textbf{Covered Actors} & \textbf{Documented Provisions} & \textbf{Implications for AI Systems} & \textbf{Implementation Status and Limitations} \\
\hline
\endfirsthead

\multicolumn{5}{c}%
{{\tablename\ \thetable{} -- continued from previous page}} \\
\hline
\textbf{Policy Instrument} & \textbf{Covered Actors} & \textbf{Documented Provisions} & \textbf{Implications for AI Systems} & \textbf{Implementation Status and Limitations} \\
\hline
\endhead

\hline \multicolumn{5}{|r|}{{Continued on next page}} \\ \hline
\endfoot

\hline
\caption{China AI Policy Overview} \label{tab:china_ai_policy} \\
\endlastfoot

Interim Measures for the Management of Generative AI Services, 2023 \cite{cac2023generativeai} & Providers of generative AI services offered to the public in mainland China &
\begin{enumerate}[leftmargin=*]
    \item Requires lawful training data and protection of intellectual property and personal information
    \item Requires providers to address prohibited content and protect users
    \item Connects services with public-opinion attributes or social mobilization capacity to security assessment and algorithm filing rules
\end{enumerate} &
\begin{enumerate}[leftmargin=*]
    \item Applies to public generative AI services rather than internal research and use
    \item Requires measures on content, complaints, data, and service security
\end{enumerate} &
\begin{enumerate}[leftmargin=*]
    \item Effective from August 15, 2023
    \item Binding administrative measure jointly issued by seven national authorities
\end{enumerate} \\
\hline

Measures for the Labeling of AI-Generated and Synthetic Content, 2025 \cite{cac2025aigclabeling} & Generative AI providers, content distribution platforms, application distributors, and users &
\begin{enumerate}[leftmargin=*]
    \item Requires explicit labels for specified generated or synthetic content
    \item Requires implicit identifiers in file metadata
    \item Sets duties for platforms that distribute or detect covered content
\end{enumerate} &
\begin{enumerate}[leftmargin=*]
    \item Adds provenance information to the production and distribution of synthetic content
    \item Prohibits malicious removal, alteration, or concealment of required labels
\end{enumerate} &
\begin{enumerate}[leftmargin=*]
    \item Effective from September 1, 2025
    \item Complements the broader AI Plus development initiative \cite{chinastatecouncil2025aiplus}
\end{enumerate} \\
\hline

Administrative Provisions on Deep Synthesis Internet Information Services, 2022 \cite{cac2022aiguidelines} & Deep synthesis service providers, technical supporters, and users &
\begin{enumerate}[leftmargin=*]
    \item Requires identity verification and information security controls
    \item Requires consent when editing biometric information such as faces or voices
    \item Requires prominent labels for content that could confuse or mislead the public
\end{enumerate} &
\begin{enumerate}[leftmargin=*]
    \item Regulates covered synthesis services rather than banning the underlying techniques
    \item Connects high-risk services to filing and security assessment duties
\end{enumerate} &
\begin{enumerate}[leftmargin=*]
    \item Effective from January 10, 2023
    \item Binding administrative provisions
\end{enumerate} \\
\hline

Provisions on the Administration of Algorithmic Recommendations in Internet Information Services \cite{cac2021recommendation} & Algorithmic recommendation service providers and users &
\begin{enumerate}[leftmargin=*]
    \item Requires disclosure of basic service principles, purposes, and operating mechanisms
    \item Gives users a non-personalized option or a convenient way to disable recommendation services
    \item Requires filing for services with public-opinion attributes or social mobilization capacity
\end{enumerate} &
\begin{enumerate}[leftmargin=*]
    \item Regulates personalization, ranking, search filtering, dispatching, and related recommendation techniques
    \item Sets additional protections for minors, older people, workers, and consumers
\end{enumerate} &
\begin{enumerate}[leftmargin=*]
    \item Effective from March 1, 2022
    \item Binding rules jointly issued by four national authorities
\end{enumerate} \\
\hline

Cybersecurity Law, Data Security Law, and Personal Information Protection Law \cite{china2016cybersecuritylaw, china2021datasecuritylaw, china2021pipl} & Network operators, data processors, personal information processors, and affected people &
\begin{enumerate}[leftmargin=*]
    \item Establishes network security and critical infrastructure duties
    \item Requires classified data protection and safeguards for important data
    \item Sets lawful bases, individual rights, processor duties, and cross-border rules for personal information
\end{enumerate} &
\begin{enumerate}[leftmargin=*]
    \item Applies to AI developers and providers when their network, data, or personal information processing falls within scope
    \item Does not amount to a single AI-specific statute
\end{enumerate} &
\begin{enumerate}[leftmargin=*]
    \item Binding national laws with distinct scopes and enforcement mechanisms
    \item The Cybersecurity Law was amended in 2025 with effect from January 1, 2026
\end{enumerate} \\
\hline

Regulations on Network Data Security Management, 2024 \cite{china2024networkdata} & Network data processors and other covered organizations &
\begin{enumerate}[leftmargin=*]
    \item Implements duties under the Cybersecurity Law, Data Security Law, and Personal Information Protection Law
    \item Addresses personal information, important data, platform duties, and cross-border transfers
\end{enumerate} &
\begin{enumerate}[leftmargin=*]
    \item Applies to AI-related data processing when it meets the regulation's scope
    \item Does not classify or approve AI models as such
\end{enumerate} &
\begin{enumerate}[leftmargin=*]
    \item Promulgated on September 24, 2024
    \item Effective from January 1, 2025
\end{enumerate} \\
\hline
\end{longtable}

\clearpage
\section{Existing Policy Evaluation Framework Review}
\label{app:frameworkreview}

This appendix records the frameworks that contributed directly to the design synthesis. It is a selected methodological comparison, not a systematic literature review.

\begin{longtable}{|p{3cm}|p{5.5cm}|p{5.5cm}|}
\hline
\textbf{Paper} & \textbf{Summary} & \textbf{Limitations} \\
\hline
\endfirsthead

\multicolumn{3}{c}%
{{\tablename\ \thetable{} -- continued from previous page}} \\
\hline
\textbf{Paper} & \textbf{Summary} & \textbf{Limitations} \\
\hline
\endhead

\hline \multicolumn{3}{|r|}{{Continued on next page}} \\ \hline
\endfoot

\hline
\caption{Literature Review of AI Policy Frameworks}
\endlastfoot

UNESCO Readiness Assessment Methodology \cite{unesco2023ram} & The methodology supports a national, multi-stakeholder assessment of legal, social, economic, scientific, educational, and technical conditions for ethical AI. It combines document review, stakeholder participation, and recommendations for policy reform and institutional capacity. & It is a diagnostic country process rather than a fixed rubric for comparing the content of versioned policy portfolios. \\
\hline

Global Index on Responsible AI, 2026 \cite{adams2026girai} & The index assesses 135 countries and jurisdictions through 38 indicators that cover policy and implementation, civil-society engagement, enabling conditions, and documented government uses. & It combines policy content with implementation evidence and country conditions, while the present framework isolates the documented design of a declared policy portfolio. \\
\hline

AI Governance InternationaL Evaluation Index, 2025 \cite{zeng2025agile} & The index evaluates 40 countries through four pillars, 17 dimensions, and 43 indicators drawn from policy documents, governance practices, research outputs, and risk exposure. & Its unit is national governance capacity rather than the content and legal force of a versioned policy portfolio. \\
\hline

Government AI Readiness Index, 2025 \cite{oxfordinsights2025gari} & The index evaluates 195 governments on their capacity to use, enable, and govern AI for public benefit. & Readiness includes institutions, infrastructure, skills, and private-sector conditions that lie outside the present rubric. \\
\hline

A Grading Rubric for AI Safety Frameworks \cite{alaga2024gradingrubricaisafety} & The paper grades company frontier AI safety frameworks with seven criteria, 21 indicators, and letter grades from A to F. It also proposes surveys, Delphi studies, and audits as application methods. & The unit is a company's voluntary safety framework rather than a national portfolio of laws, strategies, budgets, and guidance. \\
\hline

Evaluating AI Providers' Frontier Safety Frameworks \cite{saferai2025evaluating} & The study applies 65 weighted criteria across four risk-management dimensions to 12 company frameworks. It shows how a detailed rubric can expose vague or missing commitments. & The criteria address catastrophic-risk management by frontier AI providers rather than national governance across sectors and institutions. \\
\hline

Artificial Intelligence Regulation: A Meta-Framework for Formulation and Governance \cite{Related4} & The meta-framework connects ethical principles, accountability, risk management, stakeholder roles, and feedback between technology and regulation. & It organizes policy formation and governance but does not provide an operational scoring protocol for cross-national comparison. \\
\hline

Assessing Frameworks for Eliciting Privacy and Security Requirements from Laws and Regulations \cite{priv} & The paper assesses methods for translating legal rules into privacy and security requirements and examines their coverage beyond the GDPR. & The unit is a software requirements method, so the analysis informs individual criteria but not the structure of a national policy score. \\
\hline

An AI System Evaluation Framework for Advancing AI Safety: Terminology, Taxonomy, Lifecycle Mapping \cite{mapping} & The framework harmonizes evaluation terminology, identifies elements of an AI system, and maps evaluations to lifecycle stages and supply-chain roles. & It evaluates AI systems and their environments rather than national policy portfolios. \\
\hline

Evaluation Framework to Guide Implementation of AI Systems into Healthcare Settings \cite{tehai} & TEHAI evaluates capability, utility, and adoption throughout the development and deployment of clinical AI systems. & Its clinical and system-level unit is narrower than national policy, although its attention to implementation and stakeholders informs several rubric items. \\
\hline

From AI to Digital Transformation: The AI Readiness Framework \cite{digital} & The framework assesses organizational AI readiness across technologies, activities, boundaries, and goals with a self-assessment scored from 0 to 4. & It was illustrated in one insurance organization and measures organizational readiness rather than policy design. \\
\hline

Implications of the Use of Artificial Intelligence in Public Governance \cite{impli} & The systematic review organizes reported opportunities and risks of AI in government and sets an empirical research agenda for public governance. & It synthesizes research on government use of AI rather than evaluating the content of national AI policy. \\
\hline

AI Governance and the Policymaking Process \cite{perry2019ai} & The paper connects AI governance with policy-process research and explains how political feasibility, timing, sequencing, and administrative capacity affect policy development. & It provides a conceptual agenda rather than a rubric or comparative measurement procedure. \\

\end{longtable}

\clearpage
\section{Complete National AI Policy Evaluation Framework}
\label{complete-eval-framework}
The weights below implement the default rule in Section~\ref{scoring-methodology}. Each category receives 20\%. Criteria are equally weighted within each category, and sub-criteria are equally weighted within each criterion.

\subsection*{Category 1. AI Risk Governance and Safety. Weight 20\%}

\begin{longtable}{|>{\raggedright\arraybackslash}p{0.2\textwidth}|>{\raggedright\arraybackslash}p{0.47\textwidth}|>{\raggedright\arraybackslash}p{0.08\textwidth}|}
\hline
\textbf{Criterion} & \textbf{Definition} & \textbf{Weight} \\
\hline
\endfirsthead

\multicolumn{3}{c}%
{{\tablename\ \thetable{} -- continued from previous page}} \\
\hline
\textbf{Criterion} & \textbf{Definition} & \textbf{Weight} \\
\hline
\endhead

\hline \multicolumn{3}{|r|}{{Continued on next page}} \\ \hline
\endfoot
\caption{AI Risk Governance and Safety Criteria} \label{tab:risk_governance_criteria}
\endlastfoot

1. Safety Standards & The policy includes specific and adaptable safety standards for AI systems, particularly for high-risk applications and advanced or general-purpose AI risks. & 20\% \\
\hline
2. Red Teaming and Auditing & The policy mandates or strongly encourages independent audits, red teaming, and other forms of external scrutiny to identify vulnerabilities and ensure compliance. & 20\% \\
\hline
3. Human Oversight and Interventions & The policy addresses the need for meaningful human control over AI systems, including mechanisms for intervention, override, or deactivation. & 20\% \\
\hline
4. System Interfacing and Delimitation & The policy considers measures for containing or isolating potentially harmful AI systems to prevent unintended consequences or widespread damage. & 20\% \\
\hline
5. Long-Term AI Governance and Infrastructure Resilience & The policy demonstrates an understanding of the long-term issues associated with governing advanced AI and infrastructure development. & 20\% \\
\hline
\end{longtable}
\textbf{(1) Safety Standards.}

\begin{longtable}{|>{\raggedright\arraybackslash}p{0.15\textwidth}|>{\raggedright\arraybackslash}p{0.25\textwidth}|>{\raggedright\arraybackslash}p{0.4\textwidth}|}

\hline
\textbf{Sub-criterion} & \textbf{Definition} & \textbf{Indicators} \\
\hline
\endfirsthead

\multicolumn{3}{c}%
{{\tablename\ \thetable{} -- continued from previous page}} \\
\hline
\textbf{Sub-criterion} & \textbf{Definition} & \textbf{Indicators} \\
\hline
\endhead

\hline \multicolumn{3}{|r|}{{Continued on next page}} \\ \hline
\endfoot

\caption{Safety Standards Sub-criteria and Indicators} \label{tab:safety_standards_subcriteria}
\endlastfoot

1.1. Specificity of Safety Frameworks or Standards & Are the safety standards clearly defined with specific technical requirements? &
\begin{enumerate}[label=\arabic*., before=\raggedright, itemsep=0pt, topsep=1pt, partopsep=0pt, leftmargin=*] 
    \item No specific safety standards mentioned.
    \item Vague safety standards with little technical detail, deferred to future development.
    \item Some specific safety standards are mentioned, but lack in detail or coverage, with concrete technical requirements and processes outlined for future work.
    \item Clearly defined safety standards with considerable technical detail.
    \item Comprehensive and highly specific safety standards with detailed technical requirements.
\end{enumerate} \\
\hline 
1.2. Coverage of Advanced AI Risks & Do the safety standards address risks from advanced or general-purpose AI systems regardless of the terminology used? &
\begin{enumerate}[label=\arabic*., before=\raggedright, itemsep=0pt, topsep=1pt, partopsep=0pt, leftmargin=*]
    \item No standards address risks from advanced or general-purpose AI systems.
    \item The policy recognizes these risks but gives no concrete standard.
    \item A concrete standard covers some relevant risks or systems.
    \item Clear standards cover the main risks and state when they apply.
    \item Detailed and adaptable standards cover capability growth, monitoring, and review.
\end{enumerate} \\
\hline 
1.3. High-Risk Application Focus & Are there specific standards for high-risk applications (e.g., autonomous weapons, critical infrastructure)? &
\begin{enumerate}[label=\arabic*., before=\raggedright, itemsep=0pt, topsep=1pt, partopsep=0pt, leftmargin=*]
    \item No differentiation of standards based on risk level.
    \item Vague acknowledgment of high-risk applications.
    \item Some specific standards for high-risk applications.
    \item Clear and stringent standards for high-risk applications.
    \item Comprehensive standards for high-risk applications with detailed provisions for risk mitigation.
\end{enumerate} \\
\hline 
\end{longtable}

\textbf{(2) Red Teaming and Auditing. }

\begin{longtable}{|>{\raggedright\arraybackslash}p{0.15\textwidth}|>{\raggedright\arraybackslash}p{0.25\textwidth}|>{\raggedright\arraybackslash}p{0.4\textwidth}|}

\hline
\textbf{Sub-criterion} & \textbf{Definition} & \textbf{Indicators} \\
\hline
\endfirsthead

\multicolumn{3}{c}%
{{\tablename\ \thetable{} -- continued from previous page}} \\
\hline
\textbf{Sub-criterion} & \textbf{Definition} & \textbf{Indicators} \\
\hline
\endhead

\hline \multicolumn{3}{|r|}{{Continued on next page}} \\ \hline
\endfoot
\caption{Red Teaming and Auditing Sub-criteria and Indicators} \label{tab:red_teaming_subcriteria}
\endlastfoot

2.1. Mandate for Independent Audits & Does the policy mandate independent audits for AI systems, particularly for high-risk applications? &
\begin{enumerate}[label=\arabic*., before=\raggedright, itemsep=0pt, topsep=1pt, partopsep=0pt, leftmargin=*]
    \item No mention of independent audits.
    \item Audits encouraged but not mandated.
    \item Audits mandated for some high-risk applications.
    \item Audits mandated for most high-risk applications.
    \item Audits mandated for all high-risk applications and strongly encouraged for others.
\end{enumerate} \\
\hline
2.2. Misuse and Dual-Use Prevention & Does the policy address the potential for AI technologies and systems to be misused or used for harmful purposes, considering their dual-use nature and the need for security measures to prevent unauthorized access, modification, or malicious use? &
\begin{enumerate}[label=\arabic*., before=\raggedright, itemsep=0pt, topsep=1pt, partopsep=0pt, leftmargin=*]
    \item No consideration of the potential for misuse, dual-use risks, or the need for security measures to prevent unauthorized use.
    \item Misuse and dual-use risks are acknowledged but not assessed or addressed through any concrete measures.
    \item Some assessment of potential misuse and dual-use risks is mentioned, and some general security measures are outlined.
    \item Clear guidelines for assessing and addressing misuse and dual-use risks are provided, along with clear and specific security requirements.
    \item A comprehensive framework is in place for managing misuse and dual-use risks, including detailed risk assessment processes, guidelines and restrictions related to dual-use technologies, and comprehensive security requirements.
\end{enumerate} \\
\hline
2.3. Red Teaming Requirements & Does the policy specify requirements or guidelines for red teaming exercises? &
\begin{enumerate}[label=\arabic*., before=\raggedright, itemsep=0pt, topsep=1pt, partopsep=0pt, leftmargin=*]
    \item No mention of red teaming.
    \item Red teaming encouraged but not defined.
    \item Red teaming mentioned with some guidelines.
    \item Red teaming requirements are clearly defined.
    \item Comprehensive red teaming requirements, including specific methodologies and scenarios.
\end{enumerate} \\
\hline
2.4. External Scrutiny Mechanisms & Does the policy establish mechanisms for external scrutiny beyond audits and red teaming (e.g., public reporting, expert panels)? &
\begin{enumerate}[label=\arabic*., before=\raggedright, itemsep=0pt, topsep=1pt, partopsep=0pt, leftmargin=*]
    \item No mechanisms for external scrutiny.
    \item Limited opportunities for external input.
    \item Some mechanisms for external scrutiny.
    \item Well-defined mechanisms for external scrutiny.
    \item Multiple, robust mechanisms for external scrutiny, ensuring transparency and accountability.
\end{enumerate} \\
\hline
\end{longtable}

\textbf{(3) Human Oversight and Interventions. }

\begin{longtable}{|>{\raggedright\arraybackslash}p{0.15\textwidth}| 
                >{\raggedright\arraybackslash}p{0.25\textwidth}| 
                >{\raggedright\arraybackslash}p{0.4\textwidth}|} 

\hline
\textbf{Sub-criterion} & \textbf{Definition} & \textbf{Indicators} \\
\hline
\endfirsthead

\multicolumn{3}{c}%
{{\tablename\ \thetable{} -- continued from previous page}} \\
\hline
\textbf{Sub-criterion} & \textbf{Definition} & \textbf{Indicators} \\
\hline
\endhead

\hline \multicolumn{3}{|r|}{{Continued on next page}} \\ \hline
\endfoot
\hline
\caption{Human Oversight and Interventions Sub-criteria and Indicators} \label{tab:intervention_subcriteria}
\endlastfoot

3.1. Human-in-the-Loop Requirements & Does the policy mandate human-in-the-loop systems for critical decision-making, especially in high-risk applications? &
\begin{enumerate}[label=\arabic*., before=\raggedright, itemsep=0pt, topsep=1pt, partopsep=0pt, leftmargin=*]
    \item No requirements for human oversight.
    \item Human oversight is encouraged but not mandated.
    \item Human-in-the-loop required for some critical decisions.
    \item Human-in-the-loop required for most critical decisions.
    \item Human-in-the-loop mandated for all critical decisions in high-risk applications.
\end{enumerate} \\
\hline 
3.2. Meaningful Human Control & Does the policy define what constitutes meaningful human control over AI systems and how it can be maintained? &
\begin{enumerate}[label=\arabic*., before=\raggedright, itemsep=0pt, topsep=1pt, partopsep=0pt, leftmargin=*]
    \item No definition of meaningful human control.
    \item Meaningful human control is mentioned but not clearly defined.
    \item Some aspects of meaningful human control are addressed.
    \item Clear definition of meaningful human control.
    \item Comprehensive framework for ensuring meaningful human control, including provisions for human override and intervention.
\end{enumerate} \\
\hline 
3.3. Oversight of Autonomous Systems & Does the policy address the specific challenges of overseeing increasingly autonomous AI systems? &
\begin{enumerate}[label=\arabic*., before=\raggedright, itemsep=0pt, topsep=1pt, partopsep=0pt, leftmargin=*]
    \item No consideration of oversight for autonomous systems.
    \item Oversight of autonomous systems is mentioned but not addressed in detail.
    \item Some guidelines for overseeing autonomous systems.
    \item Clear guidelines and requirements for overseeing autonomous systems.
    \item Comprehensive framework for overseeing autonomous systems, including provisions for monitoring, auditing, and accountability.
\end{enumerate} \\
\hline 
3.4. Override and Deactivation Mechanisms
 & Does the policy specify clear mechanisms for human operators to intervene in AI system operations, including overriding decisions and safely deactivating or shutting down systems in case of emergency, malfunction, or other critical situations? &
\begin{enumerate}[label=\arabic*., before=\raggedright, itemsep=0pt, topsep=1pt, partopsep=0pt, leftmargin=*]
     \item The policy does not mention override, intervention, or deactivation mechanisms.
    \item Mechanisms for override or deactivation are alluded to or mentioned as possible but are not clearly defined or detailed.
    \item Mechanisms for human override of AI decisions or system deactivation are defined for some critical scenarios, but not comprehensively for both, or are not easily accessible.
    \item Clear mechanisms for both human override of AI decisions and system deactivation (including emergency shutdown) are defined for most critical scenarios, with reasonable accessibility.
    \item Comprehensive, easily accessible, and robust mechanisms for both human override of AI decisions and system deactivation (including emergency shutdown procedures) are clearly defined, well-documented, and readily available for all relevant scenarios.
\end{enumerate} \\
\end{longtable}

\textbf{(4) System Interfacing and Delimitation. }

\begin{longtable}{|>{\raggedright\arraybackslash}p{0.15\textwidth}|>{\raggedright\arraybackslash}p{0.25\textwidth}|>{\raggedright\arraybackslash}p{0.4\textwidth}|}

\hline
\textbf{Sub-criterion} & \textbf{Definition} & \textbf{Indicators} \\
\hline
\endfirsthead

\multicolumn{3}{c}%
{{\tablename\ \thetable{} -- continued from previous page}} \\
\hline
\textbf{Sub-criterion} & \textbf{Definition} & \textbf{Indicators} \\
\hline
\endhead

\hline \multicolumn{3}{|r|}{{Continued on next page}} \\ \hline
\endfoot
\caption{System Interfacing and Delimitation Sub-criteria and Indicators} \label{tab:containment_subcriteria}
\endlastfoot

4.1. Research and Development Guidelines & Does the policy provide guidelines for research and development of containment or isolation techniques for advanced AI? &
\begin{enumerate}[label=\arabic*., before=\raggedright, itemsep=0pt, topsep=1pt, partopsep=0pt, leftmargin=*]
    \item No mention of containment or isolation research.
    \item Containment research is encouraged but not prioritized.
    \item Some guidelines for containment research.
    \item Clear guidelines for containment research and development.
    \item Comprehensive guidelines and funding for containment research, recognizing its importance.
\end{enumerate} \\
\hline 
4.2. Deployment Restrictions & Does the policy restrict the deployment of AI systems that cannot be adequately contained or isolated? &
\begin{enumerate}[label=\arabic*., before=\raggedright, itemsep=0pt, topsep=1pt, partopsep=0pt, leftmargin=*]
    \item No restrictions on deployment based on containment.
    \item Vague restrictions with unclear criteria.
    \item Some restrictions based on containment feasibility.
    \item Clear restrictions based on containment feasibility for high-risk applications.
    \item Comprehensive restrictions for any AI system that cannot be adequately contained or isolated.
\end{enumerate} \\
\hline 
4.3. Emergency Protocols & Does the policy include provisions for emergency protocols in case of containment or isolation failures? &
\begin{enumerate}[label=\arabic*., before=\raggedright, itemsep=0pt, topsep=1pt, partopsep=0pt, leftmargin=*]
    \item No mention of emergency protocols.
    \item Emergency protocols are underdeveloped.
    \item Some emergency protocols are defined.
    \item Clear emergency protocols are established.
    \item Comprehensive emergency protocols, including backup containment and mitigation strategies.
\end{enumerate} \\
\hline 
\end{longtable}

\textbf{(5) Long-Term AI Governance and Infrastructure Resilience. }

\begin{longtable}{|>{\raggedright\arraybackslash}p{0.15\textwidth}|>{\raggedright\arraybackslash}p{0.25\textwidth}|>{\raggedright\arraybackslash}p{0.4\textwidth}|}

\hline
\textbf{Sub-criterion} & \textbf{Definition} & \textbf{Indicators} \\
\hline
\endfirsthead

\multicolumn{3}{c}%
{{\tablename\ \thetable{} -- continued from previous page}} \\
\hline
\textbf{Sub-criterion} & \textbf{Definition} & \textbf{Indicators} \\
\hline
\endhead

\hline \multicolumn{3}{|r|}{{Continued on next page}} \\ \hline
\endfoot
\caption{Long-Term AI Governance and Infrastructure Resilience Sub-criteria and Indicators} \label{tab:long_term_subcriteria}
\endlastfoot

5.1. Adaptive Governance & Does the policy include mechanisms for adapting governance structures and regulations as AI technology, particularly AGI, advances? &
\begin{enumerate}[label=\arabic*., before=\raggedright, itemsep=0pt, topsep=1pt, partopsep=0pt, leftmargin=*]
    \item No mechanisms for adaptation.
    \item Adaptation is possible but not clearly defined.
    \item Some mechanisms for adaptation based on new information.
    \item Clear process for reviewing and updating policies in response to AGI advancements.
    \item Flexible and adaptive governance framework with built-in mechanisms for continuous learning and adjustment.
\end{enumerate} \\
\hline
5.2. AI Supply Chain Security and Resilience & Does the policy address security risks within the AI supply chain (hardware, software, data, vendors), promotes resilience against disruptions or compromises, and includes strategic measures to foster domestic capabilities, aiming to reduce critical dependencies and enhance national self-sufficiency in key AI technologies. &
\begin{enumerate}[label=\arabic*., before=\raggedright, itemsep=0pt, topsep=1pt, partopsep=0pt, leftmargin=*]
    \item No mention of AI supply chain risks, resilience, or national dependencies.
    \item Vague acknowledgement of supply chain risks with no specific requirements or strategies for domestic capacity.
    \item Identifies some AI supply chain risks and recommends basic security practices or domestic sourcing preference, but lacks concrete strategic plans, dedicated funding, or specific incentives for AI self-sufficiency.
    \item Requires specific risk management practices for critical AI supply chain elements and includes targeted initiatives with dedicated funding or strong incentives to bolster domestic capacity and reduce dependencies in key AI technology areas.
    \item Mandates a comprehensive national strategy for AI supply chain security and resilience, integrating robust risk management, resilience plans against disruptions, significant investment in domestic self-sufficiency, and clear goals to reduce critical foreign dependencies.
\end{enumerate} \\
\hline
\end{longtable}

\subsection*{Category 2. Effectiveness and Feasibility. Weight 20\%}

\begin{longtable}{|>{\raggedright\arraybackslash}p{0.2\textwidth}|>{\raggedright\arraybackslash}p{0.47\textwidth}|>{\raggedright\arraybackslash}p{0.08\textwidth}|}

\hline
\textbf{Criterion} & \textbf{Definition} & \textbf{Weight} \\
\hline
\endfirsthead

\multicolumn{3}{c}%
{{\tablename\ \thetable{} -- continued from previous page}} \\
\hline
\textbf{Criterion} & \textbf{Definition} & \textbf{Weight} \\
\hline
\endhead

\hline \multicolumn{3}{|r|}{{Continued on next page}} \\ \hline
\endfoot
\hline
\caption{Effectiveness and Feasibility Criteria} \label{tab:effectiveness_criteria}
\endlastfoot

6. Institutional Capacity and Authority & The policy establishes appropriately empowered, independent, resourced, and coordinated institutions to implement, oversee, and enforce it. & 16.67\% \\
\hline
7. Clarity and Specificity & The policy provisions are clearly defined, specific, and actionable, avoiding ambiguity and leaving little room for misinterpretation. & 16.67\% \\
\hline
8. Enforceability & The policy includes mechanisms for monitoring compliance, detecting violations, and enforcing its provisions, including appropriate penalties for non-compliance. & 16.67\% \\
\hline
9. Measurability & The policy objectives and intended outputs are measurable, allowing later assessment of implementation and results. & 16.67\% \\
\hline
10. Resource Requirements & The resources required for implementing and enforcing the policy are realistic, feasible, and proportionate to the policy's objectives. & 16.67\% \\
\hline
11. Incentive Alignment & The policy aligns incentives for developers, deployers, users, and other affected actors to promote responsible AI development and deployment. & 16.67\% \\
\end{longtable}

\textbf{(6) Institutional Capacity and Authority. }

\begin{longtable}{|>{\raggedright\arraybackslash}p{0.15\textwidth}|>{\raggedright\arraybackslash}p{0.25\textwidth}|>{\raggedright\arraybackslash}p{0.40\textwidth}|}

\hline
\textbf{Sub-criterion} & \textbf{Definition} & \textbf{Indicators} \\
\hline
\endfirsthead

\multicolumn{3}{c}%
{{\tablename\ \thetable{} -- continued from previous page}} \\
\hline
\textbf{Sub-criterion} & \textbf{Definition} & \textbf{Indicators} \\
\hline
\endhead

\hline \multicolumn{3}{|r|}{{Continued on next page}} \\ \hline
\endfoot

\caption{Institutional Capacity and Authority Sub-criteria and Indicators} \label{tab:institutional_subcriteria}
\endlastfoot

6.1. Dedicated Oversight Body & Existence and specific mandate of a body (or bodies) primarily responsible for AI policy implementation, oversight, and enforcement. &
\begin{enumerate}[label=\arabic*., before=\raggedright, itemsep=0pt, topsep=1pt, partopsep=0pt, leftmargin=*]
    \item No specific body identified or proposed for AI oversight. Responsibility is unassigned.
    \item Mentions the need for oversight but relies on existing bodies without clearly expanding their mandate or resources for AI.
    \item Designates an existing body or proposes a new body but with an unclear, limited, or overlapping mandate specific to AI governance.
    \item Clearly designates or establishes a specific body with a defined mandate covering key aspects of the AI policy implementation and oversight.
    \item Establishes or clearly empowers a dedicated body with a comprehensive, clearly defined, and primary mandate covering all major policy areas, including monitoring, guidance, and coordinating enforcement.
\end{enumerate} \\
\hline
6.2. Independence and Authority & Level of operational independence from political or corporate influence and the legal authority granted to the oversight body. &
\begin{enumerate}[label=\arabic*., before=\raggedright, itemsep=0pt, topsep=1pt, partopsep=0pt, leftmargin=*]
    \item The oversight body (if any) has no specified independence or distinct legal authority for AI matters.
    \item Vague statements about needing authority or impartiality, but no structural guarantees or specific powers defined.
    \item Some measures for independence are outlined (e.g., reporting structure), but legal authority for key actions (e.g., investigation, binding guidance) is limited or unclear.
    \item Clear provisions ensuring significant operational independence and defined legal authority for key functions (e.g., investigation, issuing fines, setting standards within delegated powers).
    \item Strong structural/statutory independence guaranteed, coupled with comprehensive legal authority (e.g., independent rule-making within its mandate, broad investigative powers, direct enforcement capabilities).
\end{enumerate} \\
\hline
6.3. Resources and Expertise & Adequacy of dedicated funding, staffing, and necessary technical, legal, and ethical expertise within the oversight body. &
\begin{enumerate}[label=\arabic*., before=\raggedright, itemsep=0pt, topsep=1pt, partopsep=0pt, leftmargin=*]
    \item No mention of specific resource allocation or expertise requirements for the oversight body.
    \item Acknowledges the need for resources/expertise but provides no concrete commitments or plans.
    \item Some resource allocation is mentioned but does not match the mandate. The policy refers generally to a need for diverse expertise.
    \item Specifies dedicated funding streams or significant budget allocation and mandates the inclusion of specific technical, ethical, and legal expertise.
    \item Guarantees substantial, ring-fenced, long-term funding, mandates specific high-level expertise across relevant domains, and includes plans for talent recruitment, retention, and training.
\end{enumerate} \\
\hline
6.4. Interagency Coordination Mechanisms & Existence and operation of formal coordination and information-sharing procedures among government bodies involved in AI governance. &
\begin{enumerate}[label=\arabic*., before=\raggedright, itemsep=0pt, topsep=1pt, partopsep=0pt, leftmargin=*]
    \item No mention of coordination between different government agencies on AI.
    \item Vague mention of the need for collaboration without establishing specific mechanisms.
    \item Establishes informal or ad-hoc coordination mechanisms (e.g., occasional meetings).
    \item Establishes formal coordination structures (e.g., standing inter-agency committee or task force) with defined roles and responsibilities for AI policy.
    \item Establishes robust, permanent coordination mechanisms with clear mandates, leadership, regular reporting, dedicated resources, and clear protocols for information sharing and joint action on AI governance.
\end{enumerate} \\
\hline
\end{longtable}

\textbf{(7) Clarity and Specificity. }

\begin{longtable}{|>{\raggedright\arraybackslash}p{0.15\textwidth}|>{\raggedright\arraybackslash}p{0.25\textwidth}|>{\raggedright\arraybackslash}p{0.40\textwidth}|}

\hline
\textbf{Sub-criterion} & \textbf{Definition} & \textbf{Indicators} \\
\hline
\endfirsthead

\multicolumn{3}{c}%
{{\tablename\ \thetable{} -- continued from previous page}} \\
\hline
\textbf{Sub-criterion} & \textbf{Definition} & \textbf{Indicators} \\
\hline
\endhead

\hline \multicolumn{3}{|r|}{{Continued on next page}} \\ \hline
\endfoot

\caption{Clarity and Specificity Sub-criteria and Indicators} \label{tab:clarity_subcriteria}
\endlastfoot

7.1. Implementation Plan & Clarity, feasibility, and level of detail provided in the policy regarding its implementation plan, including phases, timelines, milestones, and responsible parties. &
\begin{enumerate}[label=\arabic*., before=\raggedright, itemsep=0pt, topsep=1pt, partopsep=0pt, leftmargin=*]
    \item No implementation plan or timeline outlined in the policy.
    \item Vague statements about future implementation or gradual rollout without specifics.
    \item Outlines some implementation steps or phases but lacks clear timelines, measurable milestones, or specific assignment of responsibilities.
    \item Provides a clear implementation plan with distinct phases, key milestones, assigned responsibilities for major actions, and indicative timelines.
    \item Provides a detailed, phased implementation plan with specific and measurable milestones, concrete timelines, clearly assigned responsibilities across agencies, and links to resource allocation.
\end{enumerate} \\
\hline
\end{longtable}

\textbf{(8) Enforceability. } 

\begin{longtable}{|>{\raggedright\arraybackslash}p{0.15\textwidth}|>{\raggedright\arraybackslash}p{0.25\textwidth}|>{\raggedright\arraybackslash}p{0.40\textwidth}|}

\hline
\textbf{Sub-criterion} & \textbf{Definition} & \textbf{Indicators} \\
\hline
\endfirsthead

\multicolumn{3}{c}%
{{\tablename\ \thetable{} -- continued from previous page}} \\
\hline
\textbf{Sub-criterion} & \textbf{Definition} & \textbf{Indicators} \\
\hline
\endhead

\hline \multicolumn{3}{|r|}{{Continued on next page}} \\ \hline
\endfoot

\caption{Enforceability Sub-criteria and Indicators} \label{tab:enforceability_subcriteria} \\
\endlastfoot

8.1. Technical Feasibility & Are the technical requirements of the policy feasible to implement with current or near-future technology? &
\begin{enumerate}[label=\arabic*., before=\raggedright, itemsep=0pt, topsep=1pt, partopsep=0pt, leftmargin=*]
    \item Policy requirements are largely unimplementable with current technology.
    \item Most requirements are technically challenging or impractical.
    \item A notable number of requirements are technically feasible, but others are challenging.
    \item Most requirements are technically feasible.
    \item All requirements are technically feasible with current or near-future technology.
\end{enumerate} \\
\hline
8.2. Compliance Monitoring & Does the policy establish clear mechanisms for monitoring compliance? &
\begin{enumerate}[label=\arabic*., before=\raggedright, itemsep=0pt, topsep=1pt, partopsep=0pt, leftmargin=*]
    \item No mechanisms for monitoring compliance.
    \item Monitoring mechanisms are vaguely defined or inadequate.
    \item Some mechanisms for monitoring compliance are outlined.
    \item Well-defined mechanisms for monitoring compliance.
    \item Comprehensive and robust mechanisms for monitoring compliance, including regular audits and reporting.
\end{enumerate} \\
\hline
8.3. Enforcement Mechanisms & Does the policy specify enforcement mechanisms and penalties for non-compliance? &
\begin{enumerate}[label=\arabic*., before=\raggedright, itemsep=0pt, topsep=1pt, partopsep=0pt, leftmargin=*]
    \item No enforcement mechanisms or penalties specified.
    \item Enforcement mechanisms are weak or unclear.
    \item Some enforcement mechanisms and penalties are defined.
    \item Clear enforcement mechanisms and appropriate penalties are specified.
    \item Comprehensive enforcement mechanisms with a range of penalties, including provisions for addressing serious violations.
\end{enumerate} \\
\hline
8.4. Legal Status and Force & Assessment of the policy document's formal legal standing and its binding nature on relevant actors. &
\begin{enumerate}[label=\arabic*., before=\raggedright, itemsep=0pt, topsep=1pt, partopsep=0pt, leftmargin=*]
    \item Policy is purely informational, aspirational, or strategic (e.g., national strategy document, white paper) with no legal obligations.
    \item Policy takes the form of voluntary guidelines, recommendations, or a code of conduct with no direct legal enforceability.
    \item Policy uses soft regulatory tools (e.g., mandatory for government procurement only, relies on certification schemes) or has limited binding applicability (e.g., specific narrow sector).
    \item Policy is a binding regulation or executive order creating clear legal obligations for defined actors/sectors, with established legal consequences for non-compliance.
    \item Policy is enshrined in comprehensive, legally binding primary legislation (Statute/Act) with broad applicability across relevant sectors and strong, clearly defined enforcement powers and penalties.
\end{enumerate} \\
\hline
\end{longtable}

\textbf{(9) Measurability. } 

\begin{longtable}{|>{\raggedright\arraybackslash}p{0.15\textwidth}|>{\raggedright\arraybackslash}p{0.25\textwidth}|>{\raggedright\arraybackslash}p{0.40\textwidth}|}

\hline
\textbf{Sub-criterion} & \textbf{Definition} & \textbf{Indicators} \\
\hline
\endfirsthead

\multicolumn{3}{c}%
{{\tablename\ \thetable{} -- continued from previous page}} \\
\hline
\textbf{Sub-criterion} & \textbf{Definition} & \textbf{Indicators} \\
\hline
\endhead

\hline \multicolumn{3}{|r|}{{Continued on next page}} \\ \hline
\endfoot

\caption{Measurability Sub-criteria and Indicators} \label{tab:measurability_subcriteria}
\endlastfoot

9.1. Quantifiable Objectives & Does the policy define quantifiable objectives that can be used to measure its success? &
\begin{enumerate}[label=\arabic*., before=\raggedright, itemsep=0pt, topsep=1pt, partopsep=0pt, leftmargin=*]
    \item No quantifiable objectives defined.
    \item Objectives are vaguely defined and difficult to measure.
    \item Some objectives are quantifiable, but others are not.
    \item Most objectives are quantifiable.
    \item All key objectives are clearly defined and quantifiable.
\end{enumerate} \\
\hline
9.2. Data Collection and Reporting & Does the policy include provisions for collecting and reporting data relevant to the policy's objectives and outcomes? &
\begin{enumerate}[label=\arabic*., before=\raggedright, itemsep=0pt, topsep=1pt, partopsep=0pt, leftmargin=*]
    \item No provisions for data collection or reporting.
    \item Data collection and reporting are vaguely defined.
    \item Some data collection and reporting requirements are outlined.
    \item Clear data collection and reporting requirements are specified.
    \item Comprehensive data collection and reporting requirements, including provisions for data quality and transparency.
\end{enumerate} \\
\hline
9.3. Performance Metrics & Does the policy define specific performance metrics to assess the effectiveness of AI systems in meeting the policy's objectives? &
\begin{enumerate}[label=\arabic*., before=\raggedright, itemsep=0pt, topsep=1pt, partopsep=0pt, leftmargin=*]
    \item No performance metrics defined.
    \item Performance metrics are vaguely defined or inadequate.
    \item Some performance metrics are defined.
    \item Clear and relevant performance metrics are specified.
    \item Comprehensive set of performance metrics aligned with policy objectives.
\end{enumerate} \\
\hline
\end{longtable}

\textbf{(10) Resource Requirements. }

\begin{longtable}{|>{\raggedright\arraybackslash}p{0.15\textwidth}|>{\raggedright\arraybackslash}p{0.25\textwidth}|>{\raggedright\arraybackslash}p{0.40\textwidth}|}

\hline
\textbf{Sub-criterion} & \textbf{Definition} & \textbf{Indicators} \\
\hline
\endfirsthead

\multicolumn{3}{c}%
{{\tablename\ \thetable{} -- continued from previous page}} \\
\hline
\textbf{Sub-criterion} & \textbf{Definition} & \textbf{Indicators} \\
\hline
\endhead

\hline \multicolumn{3}{|r|}{{Continued on next page}} \\ \hline
\endfoot

\caption{Resource Requirements Sub-criteria and Indicators} \label{tab:resources_subcriteria}
\endlastfoot

10.1. Financial Resources & Does the policy consider the financial resources required for implementation and enforcement? &
\begin{enumerate}[label=\arabic*., before=\raggedright, itemsep=0pt, topsep=1pt, partopsep=0pt, leftmargin=*]
    \item No consideration of financial resource requirements.
    \item Financial resource requirements are underestimated or unrealistic.
    \item Some consideration of financial resource requirements.
    \item Realistic assessment of financial resource requirements.
    \item Comprehensive analysis of financial resource requirements, including provisions for funding and cost-effectiveness.
\end{enumerate} \\
\hline
10.2. Human Resources & Does the policy consider the human resources (e.g., expertise, personnel) required for implementation and enforcement? &
\begin{enumerate}[label=\arabic*., before=\raggedright, itemsep=0pt, topsep=1pt, partopsep=0pt, leftmargin=*]
    \item No consideration of human resource requirements.
    \item Human resource requirements are underestimated or unrealistic.
    \item Some consideration of human resource requirements.
    \item Realistic assessment of human resource requirements.
    \item Comprehensive analysis of human resource requirements, including provisions for training and capacity building.
\end{enumerate} \\
\hline
10.3. Technological Resources & Does the policy consider the technological resources (e.g., infrastructure, tools, data) required for implementation and enforcement? &
\begin{enumerate}[label=\arabic*., before=\raggedright, itemsep=0pt, topsep=1pt, partopsep=0pt, leftmargin=*]
    \item No consideration of technological resource requirements.
    \item Technological resource requirements are underestimated or unrealistic.
    \item Some consideration of technological resource requirements.
    \item Realistic assessment of technological resource requirements.
    \item Comprehensive analysis of technological resource requirements, including provisions for infrastructure and tools.
\end{enumerate} \\
\hline
\end{longtable}

\textbf{(11) Incentive Alignment. }

\begin{longtable}{|>{\raggedright\arraybackslash}p{0.15\textwidth}|>{\raggedright\arraybackslash}p{0.25\textwidth}|>{\raggedright\arraybackslash}p{0.40\textwidth}|}

\hline
\textbf{Sub-criterion} & \textbf{Definition} & \textbf{Indicators} \\
\hline
\endfirsthead

\multicolumn{3}{c}%
{{\tablename\ \thetable{} -- continued from previous page}} \\
\hline
\textbf{Sub-criterion} & \textbf{Definition} & \textbf{Indicators} \\
\hline
\endhead

\hline \multicolumn{3}{|r|}{{Continued on next page}} \\ \hline
\endfoot

\caption{Incentive Alignment Sub-criteria and Indicators} \label{tab:incentive_subcriteria}
\endlastfoot

11.1. Positive Incentives & Does the policy provide positive incentives (e.g., rewards, recognition) for responsible AI practices? &
\begin{enumerate}[label=\arabic*., before=\raggedright, itemsep=0pt, topsep=1pt, partopsep=0pt, leftmargin=*]
    \item No positive incentives provided.
    \item Positive incentives are vaguely defined or weak.
    \item Some positive incentives are provided.
    \item Clear and meaningful positive incentives are offered.
    \item Comprehensive set of positive incentives that effectively promote responsible AI practices.
\end{enumerate} \\
\hline
11.2. Negative Incentives & Does the policy include negative incentives (e.g., penalties, sanctions) for irresponsible AI practices? &
\begin{enumerate}[label=\arabic*., before=\raggedright, itemsep=0pt, topsep=1pt, partopsep=0pt, leftmargin=*]
    \item No negative incentives specified.
    \item Negative incentives are weak or unclear.
    \item Some negative incentives are defined.
    \item Clear and proportionate negative incentives are specified.
    \item Comprehensive set of negative incentives that effectively deter irresponsible AI practices.
\end{enumerate} \\
\hline
11.3. Incentive Compatibility & Are the incentives designed to be compatible with each other and with the overall objectives of the policy? &
\begin{enumerate}[label=\arabic*., before=\raggedright, itemsep=0pt, topsep=1pt, partopsep=0pt, leftmargin=*]
    \item Incentives are contradictory or misaligned.
    \item Incentives are weakly aligned with each other and with policy objectives.
    \item Some incentives are aligned, but others are not.
    \item Most incentives are aligned with each other and with policy objectives.
    \item Incentives are fully compatible with each other and strongly aligned with the overall objectives of the policy.
\end{enumerate} \\
\hline
\end{longtable}

\subsection*{Category 3. Comprehensiveness and Scope. Weight 20\%}

\begin{longtable}{|>{\raggedright\arraybackslash}p{0.2\textwidth}|>{\raggedright\arraybackslash}p{0.47\textwidth}|>{\raggedright\arraybackslash}p{0.08\textwidth}|}

\hline
\textbf{Criterion} & \textbf{Definition} & \textbf{Weight} \\
\hline
\endfirsthead

\multicolumn{3}{c}%
{{\tablename\ \thetable{} -- continued from previous page}} \\
\hline
\textbf{Criterion} & \textbf{Definition} & \textbf{Weight} \\
\hline
\endhead

\hline \multicolumn{3}{|r|}{{Continued on next page}} \\ \hline
\endfoot

\caption{Comprehensiveness and Scope Criteria} \label{tab:comprehensiveness_criteria} \\
\endlastfoot

12. Coverage of AI Lifecycle & The policy addresses the full lifecycle of AI systems, from initial design and data collection to development, deployment, monitoring, and eventual decommissioning. & 20\% \\
\hline
13. Range of Risks Addressed & The policy covers a wide range of risks associated with AI systems, including technical, ethical, societal, and governance risks. & 20\% \\
\hline
14. Stakeholder Inclusion & The policy considers the perspectives and roles of developers, deployers, users, regulators, and affected communities. & 20\% \\
\hline
15. Adaptability to Technological Advancements & The policy includes mechanisms for adapting to new and emerging AI technologies, including provisions for ongoing research, monitoring, and updates. & 20\% \\
\hline
16. International Harmonization & The policy aligns with or considers international standards, practices, and agreements related to AI governance, promoting consistency and cooperation across borders. & 20\% \\
\hline
\end{longtable}

\textbf{(12) Coverage of AI Lifecycle.}

\begin{longtable}{|>{\raggedright\arraybackslash}p{0.15\textwidth}|>{\raggedright\arraybackslash}p{0.25\textwidth}|>{\raggedright\arraybackslash}p{0.40\textwidth}|}

\hline
\textbf{Sub-criterion} & \textbf{Definition} & \textbf{Indicators} \\
\hline
\endfirsthead

\multicolumn{3}{c}%
{{\tablename\ \thetable{} -- continued from previous page}} \\
\hline
\textbf{Sub-criterion} & \textbf{Definition} & \textbf{Indicators} \\
\hline
\endhead

\hline \multicolumn{3}{|r|}{{Continued on next page}} \\ \hline
\endfoot

\caption{Coverage of AI Lifecycle Sub-criteria and Indicators} \label{tab:lifecycle_subcriteria}

\endlastfoot

12.1. Design Phase & Does the policy include provisions for responsible AI design principles and practices? &
\begin{enumerate}[label=\arabic*., before=\raggedright, itemsep=0pt, topsep=1pt, partopsep=0pt, leftmargin=*]
    \item No consideration of the design phase.
    \item Design phase is mentioned but not addressed in detail.
    \item Some guidelines for responsible AI design.
    \item Clear guidelines and requirements for responsible AI design.
    \item Comprehensive framework for responsible AI design, including ethical considerations and risk assessment.
\end{enumerate} \\
\hline
12.2. Data Collection and Preprocessing & Does the policy address data collection, preprocessing, and governance practices? &
\begin{enumerate}[label=\arabic*., before=\raggedright, itemsep=0pt, topsep=1pt, partopsep=0pt, leftmargin=*]
    \item No consideration of data practices.
    \item Data practices are mentioned but not addressed in detail.
    \item Some guidelines for responsible data practices.
    \item Clear guidelines and requirements for data collection, preprocessing, and governance.
    \item Comprehensive framework for responsible data practices, including provisions for data quality, fairness, and privacy.
\end{enumerate} \\
\hline
12.3. Development and Testing & Does the policy include provisions for responsible AI development and testing methodologies? &
\begin{enumerate}[label=\arabic*., before=\raggedright, itemsep=0pt, topsep=1pt, partopsep=0pt, leftmargin=*]
    \item No consideration of development and testing.
    \item Development and testing are mentioned but not addressed in detail.
    \item Some guidelines for responsible development and testing.
    \item Clear guidelines and requirements for development and testing, including validation and verification.
    \item Comprehensive framework for responsible development and testing, including provisions for robustness, security, and explainability.
\end{enumerate} \\
\hline
12.4. Deployment and Monitoring & Does the policy address the deployment and ongoing monitoring of AI systems? &
\begin{enumerate}[label=\arabic*., before=\raggedright, itemsep=0pt, topsep=1pt, partopsep=0pt, leftmargin=*]
    \item No consideration of deployment and monitoring.
    \item Deployment and monitoring are mentioned but not addressed in detail.
    \item Some guidelines for responsible deployment and monitoring.
    \item Clear guidelines and requirements for deployment and monitoring, including performance tracking and risk assessment.
    \item Comprehensive framework for responsible deployment and monitoring, including provisions for continuous evaluation and adaptation.
\end{enumerate} \\
\hline
12.5. Decommissioning & Does the policy address the eventual decommissioning or retirement of AI systems? &
\begin{enumerate}[label=\arabic*., before=\raggedright, itemsep=0pt, topsep=1pt, partopsep=0pt, leftmargin=*]
    \item No consideration of decommissioning.
    \item Decommissioning is mentioned but not addressed in detail.
    \item Some guidelines for responsible decommissioning.
    \item Clear guidelines and requirements for decommissioning.
    \item Comprehensive framework for responsible decommissioning, including provisions for data retention, model updates, and system retirement.
\end{enumerate} \\
\hline
\end{longtable}

\textbf{(13) Range of Risks Addressed.}

\begin{longtable}{|>{\raggedright\arraybackslash}p{0.15\textwidth}|>{\raggedright\arraybackslash}p{0.25\textwidth}|>{\raggedright\arraybackslash}p{0.40\textwidth}|}

\hline
\textbf{Sub-criterion} & \textbf{Definition} & \textbf{Indicators} \\
\hline
\endfirsthead

\multicolumn{3}{c}%
{{\tablename\ \thetable{} -- continued from previous page}} \\
\hline
\textbf{Sub-criterion} & \textbf{Definition} & \textbf{Indicators} \\
\hline
\endhead

\hline \multicolumn{3}{|r|}{{Continued on next page}} \\ \hline
\endfoot

\caption{Range of Risks Addressed Sub-criteria and Indicators} \label{tab:range_risks_subcriteria}
\endlastfoot

13.1. Technical Risks & Does the policy address technical risks such as safety, security, robustness, reliability, and integration with existing non-AI systems and infrastructure? &
\begin{enumerate}[label=\arabic*., before=\raggedright, itemsep=0pt, topsep=1pt, partopsep=0pt, leftmargin=*]
    \item No consideration of technical risks.
    \item Technical risks are mentioned but not addressed in detail.
    \item Some guidelines for addressing technical risks.
    \item Clear guidelines and requirements for mitigating technical risks.
    \item Comprehensive framework for managing technical risks, including specific standards and best practices.
\end{enumerate} \\
\hline
13.2. Ethical Risks & Does the policy address ethical risks such as bias, fairness, transparency, accountability, and privacy? &
\begin{enumerate}[label=\arabic*., before=\raggedright, itemsep=0pt, topsep=1pt, partopsep=0pt, leftmargin=*]
    \item No consideration of ethical risks.
    \item Ethical risks are mentioned but not addressed in detail.
    \item Some guidelines for addressing ethical risks.
    \item Clear guidelines and requirements for mitigating ethical risks.
    \item Comprehensive framework for managing ethical risks, including provisions for ethical audits and impact assessments.
\end{enumerate} \\
\hline
13.3. Societal Risks & Does the policy address societal risks such as job displacement, inequality, social manipulation, and erosion of trust? &
\begin{enumerate}[label=\arabic*., before=\raggedright, itemsep=0pt, topsep=1pt, partopsep=0pt, leftmargin=*]
    \item No consideration of societal risks.
    \item Societal risks are mentioned but not addressed in detail.
    \item Some guidelines for addressing societal risks.
    \item Clear guidelines and requirements for mitigating societal risks.
    \item Comprehensive framework for managing societal risks, including provisions for public engagement and social impact assessments.
\end{enumerate} \\
\hline
13.4. Governance Risks & Does the policy address governance risks such as lack of accountability, regulatory gaps, and challenges to international cooperation? &
\begin{enumerate}[label=\arabic*., before=\raggedright, itemsep=0pt, topsep=1pt, partopsep=0pt, leftmargin=*]
    \item No consideration of governance risks.
    \item Governance risks are mentioned but not addressed in detail.
    \item Some guidelines for addressing governance risks.
    \item Clear guidelines and requirements for mitigating governance risks.
    \item Comprehensive framework for managing governance risks, including provisions for regulatory oversight and international collaboration.
\end{enumerate} \\
\hline
\end{longtable}

\textbf{(14) Stakeholder Inclusion.}

\begin{longtable}{|>{\raggedright\arraybackslash}p{0.15\textwidth}|>{\raggedright\arraybackslash}p{0.25\textwidth}|>{\raggedright\arraybackslash}p{0.40\textwidth}|}

\hline
\textbf{Sub-criterion} & \textbf{Definition} & \textbf{Indicators} \\
\hline
\endfirsthead

\multicolumn{3}{c}%
{{\tablename\ \thetable{} -- continued from previous page}} \\
\hline
\textbf{Sub-criterion} & \textbf{Definition} & \textbf{Indicators} \\
\hline
\endhead

\hline \multicolumn{3}{|r|}{{Continued on next page}} \\ \hline
\endfoot

\caption{Stakeholder Inclusion Sub-criteria and Indicators} \label{tab:stakeholder_subcriteria}
\endlastfoot

14.1. Stakeholder Identification & Does the policy clearly identify all relevant stakeholders affected by AI systems? &
\begin{enumerate}[label=\arabic*., before=\raggedright, itemsep=0pt, topsep=1pt, partopsep=0pt, leftmargin=*]
    \item No stakeholders are identified.
    \item Some stakeholders are identified, but with significant gaps.
    \item Most relevant stakeholders are identified.
    \item All relevant stakeholders are clearly identified.
    \item Comprehensive stakeholder identification, including marginalized and underrepresented groups.
\end{enumerate} \\
\hline
14.2. Stakeholder Engagement & Does the policy include mechanisms for engaging stakeholders in the policy development and implementation process? &
\begin{enumerate}[label=\arabic*., before=\raggedright, itemsep=0pt, topsep=1pt, partopsep=0pt, leftmargin=*]
    \item No mechanisms for stakeholder engagement.
    \item Limited opportunities for stakeholder input.
    \item Some mechanisms for stakeholder engagement.
    \item Well-defined mechanisms for stakeholder engagement.
    \item Multiple, robust mechanisms for stakeholder engagement, ensuring diverse perspectives are considered.
\end{enumerate} \\
\hline
14.3. Stakeholder Roles and Responsibilities & Does the policy clearly define the roles and responsibilities of different stakeholders in relation to AI safety and governance? &
\begin{enumerate}[label=\arabic*., before=\raggedright, itemsep=0pt, topsep=1pt, partopsep=0pt, leftmargin=*]
    \item Roles and responsibilities are not defined.
    \item Roles and responsibilities are vaguely mentioned.
    \item Some roles and responsibilities are defined, but gaps remain.
    \item Most roles and responsibilities are clearly defined.
    \item Comprehensive definition of roles and responsibilities for all relevant stakeholders, including provisions for accountability and redress.
\end{enumerate} \\
\hline
\end{longtable}

\textbf{(15) Adaptability to Technological Advancements.}

\begin{longtable}{|>{\raggedright\arraybackslash}p{0.15\textwidth}|>{\raggedright\arraybackslash}p{0.25\textwidth}|>{\raggedright\arraybackslash}p{0.40\textwidth}|}

\hline
\textbf{Sub-criterion} & \textbf{Definition} & \textbf{Indicators} \\
\hline
\endfirsthead

\multicolumn{3}{c}%
{{\tablename\ \thetable{} -- continued from previous page}} \\
\hline
\textbf{Sub-criterion} & \textbf{Definition} & \textbf{Indicators} \\
\hline
\endhead

\hline \multicolumn{3}{|r|}{{Continued on next page}} \\ \hline
\endfoot

\caption{Adaptability to Technological Advancements Sub-criteria and Indicators} \label{tab:adaptability_subcriteria}
\endlastfoot

15.1. Technology Monitoring & Does the policy establish mechanisms for monitoring advancements in AI technology and their potential implications? &
\begin{enumerate}[label=\arabic*., before=\raggedright, itemsep=0pt, topsep=1pt, partopsep=0pt, leftmargin=*]
    \item No provisions for technology monitoring.
    \item Technology monitoring is mentioned but not clearly defined.
    \item Some mechanisms for technology monitoring are outlined.
    \item Clear mechanisms for monitoring technological advancements.
    \item Comprehensive framework for monitoring technological advancements, including early warning systems for emerging risks.
\end{enumerate} \\
\hline
15.2. Policy Review and Update & Does the policy include a process for regularly reviewing and updating its provisions in response to new technological developments? &
\begin{enumerate}[label=\arabic*., before=\raggedright, itemsep=0pt, topsep=1pt, partopsep=0pt, leftmargin=*]
    \item No process for policy review or update.
    \item Policy review is possible but not clearly defined or mandated.
    \item Some provisions for policy review and update.
    \item Clear process for regular policy review and update.
    \item Comprehensive process for policy review and update, including triggers for review based on specific technological advancements.
\end{enumerate} \\
\hline
15.3. Flexibility and Adaptability & Is the policy framework designed to be flexible and adaptable to accommodate new AI technologies and applications? &
\begin{enumerate}[label=\arabic*., before=\raggedright, itemsep=0pt, topsep=1pt, partopsep=0pt, leftmargin=*]
    \item Policy framework is rigid and inflexible.
    \item Limited flexibility to accommodate new technologies.
    \item Some flexibility but may not be sufficient for all new technologies.
    \item Policy framework is designed to be adaptable to most new technologies.
    \item Highly flexible and adaptable framework that can accommodate a wide range of emerging AI technologies, applications, and sudden changes to the AI landscape of a nation.
\end{enumerate} \\
\hline
15.4. Incident reporting, review and global Monitoring & Existence of mechanisms for reporting AI incidents/failures or events and processes for analyzing these incidents to adapt policy and best practices. &
\begin{enumerate}[label=\arabic*., before=\raggedright, itemsep=0pt, topsep=1pt, partopsep=0pt, leftmargin=*]
    \item No formal mechanisms for internal incident reporting/review or systematic global AI monitoring.
    \item Informal or ad hoc approaches to incident reporting or global monitoring with no defined structure, analysis, or feedback.
    \item Establishes formal but limited/voluntary internal incident reporting or assigns responsibility for basic global monitoring, but lacks systematic analysis and policy feedback integration for either.
    \item Formal internal incident reporting/review process (may be limited in scope) and a structured process for regular global AI monitoring and reporting to inform policy reviews.
    \item Comprehensive, mandatory internal incident reporting, review, and learning system and robust, continuous global AI monitoring with dedicated analysis and a formal, rapid feedback loop for policy/strategy adaptation.
\end{enumerate} \\
\hline
\end{longtable}

\textbf{(16) International Harmonization.}

\begin{longtable}{|>{\raggedright\arraybackslash}p{0.15\textwidth}|>{\raggedright\arraybackslash}p{0.25\textwidth}|>{\raggedright\arraybackslash}p{0.40\textwidth}|}

\hline
\textbf{Sub-criterion} & \textbf{Definition} & \textbf{Indicators} \\
\hline
\endfirsthead

\multicolumn{3}{c}%
{{\tablename\ \thetable{} -- continued from previous page}} \\
\hline
\textbf{Sub-criterion} & \textbf{Definition} & \textbf{Indicators} \\
\hline
\endhead

\hline \multicolumn{3}{|r|}{{Continued on next page}} \\ \hline
\endfoot

\caption{International Harmonization Sub-criteria and Indicators} \label{tab:harmonization_subcriteria}
\endlastfoot

16.1. Alignment with International Standards & Does the policy align with relevant international standards and guidelines for AI governance? &
\begin{enumerate}[label=\arabic*., before=\raggedright, itemsep=0pt, topsep=1pt, partopsep=0pt, leftmargin=*]
    \item No consideration of international standards.
    \item Limited alignment with international standards.
    \item Some alignment with international standards.
    \item Strong alignment with international standards.
    \item Comprehensive alignment with international standards, including active participation in their development.
\end{enumerate} \\
\hline
16.2. Cross-Border Data Flows & Does the policy address the challenges of cross-border data flows and data governance in the context of AI systems? &
\begin{enumerate}[label=\arabic*., before=\raggedright, itemsep=0pt, topsep=1pt, partopsep=0pt, leftmargin=*]
    \item No consideration of cross-border data flows.
    \item Cross-border data flows are mentioned but not addressed in detail.
    \item Some guidelines for managing cross-border data flows.
    \item Clear guidelines and requirements for managing cross-border data flows.
    \item Comprehensive framework for managing cross-border data flows, including provisions for data localization, data sovereignty, and international data sharing agreements.
\end{enumerate} \\
\hline
16.3. International Cooperation & Does the policy promote international cooperation in AI governance, including mechanisms for information sharing, joint research, and coordinated policy development? &
\begin{enumerate}[label=\arabic*., before=\raggedright, itemsep=0pt, topsep=1pt, partopsep=0pt, leftmargin=*]
    \item No provisions for international cooperation.
    \item International cooperation is encouraged but not prioritized.
    \item Some mechanisms for international cooperation are outlined.
    \item Clear commitments to international cooperation, including specific initiatives and partnerships.
    \item Comprehensive framework for international cooperation, including mechanisms for joint research, information sharing, and coordinated policy development.
\end{enumerate} \\
\hline
\end{longtable}

\subsection*{Category 4. User Rights, Protection and Agency. Weight 20\%}
\begin{longtable}{|>{\raggedright\arraybackslash}p{0.2\textwidth}|>{\raggedright\arraybackslash}p{0.47\textwidth}|>{\raggedright\arraybackslash}p{0.08\textwidth}|}

\hline
\textbf{Criterion} & \textbf{Definition} & \textbf{Weight} \\
\hline
\endfirsthead

\multicolumn{3}{c}%
{{\tablename\ \thetable{} -- continued from previous page}} \\
\hline
\textbf{Criterion} & \textbf{Definition} & \textbf{Weight} \\
\hline
\endhead

\hline \multicolumn{3}{|r|}{{Continued on next page}} \\ \hline
\endfoot
\hline
\caption{User Rights, Protection and Agency Criteria} \label{tab:ethical_criteria}
\endlastfoot

17. Fairness and Non-discrimination & The policy addresses bias, fairness, and non-discrimination in AI systems, including provisions for identifying, mitigating, and monitoring bias. & 25\% \\
\hline
18. Transparency and Explainability & The policy promotes transparency and explainability in AI decision-making, enabling affected people to understand relevant outputs or decisions. & 25\% \\
\hline
19. Privacy and Data Protection & The policy includes provisions for protecting privacy and securing personal data used in AI systems under the applicable data protection rules. & 25\% \\
\hline
20. Accountability & The policy establishes responsibility for the development, deployment, and use of AI systems, including mechanisms for liability and redress when appropriate. & 25\% \\
\end{longtable}

\textbf{(17) Fairness and Non-discrimination.}

\begin{longtable}{|>{\raggedright\arraybackslash}p{0.15\textwidth}|>{\raggedright\arraybackslash}p{0.25\textwidth}|>{\raggedright\arraybackslash}p{0.40\textwidth}|}

\hline
\textbf{Sub-criterion} & \textbf{Definition} & \textbf{Indicators} \\
\hline
\endfirsthead

\multicolumn{3}{c}%
{{\tablename\ \thetable{} -- continued from previous page}} \\
\hline
\textbf{Sub-criterion} & \textbf{Definition} & \textbf{Indicators} \\
\hline
\endhead

\hline \multicolumn{3}{|r|}{{Continued on next page}} \\ \hline
\endfoot

\caption{Fairness and Non-discrimination Sub-criteria and Indicators} \label{tab:fairness_subcriteria}
\endlastfoot

17.1. Bias Detection and Mitigation & Does the policy require or encourage the use of methods for detecting and mitigating bias in AI systems? &
\begin{enumerate}[label=\arabic*., before=\raggedright, itemsep=0pt, topsep=1pt, partopsep=0pt, leftmargin=*]
    \item No consideration of bias.
    \item Bias is mentioned but not addressed in detail.
    \item Some guidelines for bias detection and mitigation.
    \item Clear guidelines and requirements for bias detection and mitigation.
    \item Comprehensive framework for bias detection and mitigation, including specific methodologies and metrics.
\end{enumerate} \\
\hline
17.2. Protected Groups & Does the policy explicitly address the potential impact of AI systems on protected groups (e.g., based on race, gender, religion)? &
\begin{enumerate}[label=\arabic*., before=\raggedright, itemsep=0pt, topsep=1pt, partopsep=0pt, leftmargin=*]
    \item No consideration of protected groups.
    \item Protected groups are mentioned but not addressed in detail.
    \item Some guidelines for protecting the rights of protected groups.
    \item Clear guidelines and requirements for protecting the rights of protected groups.
    \item Comprehensive framework for protecting the rights of protected groups, including provisions for auditing and redress.
\end{enumerate} \\
\hline
\end{longtable}

\textbf{(18) Transparency and Explainability.}

\begin{longtable}{|>{\raggedright\arraybackslash}p{0.15\textwidth}|>{\raggedright\arraybackslash}p{0.25\textwidth}|>{\raggedright\arraybackslash}p{0.40\textwidth}|}

\hline
\textbf{Sub-criterion} & \textbf{Definition} & \textbf{Indicators} \\
\hline
\endfirsthead

\multicolumn{3}{c}%
{{\tablename\ \thetable{} -- continued from previous page}} \\
\hline
\textbf{Sub-criterion} & \textbf{Definition} & \textbf{Indicators} \\
\hline
\endhead

\hline \multicolumn{3}{|r|}{{Continued on next page}} \\ \hline
\endfoot

\caption{Transparency and Explainability Sub-criteria and Indicators} \label{tab:transparency_subcriteria}
\endlastfoot

18.1. Transparency Requirements & Does the policy specify transparency requirements for AI systems, including documentation of data, algorithms, and decision-making processes? &
\begin{enumerate}[label=\arabic*., before=\raggedright, itemsep=0pt, topsep=1pt, partopsep=0pt, leftmargin=*]
    \item No transparency requirements.
    \item Transparency is encouraged but not clearly defined.
    \item Some transparency requirements are outlined.
    \item Clear and specific transparency requirements are specified.
    \item Comprehensive transparency requirements, including provisions for data provenance, model documentation, and decision-making logic.
\end{enumerate} \\
\hline
18.2. Explainability Methods & Does the policy encourage or mandate the use of explainability methods to make AI systems more understandable to humans? &
\begin{enumerate}[label=\arabic*., before=\raggedright, itemsep=0pt, topsep=1pt, partopsep=0pt, leftmargin=*]
    \item No consideration of explainability.
    \item Explainability is mentioned but not addressed in detail.
    \item Some guidelines for using explainability methods.
    \item Clear guidelines and recommendations for using explainability methods.
    \item Comprehensive framework for explainability, including provisions for different types of explanations (e.g., local, global) and different stakeholder needs.
\end{enumerate} \\
\hline
18.3. Interpretability for Non-Experts & Does the policy consider the need for interpretability by non-experts, including lay users and affected communities? &
\begin{enumerate}[label=\arabic*., before=\raggedright, itemsep=0pt, topsep=1pt, partopsep=0pt, leftmargin=*]
    \item No consideration of interpretability for non-experts.
    \item Interpretability for non-experts is mentioned but not prioritized.
    \item Some guidelines for making AI systems understandable to non-experts.
    \item Clear guidelines and requirements for interpretability by non-experts.
    \item Comprehensive framework for interpretability by non-experts, including provisions for user-friendly explanations and interfaces.
\end{enumerate} \\
\hline
\end{longtable}

\textbf{(19) Privacy and Data Protection.}

\begin{longtable}{|>{\raggedright\arraybackslash}p{0.15\textwidth}|>{\raggedright\arraybackslash}p{0.25\textwidth}|>{\raggedright\arraybackslash}p{0.40\textwidth}|}

\hline
\textbf{Sub-criterion} & \textbf{Definition} & \textbf{Indicators} \\
\hline
\endfirsthead

\multicolumn{3}{c}%
{{\tablename\ \thetable{} -- continued from previous page}} \\
\hline
\textbf{Sub-criterion} & \textbf{Definition} & \textbf{Indicators} \\
\hline
\endhead

\hline \multicolumn{3}{|r|}{{Continued on next page}} \\ \hline
\endfoot

\caption{Privacy and Data Protection Sub-criteria and Indicators} \label{tab:privacy_subcriteria}
\endlastfoot

19.1. Data Minimization and Purpose Limitation & Does the policy promote data minimization principles, limiting the collection and use of personal data to what is strictly necessary for the intended purpose? &
\begin{enumerate}[label=\arabic*., before=\raggedright, itemsep=0pt, topsep=1pt, partopsep=0pt, leftmargin=*]
    \item No consideration of data minimization.
    \item Data minimization is encouraged but not clearly defined.
    \item Some guidelines for data minimization.
    \item Clear requirements for data minimization and purpose limitation.
    \item Comprehensive framework for data minimization and purpose limitation, including provisions for data deletion and retention policies.
\end{enumerate} \\
\hline
19.2. Data Security & Does the policy mandate appropriate security measures to protect personal data from unauthorized access, use, or disclosure? &
\begin{enumerate}[label=\arabic*., before=\raggedright, itemsep=0pt, topsep=1pt, partopsep=0pt, leftmargin=*]
    \item No specific security measures mentioned.
    \item Security measures are encouraged but not defined.
    \item Some security measures are outlined.
    \item Clear and specific security requirements are specified.
    \item Comprehensive security requirements, including encryption, access controls, and regular security audits.
\end{enumerate} \\
\hline
19.3. Consent and Individual Rights & Does the policy ensure that individuals have meaningful control over their personal data, including the right to access, rectify, erase, and restrict processing? &
\begin{enumerate}[label=\arabic*., before=\raggedright, itemsep=0pt, topsep=1pt, partopsep=0pt, leftmargin=*]
    \item No consideration of individual rights.
    \item Individual rights are mentioned but not fully addressed.
    \item Some provisions for individual rights.
    \item Clear mechanisms for individuals to exercise their rights.
    \item Comprehensive framework for individual rights, including provisions for informed consent, data portability, and the right to be forgotten.
\end{enumerate} \\
\hline
\end{longtable}

\textbf{(20) Accountability.}

\begin{longtable}{|>{\raggedright\arraybackslash}p{0.15\textwidth}|>{\raggedright\arraybackslash}p{0.25\textwidth}|>{\raggedright\arraybackslash}p{0.40\textwidth}|}

\hline
\textbf{Sub-criterion} & \textbf{Definition} & \textbf{Indicators} \\
\hline
\endfirsthead

\multicolumn{3}{c}%
{{\tablename\ \thetable{} -- continued from previous page}} \\
\hline
\textbf{Sub-criterion} & \textbf{Definition} & \textbf{Indicators} \\
\hline
\endhead

\hline \multicolumn{3}{|r|}{{Continued on next page}} \\ \hline
\endfoot

\caption{Accountability Sub-criteria and Indicators} \label{tab:accountability_subcriteria}
\endlastfoot

20.1. Responsibility Allocation & Does the policy clearly define the responsibilities of different actors in the AI lifecycle (e.g., developers, deployers, users)? &
\begin{enumerate}[label=\arabic*., before=\raggedright, itemsep=0pt, topsep=1pt, partopsep=0pt, leftmargin=*]
    \item Responsibilities are not defined.
    \item Responsibilities are vaguely mentioned.
    \item Some responsibilities are defined, but gaps remain.
    \item Most responsibilities are clearly defined.
    \item Comprehensive definition of responsibilities for all relevant actors.
\end{enumerate} \\
\hline
20.2. Liability Framework & Does the policy establish a liability framework for determining who is liable when AI systems cause harm or errors? &
\begin{enumerate}[label=\arabic*., before=\raggedright, itemsep=0pt, topsep=1pt, partopsep=0pt, leftmargin=*]
    \item No liability framework established.
    \item Liability framework is unclear or inadequate.
    \item Some aspects of liability are addressed.
    \item Clear liability framework for most scenarios.
    \item Comprehensive liability framework that addresses various types of harm and different levels of AI autonomy.
\end{enumerate} \\
\hline
20.3. Redress Mechanisms & Does the policy provide mechanisms for individuals or groups to seek redress for harm caused by AI systems? &
\begin{enumerate}[label=\arabic*., before=\raggedright, itemsep=0pt, topsep=1pt, partopsep=0pt, leftmargin=*]
    \item No redress mechanisms provided.
    \item Redress mechanisms are limited or difficult to access.
    \item Some redress mechanisms are available.
    \item Clear and accessible redress mechanisms are established.
    \item Comprehensive redress mechanisms, including provisions for compensation, remediation, and appeal.
\end{enumerate} \\
\hline
\end{longtable}

\subsection*{Category 5. Socioeconomic Impact and Innovation. Weight 20\%}
\begin{longtable}{|>{\raggedright\arraybackslash}p{0.2\textwidth}|>{\raggedright\arraybackslash}p{0.47\textwidth}|>{\raggedright\arraybackslash}p{0.08\textwidth}|}

\hline
\textbf{Criterion} & \textbf{Definition} & \textbf{Weight} \\
\hline
\endfirsthead

\multicolumn{3}{c}%
{{\tablename\ \thetable{} -- continued from previous page}} \\
\hline
\textbf{Criterion} & \textbf{Definition} & \textbf{Weight} \\
\hline
\endhead

\hline \multicolumn{3}{|r|}{{Continued on next page}} \\ \hline
\endfoot

\caption{Socioeconomic Impact and Innovation Criteria} \label{tab:socioeconomic_criteria}
\endlastfoot

21. Societal Benefit & The policy promotes the development and use of AI systems that provide public benefits in areas such as healthcare, education, environmental sustainability, and economic activity. & 20\% \\
\hline
22. Innovation & The policy fosters responsible AI research, development, and deployment while addressing relevant risks. & 20\% \\
\hline
23. Economic Growth & The policy supports economic development through AI adoption while addressing challenges such as job displacement and market disruption. & 20\% \\
\hline
24. Social Equity & The policy addresses how AI benefits and harms are distributed, including effects on marginalized or vulnerable groups. & 20\% \\
\hline
25. Public Trust & The policy supports warranted public trust through transparency, accountability, public education, and opportunities for participation. & 20\% \\
\hline
\end{longtable}

\textbf{(21) Societal Benefit. }

\begin{longtable}{|>{\raggedright\arraybackslash}p{0.15\textwidth}|>{\raggedright\arraybackslash}p{0.25\textwidth}|>{\raggedright\arraybackslash}p{0.40\textwidth}|}

\hline
\textbf{Sub-criterion} & \textbf{Definition} & \textbf{Indicators} \\
\hline
\endfirsthead

\multicolumn{3}{c}%
{{\tablename\ \thetable{} -- continued from previous page}} \\
\hline
\textbf{Sub-criterion} & \textbf{Definition} & \textbf{Indicators} \\
\hline
\endhead

\hline \multicolumn{3}{|r|}{{Continued on next page}} \\ \hline
\endfoot

\caption{Societal Benefit Sub-criteria and Indicators} \label{tab:societal_benefit_subcriteria}
\endlastfoot

21.1. Identification of Beneficial Applications & Does the policy identify specific areas where AI can provide significant societal benefits? &
\begin{enumerate}[label=\arabic*., before=\raggedright, itemsep=0pt, topsep=1pt, partopsep=0pt, leftmargin=*]
    \item No identification of beneficial applications.
    \item Beneficial applications are mentioned but not prioritized.
    \item Some areas of societal benefit are identified.
    \item Clear identification of key areas where AI can provide significant benefits.
    \item Comprehensive analysis of potential societal benefits across various sectors.
\end{enumerate} \\
\hline
21.2. Incentives for Beneficial AI & Does the policy provide incentives (e.g., funding, grants, recognition) for the development and deployment of AI systems that address societal needs? &
\begin{enumerate}[label=\arabic*., before=\raggedright, itemsep=0pt, topsep=1pt, partopsep=0pt, leftmargin=*]
    \item No incentives for beneficial AI.
    \item Incentives are weak or unclear.
    \item Some incentives for beneficial AI are provided.
    \item Clear and meaningful incentives for beneficial AI.
    \item Comprehensive set of incentives that effectively promote the development and deployment of beneficial AI.
\end{enumerate} \\
\hline
21.3. Measurement of Societal Impact & Does the policy include mechanisms for measuring and evaluating the societal impact of AI systems? &
\begin{enumerate}[label=\arabic*., before=\raggedright, itemsep=0pt, topsep=1pt, partopsep=0pt, leftmargin=*]
    \item No mechanisms for measuring societal impact.
    \item Societal impact is mentioned but not measured.
    \item Some metrics for measuring societal impact are proposed.
    \item Clear metrics for measuring societal impact.
    \item Comprehensive framework for measuring and evaluating the societal impact of AI, including both positive and negative impacts.
\end{enumerate} \\
\hline
\end{longtable}

\textbf{(22) Innovation. }

\begin{longtable}{|>{\raggedright\arraybackslash}p{0.15\textwidth}|>{\raggedright\arraybackslash}p{0.25\textwidth}|>{\raggedright\arraybackslash}p{0.40\textwidth}|}

\hline
\textbf{Sub-criterion} & \textbf{Definition} & \textbf{Indicators} \\
\hline
\endfirsthead

\multicolumn{3}{c}%
{{\tablename\ \thetable{} -- continued from previous page}} \\
\hline
\textbf{Sub-criterion} & \textbf{Definition} & \textbf{Indicators} \\
\hline
\endhead

\hline \multicolumn{3}{|r|}{{Continued on next page}} \\ \hline
\endfoot

\caption{Innovation Sub-criteria and Indicators} \label{tab:innovation_subcriteria}
\endlastfoot

22.1. Support for Research and Development & Does the policy provide support for research and development in AI, including funding, infrastructure, and talent development? &
\begin{enumerate}[label=\arabic*., before=\raggedright, itemsep=0pt, topsep=1pt, partopsep=0pt, leftmargin=*]
    \item No support for AI research and development.
    \item Limited support for AI research and development.
    \item Some funding and support for AI research and development.
    \item Significant funding and support for AI research and development.
    \item Comprehensive support for AI research and development, including long-term investments and strategic initiatives.
\end{enumerate} \\
\hline
22.2. Regulatory Sandboxes and Testbeds & Does the policy promote the use of regulatory sandboxes or testbeds to experiment with new AI technologies and applications in a controlled environment? &
\begin{enumerate}[label=\arabic*., before=\raggedright, itemsep=0pt, topsep=1pt, partopsep=0pt, leftmargin=*]
    \item No mention of regulatory sandboxes or testbeds.
    \item Sandboxes or testbeds are mentioned but not clearly defined.
    \item Some provisions for regulatory sandboxes or testbeds.
    \item Clear guidelines and procedures for establishing and operating regulatory sandboxes or testbeds.
    \item Comprehensive framework for regulatory sandboxes or testbeds, including provisions for data sharing, risk assessment, and stakeholder engagement.
\end{enumerate} \\
\hline
22.3. Innovation-Friendly Regulation & Is the policy framework designed to be flexible and adaptable to accommodate new AI innovations without stifling their development? &
\begin{enumerate}[label=\arabic*., before=\raggedright, itemsep=0pt, topsep=1pt, partopsep=0pt, leftmargin=*]
    \item Policy framework is rigid and inflexible, hindering innovation.
    \item Limited flexibility to accommodate new innovations.
    \item Some flexibility but may not be sufficient for all new innovations.
    \item Policy framework is designed to be adaptable to most new innovations.
    \item Highly flexible and adaptable framework that encourages responsible innovation while mitigating risks.
\end{enumerate} \\
\hline
\end{longtable}

\textbf{(23) Economic Growth. }

\begin{longtable}{|>{\raggedright\arraybackslash}p{0.15\textwidth}|>{\raggedright\arraybackslash}p{0.25\textwidth}|>{\raggedright\arraybackslash}p{0.40\textwidth}|}

\hline
\textbf{Sub-criterion} & \textbf{Definition} & \textbf{Indicators} \\
\hline
\endfirsthead

\multicolumn{3}{c}%
{{\tablename\ \thetable{} -- continued from previous page}} \\
\hline
\textbf{Sub-criterion} & \textbf{Definition} & \textbf{Indicators} \\
\hline
\endhead

\hline \multicolumn{3}{|r|}{{Continued on next page}} \\ \hline
\endfoot

\caption{Economic Growth Sub-criteria and Indicators} \label{tab:economic_growth_subcriteria}
\endlastfoot

23.1. Investment in AI Ecosystem & Does the policy promote investment in the AI ecosystem, including startups, businesses, and research institutions? &
\begin{enumerate}[label=\arabic*., before=\raggedright, itemsep=0pt, topsep=1pt, partopsep=0pt, leftmargin=*]
    \item No provisions for promoting investment in the AI ecosystem.
    \item Investment is encouraged but not prioritized.
    \item Some incentives for investing in the AI ecosystem.
    \item Clear incentives and support for investment in the AI ecosystem.
    \item Comprehensive strategy for promoting investment in the AI ecosystem, including tax breaks, grants, and venture capital initiatives.
\end{enumerate} \\
\hline
23.2. Workforce Development & Does the policy address the need for workforce development and retraining to prepare workers for the changing job market due to AI adoption? &
\begin{enumerate}[label=\arabic*., before=\raggedright, itemsep=0pt, topsep=1pt, partopsep=0pt, leftmargin=*]
    \item No consideration of workforce development.
    \item Workforce development is mentioned but not addressed in detail.
    \item Some guidelines for workforce development and retraining.
    \item Clear guidelines and programs for workforce development and retraining.
    \item Comprehensive framework for workforce development, including provisions for education, training, and job placement.
\end{enumerate} \\
\hline
23.3. Market Competitiveness & Does the policy promote fair competition in the AI market, preventing monopolies and ensuring a level playing field for businesses of all sizes? &
\begin{enumerate}[label=\arabic*., before=\raggedright, itemsep=0pt, topsep=1pt, partopsep=0pt, leftmargin=*]
    \item No consideration of market competitiveness.
    \item Market competitiveness is mentioned but not addressed in detail.
    \item Some guidelines for promoting fair competition.
    \item Clear guidelines and regulations to prevent monopolies and promote fair competition.
    \item Comprehensive framework for ensuring market competitiveness, including antitrust measures and support for small and medium-sized enterprises.
\end{enumerate} \\
\hline
\end{longtable}

\textbf{(24) Social Equity. }

\begin{longtable}{|>{\raggedright\arraybackslash}p{0.15\textwidth}|>{\raggedright\arraybackslash}p{0.25\textwidth}|>{\raggedright\arraybackslash}p{0.40\textwidth}|}

\hline
\textbf{Sub-criterion} & \textbf{Definition} & \textbf{Indicators} \\
\hline
\endfirsthead

\multicolumn{3}{c}%
{{\tablename\ \thetable{} -- continued from previous page}} \\
\hline
\textbf{Sub-criterion} & \textbf{Definition} & \textbf{Indicators} \\
\hline
\endhead

\hline \multicolumn{3}{|r|}{{Continued on next page}} \\ \hline
\endfoot

\caption{Social Equity Sub-criteria and Indicators} \label{tab:social_equity_subcriteria}
\endlastfoot

24.1. Access to AI Benefits & Does the policy promote equitable access to the benefits of AI technologies, regardless of socioeconomic status, geographic location, or other factors? &
\begin{enumerate}[label=\arabic*., before=\raggedright, itemsep=0pt, topsep=1pt, partopsep=0pt, leftmargin=*]
    \item No consideration of equitable access to AI benefits.
    \item Equitable access is mentioned but not prioritized.
    \item Some initiatives to promote equitable access.
    \item Clear guidelines and programs to ensure equitable access.
    \item Comprehensive framework for promoting equitable access to AI benefits, including targeted interventions for marginalized communities.
\end{enumerate} \\
\hline
24.2. Social Safety Nets & Does the policy consider the need for social safety nets like Universal Basic Income (UBI) to support individuals and communities affected by job displacement or other negative consequences of AI adoption? &
\begin{enumerate}[label=\arabic*., before=\raggedright, itemsep=0pt, topsep=1pt, partopsep=0pt, leftmargin=*]
    \item No consideration of social safety nets.
    \item Social safety nets are mentioned but not addressed in detail.
    \item Some provisions for social safety nets.
    \item Clear guidelines and programs for social safety nets.
    \item Comprehensive framework for social safety nets, including provisions for retraining, job placement, and income support (UBI).
\end{enumerate} \\
\hline
24.3. Digital Divide & Does the policy address the digital divide and ensure that access to AI technologies and their benefits is not limited by factors such as internet access, digital literacy, or device availability? &
\begin{enumerate}[label=\arabic*., before=\raggedright, itemsep=0pt, topsep=1pt, partopsep=0pt, leftmargin=*]
    \item No consideration of the digital divide.
    \item Digital divide is mentioned but not prioritized.
    \item Some initiatives to address the digital divide.
    \item Clear guidelines and programs to bridge the digital divide.
    \item Comprehensive framework for addressing the digital divide, including provisions for infrastructure development, digital literacy training, and affordable access to devices and internet.
\end{enumerate} \\
\hline
\end{longtable}

\textbf{(25) Public Trust. }

\begin{longtable}{|>{\raggedright\arraybackslash}p{0.15\textwidth}|>{\raggedright\arraybackslash}p{0.25\textwidth}|>{\raggedright\arraybackslash}p{0.40\textwidth}|}

\hline
\textbf{Sub-criterion} & \textbf{Definition} & \textbf{Indicators} \\
\hline
\endfirsthead

\multicolumn{3}{c}%
{{\tablename\ \thetable{} -- continued from previous page}} \\
\hline
\textbf{Sub-criterion} & \textbf{Definition} & \textbf{Indicators} \\
\hline
\endhead

\hline \multicolumn{3}{|r|}{{Continued on next page}} \\ \hline
\endfoot
\hline
\caption{Public Trust Sub-criteria and Indicators} \label{tab:public_trust_subcriteria}
\endlastfoot

25.1. Public Awareness and Education & Does the policy include initiatives to raise public awareness and understanding of AI technologies, their potential benefits and risks? &
\begin{enumerate}[label=\arabic*., before=\raggedright, itemsep=0pt, topsep=1pt, partopsep=0pt, leftmargin=*]
    \item No provisions for public awareness or education.
    \item Public awareness is mentioned but not prioritized.
    \item Some initiatives to raise public awareness.
    \item Clear guidelines and programs for public education on AI.
    \item Comprehensive public awareness and education campaigns, including resources for different stakeholder groups.
\end{enumerate} \\
\hline
25.2. Public Engagement & Does the policy provide opportunities for public engagement and consultation in the development and implementation of AI policies? &
\begin{enumerate}[label=\arabic*., before=\raggedright, itemsep=0pt, topsep=1pt, partopsep=0pt, leftmargin=*]
    \item No mechanisms for public engagement.
    \item Limited opportunities for public input.
    \item Some mechanisms for public consultation.
    \item Well-defined mechanisms for public engagement.
    \item Multiple, robust mechanisms for public engagement, ensuring diverse perspectives are considered in policy-making.
\end{enumerate} \\
\hline
25.3. Trust-Building Measures & Does the policy promote trust-building measures, such as transparency, explainability, and accountability, in the design and deployment of AI systems? &
\begin{enumerate}[label=\arabic*., before=\raggedright, itemsep=0pt, topsep=1pt, partopsep=0pt, leftmargin=*]
    \item No explicit trust-building measures.
    \item Trust is mentioned but not addressed in detail.
    \item Some guidelines for promoting trust.
    \item Clear requirements for transparency, explainability, and accountability.
    \item Comprehensive framework for building and maintaining public trust in AI, including independent audits, certification schemes, and redress mechanisms.
\end{enumerate} \\

\end{longtable}

\normalsize
\fussy
\twocolumn
\bibliography{aaai25}

\end{document}